\documentstyle[galley,epsfig]{mn2e}
\title{Ba star}
\title[Ba star \thanks]
{Characterizing the companion AGBs using surface chemical composition of barium stars \thanks{Based on data collected using HCT/HESP, UVES and FEROS}}

\author[J. Shejeelammal et al.]{J. Shejeelammal$^{1}$, Aruna Goswami$^{1}$, Partha Pratim Goswami$^{1}$, 
\newauthor Rajeev Singh Rathour$^{1,2}$, Thomas Masseron$^{3,4}$  \\
$^{1}$Indian Institute of Astrophysics, Koramangala, Bangalore 560034,
India;  aruna@iiap.res.in\\ 
$^{2}$ Indian Institute of Science Education and Research, Pune, Maharashtra 411008, India\\
$^{3}$ Instituto de Astrofisica de Canarias, E-38205 La Laguna, Tenerife, Spain \\
$^{4}$ Departamento de Astrofisica, Universidad de La Laguna, E-38206 La Laguna, Tenerife, Spain. 
}

\begin{document}

\date{ Accepted ---;  Received ---;  in original
form --- \large \bf }

\pagerange{\pageref{firstpage}--\pageref{lastpage}} \pubyear{2014}

\maketitle


\begin{abstract}
Barium stars are one of the important probes to understand the origin 
and evolution of slow neutron-capture process elements in the Galaxy.
These are extrinsic stars, where the observed s-process element 
abundances are believed to have an origin in the now invisible 
companions that produced these elements  at their Asymptotic Giant 
Branch phase  of  evolution. We have attempted to understand the 
s-process nucleosynthesis, as well  as the physical properties 
of the companion stars through a detailed  comparison  of observed 
elemental abundances of 10 barium stars with the  predictions from 
AGB nucleosynthesis models, FRUITY. For these stars, we  have presented 
estimates of abundances of several elements, C, N, O, Na, Al, 
$\alpha$-elements, Fe-peak elements and neutron-capture elements 
Rb, Sr, Y, Zr, Ba, La, Ce, Pr, Nd, Sm and Eu. The abundance estimates
are based on high resolution spectral analysis. Observations of Rb in 
four of these stars have allowed us to put a limit to the mass of 
the companion AGB stars. Our analysis clearly shows that the former 
companions responsible for the surface abundance peculiarities of these  
stars are low-mass AGB stars. Kinematic analysis have shown  the stars 
to be  members of  Galactic disk population.  
\end{abstract}


\section{Introduction}
Understanding the nucleosynthesis and evolution of Asymptotic Giant 
Branch (AGB) stars are of  primary importance as they are the major 
factories of some key elements in the Universe (Busso et al. 
1999, Herwig 2005). They are the predominant sites for the slow 
neutron-capture nucleosyntesis, and major contributors of elements heavier 
than iron; upto half of all the heavy elements are produced through 
s-process (Busso et al. 1999). There are certain isotopes like 
$^{86}$Sr, $^{96}$Mo, $^{104}$Pd, $^{116}$Sn etc., which are known to be
produced only through the s-process.  It has been estimated that a third 
of the total carbon content in the Galaxy is  produced in AGB stars, which 
is about the same amount as produced in CCSNe and Wolf-Rayet stars 
(Dray et al. 2003). Besides these, the intermediate-mass AGB stars are 
the major producers of $^{14}$N in the Galaxy (Henry et al. 2000, 
Merle et al. 2016). 

The exact physical conditions and nucleosynthetic processes occuring at the
interior of AGB stars are not clearly understood that hinders a better 
understanding of the contribution of these stars to the Galactic chemical
enrichment. This demands a need for detailed chemical composition studies
for an extended sample of AGB stars. However, the spectra of the AGB 
stars are complicated as it is overwhelmed  with the molecular 
contributions arising due to their low photospheric temperature.
This makes the derivation of exact elemental abundance difficult.   
In this regard, the extrinsic stars, which are known to have received 
products of AGB phase of evolution via binary mass transfer mechanisms, 
form  vital tools to trace the AGB nucleosynthesis. The important classes 
of such extrinsic stars are barium stars as the analysis of their generally
hotter spectra is more accurate  (Bidelman \& Keenan 1951), 
CH stars (Keenan 1942) and CEMP-s stars  (Beers \& Christlieb 2005). 
Most of them are radial 
velocity variables (McClure et al. 1980, McClure 1983, 1984, 
McClure \& Woodsworth 1990, Udry et al. 1998a,b,  Lucatello et al. 2005)
associated with a now invisible white dwarf companion.

Detailed studies on  barium  stars include Allen \& Barbuy (2006a), 
Smiljanic et al. (2007), de Castro et al. (2016),  Yang et al. (2016) and 
many others. However, these studies  have not included abundances of 
several heavy elements such as Rb for all the stars  and also for C, N 
and O. In this work, we have undertaken to  carry out a detailed 
spectroscopic analysis for a sample of ten barium/CH star candidates
and  derived whenever possible the  abundances of C, N, O and the neutron 
density dependent [Rb/Zr] abundance ratio  to investigate the neutron 
source in the former companion AGB stars. There are two important 
neutron sources for the s-process in the He  intershell of AGB 
stars:  $^{13}$C($\alpha$, n)$^{16}$O reaction  during the radiative 
inter-pulse period and $^{22}$Ne($\alpha$, n)$^{25}$Mg  reaction 
during the convective thermal pulses. $^{13}$C($\alpha$, n)$^{16}$O 
reaction is the dominant neutron source in low-mass AGB stars with 
initial mass $\leq$ 3 M$_{\odot}$.  The temperature 
T $\geq$ 90 $\times$ 10$^{6}$ K required for the operation of this 
reaction is  provides a neutron density N$_{n}$ $\sim$ 10$^{8}$ cm$^{-3}$ 
in a timescale of $\geq$ 10$^{3}$ years (Straniero et al. 1995, 
Gallino et al. 1998, Goriely \& Mowlavi 2000, Busso et al. 2001).
A temperature 300$\times$10$^{6}$ K, required for the activation 
of $^{22}$Ne source  is achieved during the TPs in 
intermediate-mass AGB stars (initial mass $\geq$ 4 M$_{\odot}$). 
It produces a neutron density N$_{n}$ $\sim$ 10$^{13}$ cm$^{-3}$ 
in a timescale of $\sim$ 10 years. The temperature required for 
the $^{22}$Ne source is reached in low-mass stars during the 
last few TPs providing N$_{n}$ $\sim$ 10$^{10}$ - 10$^{11}$ cm$^{-3}$
(Iben 1975, Busso et al. 2001).  The Rb is produced only when 
the N$_{n}$ $>$ 5$\times$ 10$^{8}$ cm$^{-3}$, otherwise Sr, Y, 
Zr etc. are produced. Hence, the [Rb/Zr] ratio can be used as 
an indicator of mass of AGB stars. We could determine
Rb abundance in four of our program stars; HD~32712, HD~36650,
HD~179832 and HD~211173.

In section 2, we describe the source of the spectra used in this 
study. Section 3 describes  the methodology used for the determination 
of atmospheric parameters, elemental abundances and radial velocities. 
A discussion on the stellar mass  determination is also provided in 
the same section. A comparison of  our result with the literature 
values are presented in section 3.  In section 4, we discuss the 
procedures adopted for the abundance  determination of different 
elements. Section 5 provides a discussion on abundance uncertainties.
Section 6 provides the discussion on the elemental abundance ratios and their 
interpretations based on the existing nucleosynthesis theories. This section 
also provides a comparison of the observational data with the FRUITY models
of Cristallo et al. (2009, 2011, 2015b) and a parametric model
based  analysis.  A discussion on the individual stars are also 
given in the same section.   Conclusions are drawn in section 7.

\section{OBJECT SELECTION, DATA ACQUISITION AND DATA REDUCTION}
\begin{figure}
\centering
\includegraphics[width=\columnwidth]{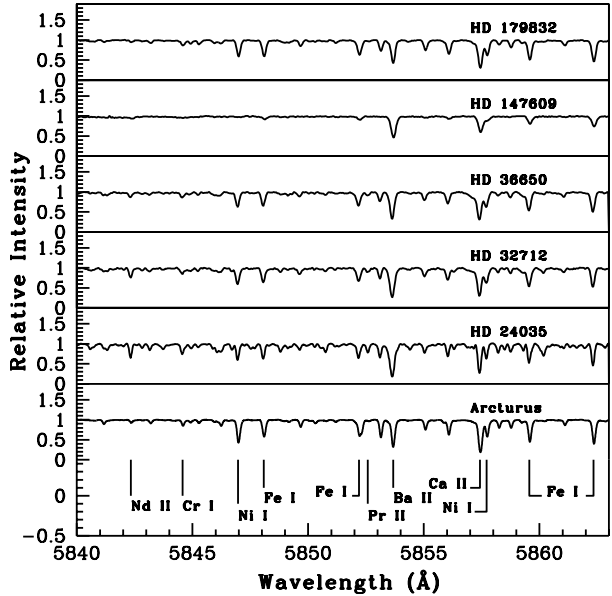}
\caption{ Sample spectra of the program stars in the  wavelength region 
5840 to 5863 {\bf  {\rm \AA}}.}\label{sample_spectra}
\end{figure}

 The objects analyzed in this study are taken from  the CH star 
catalog of Bartkevicius (1996). Six of them  are also found listed  
in the barium star catalog of L\"u (1991). These stars lie among 
the typical CH stars in the color – magnitude 
((B-V) v/s MV) diagram. The spectra of these 
objects are acquired from three different sources.  For HD~219116, 
HD~154276 and HD~147609, the high resolution spectra 
($\lambda/\delta\lambda \sim 60,000 $) were obtained on October 2015, May 2017 
and June 2017 using the high resolution fiber fed  Hanle Echelle 
SPectrograph (HESP)  attached to the 2m Himalayan Chandra Telescope 
(HCT) at the Indian Astronomical  Observatory, Hanle. The wavelength 
coverage of the HESP spectra spans from  3530 - 9970 {\rm \AA}. The 
Data are reduced following the standard procedures  using various 
tasks in Image Reduction and Analysis Facility 
(IRAF\footnote{IRAF is distributed by the National Optical Astronomical 
Observatories, which is operated by the Association for Universities 
for Research in Astronomy, Inc., under contract to the National 
Science Foundation}) software. For HD~24035 and HD~207585 high 
resolution spectra ($\lambda/\delta\lambda \sim 48,000 $)  are 
obtained with the UVES (Ultraviolet and Visual Echelle Spectrograph) 
of the 8.2m Very Large Telescope (VLT) of ESO at Cerro Paranal, Chile.
A high resolution spectrum of HD~219116 is also obtained from UVES/VLT.
The wavelength coverage of the UVES spectra spans from 3290 - 6650 {\rm \AA}.
For HD~32712, HD~36650, HD~94518, HD~211173 and HD~179832, high 
resolution spectra ($\lambda/\delta\lambda \sim 48,000 $) are obtained 
with the FEROS (Fiber-fed Extended Range Optical Spectrograph) of the 
1.52 m telescope of ESO at La Silla, Chile. The wavelength coverage 
of the FEROS spectra  spans from 3520 - 9200 {\rm \AA}. Basic data of 
the program stars along with the source of spectra are  given in the 
Table \ref{basic data of program stars}. A few sample spectra are 
shown in Figure \ref{sample_spectra}.

{\footnotesize
\begin{table*}
\caption{Basic data for the program stars.}\label{basic data of program stars}
\begin{tabular}{lcccccccccc}
\hline
Star      &RA$(2000)$       &Dec.$(2000)$    &B       &V       &J        &H        &K     &Exposure       &Date of obs.  & Source  \\
          &                 &                &        &        &         &         &      &(seconds)      &              & of spectrum\\
\hline

HD 24035  &03 43 42.53     &$-72$ 36 32.80   &9.74    &8.51    &6.567    &6.043    &5.919 &900            &05/04/2002    &UVES    \\      
HD 32712  &05 01 34.91     &$-$58 31 15.05   &9.71    &8.55    &6.634    &6.054    &5.912 &1200           &11/11/1999    &FEROS   \\
HD 36650  &05 27 42.92     &$-$68 04 27.16   &9.91    &8.79    &6.812    &6.297    &6.190 &1200           &10/11/1999    &FEROS   \\
HD 94518  &10 54 12.20     &$-$31 09 34.58   &8.95    &8.36    &7.182    &6.891    &6.824 &900            &02/01/2000    &FEROS   \\ 
HD 147609 &16 21 51.99     &+27 22 27.19     &9.69    &9.18    &8.211    &8.035    &7.948 &2400(3)        &01/06/2017    &HESP    \\
HD 154276 &17 03 49.15     &+17 11 21.08     &9.80    &9.13    &7.911    &7.624    &7.549 &2400(3)        &06/05/2017    &HESP    \\
HD 179832 &19 16 30.00     &$-$49 13 13.01   &9.46    &8.44    &6.660    &6.163    &6.031 &600            &14/07/2000    &FEROS   \\
HD 207585 &21 50 34.71     &$-$24 11 11.68   &10.50   &9.78    &8.633    &8.341    &8.301 &240            &24/04/2002    &UVES    \\
HD 211173 &22 15 57.01     &$-$31 51 38.52   &9.43    &8.49    &6.810    &6.332    &6.218 &600            &14/07/2000    &FEROS   \\
HD 219116 &23 13 30.24     &$-$17 22 08.71   &10.29   &9.25    &7.602    &7.137    &7.012 &240            &19/05/2002    &UVES    \\
          &                &                 &        &        &         &         &      &2400(3)        &30/10/2015    &HESP    \\
\hline

\end{tabular}

The numbers in the parenthesis with exposures indicate the number of 
frames taken.
\end{table*}
}

\section{STELLAR ATMOSPHERIC PARAMETERS AND RADIAL VELOCITY ESTIMATION}
We have estimated the photometric temperature of the program stars using the 
temperature calibration equations of Alonso et al. (1994, 1996) for dwarfs 
and Alonso et al. (1999, 2001) for giants, and following the detailed
procedures as described in our earlier papers (Goswami et al. 2006, 2016).
We made use of the 2MASS J, H,  K magnitudes taken from SIMBAD 
(Cutri et al. 2003) for this calculation.  The photometric temperature 
estimates had been  used as an initial guess  for deriving the 
spectroscopic effective temperature of each object. 

To determine the stellar atmospheric parameters, we have used a set of 
clean, unblended Fe I and Fe II lines with excitation potential in the 
range 0.0 - 6.0 eV and equivalent width 20 - 180 {\rm m\AA}.
IRAF software is used for the equivalent width measurement. A 
pseudo continuum (at 1) is fitted to the observed spectrum using the 
spline function. The equivalent width is measured for each spectral 
line by fitting a Gaussian profile. An initial 
model atmosphere was selected from the Kurucz grid of model atmosphere 
with no convective overshooting (http://cfaku5.cfa.hardvard.edu/) using 
the photometric temperature estimate and the guess of log g value for 
giants/dwarfs. A final model atmosphere was adopted by the iterative
method from the initially selected one, using the most recent version 
of the radiative transfer code MOOG (Sneden 1973) based on the 
assumptions of Local Thermodynamic Equilibrium (LTE).

\par The effective temperature is determined by the method of excitation 
equilibrium, forcing the slope of the abundances from Fe I lines versus 
excitation potentials of the measured lines to be zero. The micro-turbulent 
velocity at a fixed effective temperature is determined by demanding that 
there will be no dependence of the derived Fe I abundance on the reduced 
equivalent width of the corresponding lines. Micro-turbulent velocity is 
fixed at a value which gives zero slope for the plot of equivalent width 
versus abundances of Fe I lines.  The surface gravity, log\,g value is 
determined by means of ionization balance, that is by forcing the Fe I 
and Fe II lines to produce the same abundance at the selected effective 
temperature and microturbulent velocity. The estimated abundances from 
Fe I to Fe II lines as a function of excitation potential and 
equivalent widths, respectively, are shown in figures that are made 
available as on-line materials. 

 A comparison of our results with the literature values whenever 
available shows a close match well within the error limits.  

\par Radial velocities of the program stars are calculated 
using a set of clean and unblended lines of several elements. 
In Table \ref{atmospheric parameters} we present the derived atmospheric 
parameters and radial velocities of  the program stars. Our radial 
velocity estimates are used to study the kinematic properties of the stars. 
A table giving the results from kinematic analysis is presented in the Appendix (Table \ref{kinematic_analysis}).  
Three objects in our sample,  HD~24035, HD 147609 and HD 207585  
are  confirmed binaries with orbital periods of 377.83  $\pm$ 0.35 days 
(Udry et al. 1998a), 672 $\pm$ 2 days (Escorza et al. 2019), and 
1146 $\pm$ 1.5 days ( Escorza et al. 2019) respectively.  Our estimated radial 
velocity ($-$1.56 km$^{-1}$ ) for HD~24035 is slightly higher than the range 
of radial velocities found in literature  ($-$2.14 to $-$19.81) for this
object.  However, for HD~207585 ($-$65.9 km$^{-1}$) and 
HD~147609 ($-$18.17 $^{-1}$), our estimates fall well within the range of 
velocities available in literature, i.e.,  ($-$52.2 to $-$74.1) 
and ($-$19.2 to $-$11.9) respectively.

{\footnotesize
\begin{table*}
\caption{Derived atmospheric parameters for the program stars.} \label{atmospheric parameters}
\begin{tabular}{lccccccccc}
\hline
Star          &T$\rm_{eff}$  & log g &$\zeta$         & [Fe I/H]          &[Fe II/H]          & V$_{r}$             &  V$_{r}$   \\
              &    (K)    & cgs   &(km s$^{-1}$)   &                   &                   & (km s$^{-1}$)       & (km s$^{-1}$) \\
              & $\pm 100$ & $\pm 0.2$& $\pm 0.2$   &                   &                   &                     &               \\  
\hline
HD 24035      & 4750      & 2.20  & 1.58           & $-$0.51$\pm$0.19  & $-$0.50$\pm$0.16  & $-$1.56$\pm$0.25    & $-$12.51$\pm$0.13\\
HD 32712      & 4550      & 2.53  & 1.24           & $-$0.25$\pm$0.12  & $-$0.25$\pm$0.12  & +10.37$\pm$0.02     & +11.27$\pm$0.16   \\
HD 36650      & 4880      & 2.40  & 1.30           & $-$0.02$\pm$0.12  & $-$0.02$\pm$0.14  & +36.40$\pm$0.19     & +31.52$\pm$0.47  \\
HD 94518      & 5700      & 3.86  & 1.30           & $-$0.55$\pm$0.10  & $-$0.55$\pm$0.12  & +92.20$\pm$0.43     & 92.689$\pm$0.015  \\
HD 147609     & 6350      & 3.50  & 1.55           & $-$0.28$\pm$0.16  & $-$0.28$\pm$0.12  & $-$18.17$\pm$1.47   & $-$17.11$\pm$0.82\\
HD 154276     & 5820      & 4.28  & 0.63           & $-$0.09$\pm$0.13  & $-$0.10$\pm$0.14  & $-$64.17$\pm$1.42   & $-$55.94$\pm$0.17\\
HD 179832     & 4780      & 2.70  & 0.99           & +0.23$\pm$0.04    & +0.22$\pm$0.06    & +6.73$\pm$0.03      & +7.64$\pm$0.13\\
HD 207585     & 5800      & 3.80  & 1.00           & $-$0.38$\pm$0.12  & $-$0.38$\pm$0.11  & $-$65.97$\pm$0.07   & $-$60.10$\pm$1.20  \\
HD 211173     & 4900      & 2.60  & 1.15           & $-$0.17$\pm$0.10  & $-$0.17$\pm$0.09  & $-$27.84$\pm$0.25   & $-$28.19$\pm$0.63\\
HD 219116     & 5050      & 2.50  & 1.59           & $-$0.45$\pm$0.11  & $-$0.44$\pm$0.11  & $-$40.90$\pm$0.25   & $-$11.00$\pm$7.30 \\
\hline
\end{tabular}

In Columns 7 and 8 we present radial velocities from the respective spectra and SIMBAD respectively
\end{table*}
}

\par We have determined the mass of the program stars from their 
location in Hertzsprung-Russell diagram (Girardi et al. 2000 data base 
of evolutionary tracks) using the spectroscopic temperature estimate, 
T$\rm_{eff}$, and the luminosity, log$(L/L_{\odot})$.\\
log (L/L$_{\odot}$)=0.4(M$_{bol\odot}$ - V - 5 - 5log ($\pi$) + A$_{V}$ - BC) \\
The visual magnitudes V are taken from Simbad and the parallaxes $\pi$  
from Gaia DR2 (https://gea.esac.esa.int/archive/). The bolometric 
correction, BC, is calculated using  the empirical calibrations of 
Alonso et al. (1995) for dwarfs and Alonso et al. (1999) for giants. 
The interstellar extinction A$_{V}$ is calculated using the calibration 
equations given in Chen et al. (1998).  From the estimated mass, 
log\,{g} is calculated using \\
log (g/g$_{\odot}$)= log (M/M$_{\odot}$) + 4log (T$_{eff}$/T$_{eff\odot}$) - log (L/L$_{\odot}$)\\
We have adopted the solar values log g$_{\odot}$ = 4.44, T$_{eff\odot}$ = 5770K and M$_{bol\odot}$ = 4.74 mag.

\par
\begin{figure}
\centering
\includegraphics[width=\columnwidth]{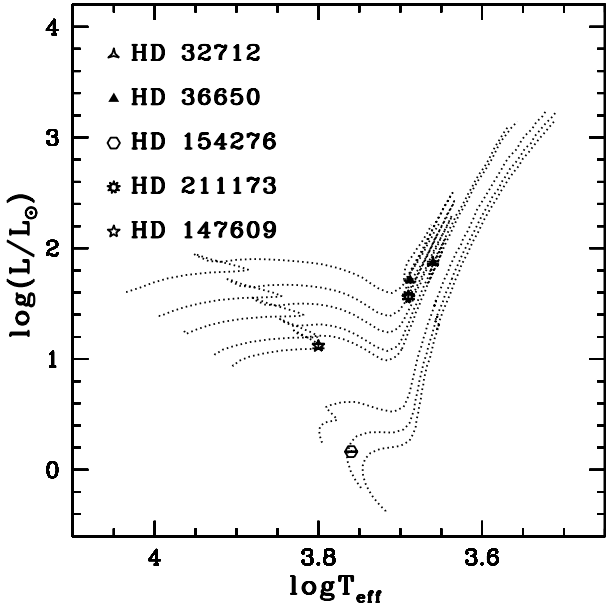}
\caption{The locations of HD~32712, HD~36650, HD~154276, HD~211173 and HD~147609. 
The evolutionary tracks for 0.9, 1.0, 1.2, 1.7, 1.8, 2.0, 2.2, and 2.5 M$_{\odot}$ 
are shown from bottom to top for z = 0.019.} \label{track_019}
\end{figure}

 We have used z = 0.004 tracks for HD~24035 and HD~94518; 
z = 0.008 for HD~207585 and HD~219116;   z = 0.019 for  HD~32712, 
HD~36650, HD~147609, HD~154276 and HD~211173, and  z = 0.030 for 
HD~179832.  As an example, the evolutionary tracks for a few objects 
are shown in  Figure \ref{track_019}. The mass estimates are presented 
in Table \ref{mass age}.

{\footnotesize
\begin{table*}
\caption{Estimates of  log\,{g} and Mass using parallax method} \label{mass age}
\begin{tabular}{lcccccc}
\hline                       
 Star name      & Parallax           & $M_{bol}$       & log(L/L$_{\odot}$) & Mass(M$_{\odot}$) & log g          & log g (spectroscopic)   \\
                & (mas)              &                 &                    &                   & (cgs)          & (cgs)                    \\
\hline
HD 24035        & 4.612$\pm$0.101    & 1.401$\pm$0.05  & 1.339$\pm$0.02     & 0.70$\pm$0.21    & 2.61$\pm$0.02  & 2.20                      \\
HD 32712        & 2.621$\pm$0.026    & 0.081$\pm$0.022 & 1.868$\pm$0.01     & 1.80$\pm$0.26    & 2.41$\pm$0.005 & 2.53                   \\
HD 36650        & 2.655$\pm$0.027    & 0.474$\pm$0.023 & 1.710$\pm$0.01     & 2.20$\pm$0.26    & 2.78$\pm$0.01  & 2.40                    \\
HD 94518        & 13.774$\pm$0.05    & 3.872$\pm$0.01  & 0.351$\pm$0.003    & 0.85$\pm$0.06    & 4.00$\pm$0.005 & 3.86                    \\
HD 147609       & 4.301$\pm$0.107    & 1.955$\pm$0.055 & 1.118$\pm$0.02     & 1.70$\pm$0.05    & 3.71$\pm$0.02  & 3.50                    \\
HD 154276       & 11.554$\pm$0.025   & 4.339$\pm$0.005 & 0.164$\pm$0.002    & 1.00$\pm$0.05    & 4.29$\pm$0.002 & 4.28                    \\
HD 179832       & 2.914$\pm$0.052    & 0.216$\pm$0.041 & 1.814$\pm$0.02     & 2.5$\pm$0.28     & 2.70$\pm$0.02  & 2.70                    \\
HD 207585       & 5.3146$\pm$0.407   & 3.313$\pm$0.165 & 0.575$\pm$0.065    & 1.05$\pm$0.05     & 3.90$\pm$0.045 & 3.80                    \\
HD 211173       & 3.387$\pm$0.066    & 0.839$\pm$0.042 & 1.564$\pm$0.02     & 2.20$\pm$0.24     & 2.93$\pm$0.02  & 2.60                    \\
HD 219116       & 1.584$\pm$0.044    & 0.002$\pm$0.06  & 1.901$\pm$0.02     & 2.35$\pm$0.17     & 2.68$\pm$0.015 & 2.50                     \\
\hline
\end{tabular}
\end{table*}
}

{\footnotesize
\begin{table*}
\caption{Comparison of  estimated stellar parameters with literature values}  \label{Comparison }
\begin{tabular}{lccccccc}
\hline
Star        & T$_{eff}$ & log g &$\zeta$       & [Fe I/H]  & [Fe II/H] & Ref.  \\
            &    (K)    &       &(km s$^{-1}$) &           &           \\
\hline
HD~24035    & 4750      & 2.20  & 1.58         & $-$0.51   & $-$0.50   & 1  \\
            & 4700      & 2.50  & 1.30         & $-$0.23   & $-$0.28   & 2 \\
            & 4500      & 2.00  & -            & $-$0.14   &  -        & 3  \\
HD~32712    & 4550      & 2.53  & 1.24         & $-$0.25   & $-$0.25   & 1  \\
            & 4600      & 2.10  & 1.30         & $-$0.24   & $-$0.25   & 2  \\
HD~36650    & 4880      & 2.40  & 1.30         & $-$0.02   & $-$0.02   & 1  \\
            & 4800      & 2.30  & 1.50         & $-$0.28   & $-$0.28   & 2  \\
HD~94518    & 5700      & 3.86  & 1.30         & $-$0.55   & $-$0.55   & 1   \\
            & 5859      & 4.20  & 4.15         & $-$0.56   & -         & 4   \\
            & 5859      & 4.15  & 1.20         & $-$0.49   & $-$0.50   & 5  \\
            & 5709      & 3.86  & 2.23         & $-$0.84   & -         & 6  \\
HD~147609   & 6350      & 3.50  & 1.55         & $-$0.28   & $-$0.28   & 1  \\
            & 6411      & 3.90  & 1.26         & $-$0.23   & -   & 7  \\
            & 5960      & 3.30  & 1.50         & $-0.45$   & $+0.08$   & 8 \\
            & 6270      & 3.50  & 1.20         &   -       &   -       & 9 \\
            & 6300      & 3.61  & 1.20         &   -       &   -       & 10 \\
HD~154276   & 5820      & 4.28  & 0.63         & $-$0.09   & $-$0.10   & 1 \\
            & 5722      & 4.28  & 0.93         & $-$0.29   & -         & 5 \\
            & 5731      & 4.35  & 1.28         & $-$0.30   & -         & 11 \\
HD~179832   & 4780      & 2.70  & 0.99         & +0.23     & +0.22     & 1\\
HD~207585   & 5800      & 3.80  & 1.00         & $-$0.38   & $-$0.38   & 1  \\
            & 5800      & 4.00  & -            & $-$0.20   & -         & 3  \\ 
            & 5400      & 3.30  & 1.80         & $-$0.57   & -         & 12  \\
            & 5400      & 3.50  & 1.50         & $-$0.50   & -         & 13 \\    
HD~211173   & 4900      & 2.60  & 1.15         & $-$0.17   & $-$0.17   & 1  \\
            & 4800      & 2.50  & -            & $-$0.12   & -         & 3  \\
HD~219116   & 5050      & 2.50  & 1.59         & $-$0.45   & $-$0.44   & 1  \\
            & 4900      & 2.30  & 1.60         & $-$0.61   & $-$0.62   & 2  \\
            & 4800      & 1.80  & -            & $-$0.34   & -         & 3   \\
            & 5300      & 3.50  & 2.00         & $-$0.30   & -         & 14  \\
            & 5300      & 3.50  & -            & $-$0.34   & -         & 15  \\
\hline
\end{tabular}

References: 1. Our work, 2. de Castro et al. (2016), 3. Masseron et al. (2010), 
4. Battistini \& Bensby (2015), 5. Bensby et al. (2014), 6. Axer et al. (1994),
7. Escorza et al. (2019) 8. Allen \& Barbuy (2006a), 9. North et al. (1994a), 
10. Th\'evenin \& Idiart (1999), 11. Ramirez et al. (2013), 12. Luck \& Bond (1991), 
13. Smith \& Lambert (1986a), 14. Smith et al. (1993),  15. Cenarro et al. (2007) \\

\end{table*}
}

\section{ABUNDANCE DETERMINATION}
Abundances of  most of the elements are determined from the measured 
equivalent width of  lines of the neutral and ionized atoms using the 
most recent version of MOOG and the adopted model atmospheres.
Absorption lines corresponding to different elements are identified 
by comparing closely the spectra of program stars with the Doppler 
corrected spectrum of the star Arcturus. The log $gf$ and the lower 
excitation potential values of the lines are taken from the Kurucz 
database of atomic line lists. The equivalent width of the spectral 
lines are measured using various tasks in IRAF. 
A master line list including all the elements was generated.  
For the elements showing hyper-fine splitting and for molecular bands, 
spectrum synthesis of MOOG was used to find the abundances. 
Elements Sc, V, Mn, Co, Cu, Ba, La and Eu are affected by Hyper-fine 
Splitting. The hyper-fine components of Sc and Mn are taken 
from Prochaska \& Mcwilliam 2000, V, Co and Cu from Prochaska et al. 2000,
Ba from Mcwilliam 1998, La from Jonsell et al. 2006, Eu from Worely 
et al. 2013.  All the abundances are found relative to the respective 
solar values (Asplund et al. 2009).

\par The abundance estimates are given in Tables \ref{abundance_table1} 
through \ref{abundance_table3} and the lines used  for the the 
abundance estimation are presented in Tables \ref{linelist1} and 
\ref{linelist2}. The detailed abundance analyses and discussion  
are given in the  section 6.

\begin{figure}
\centering
\includegraphics[width=\columnwidth, height= \columnwidth]{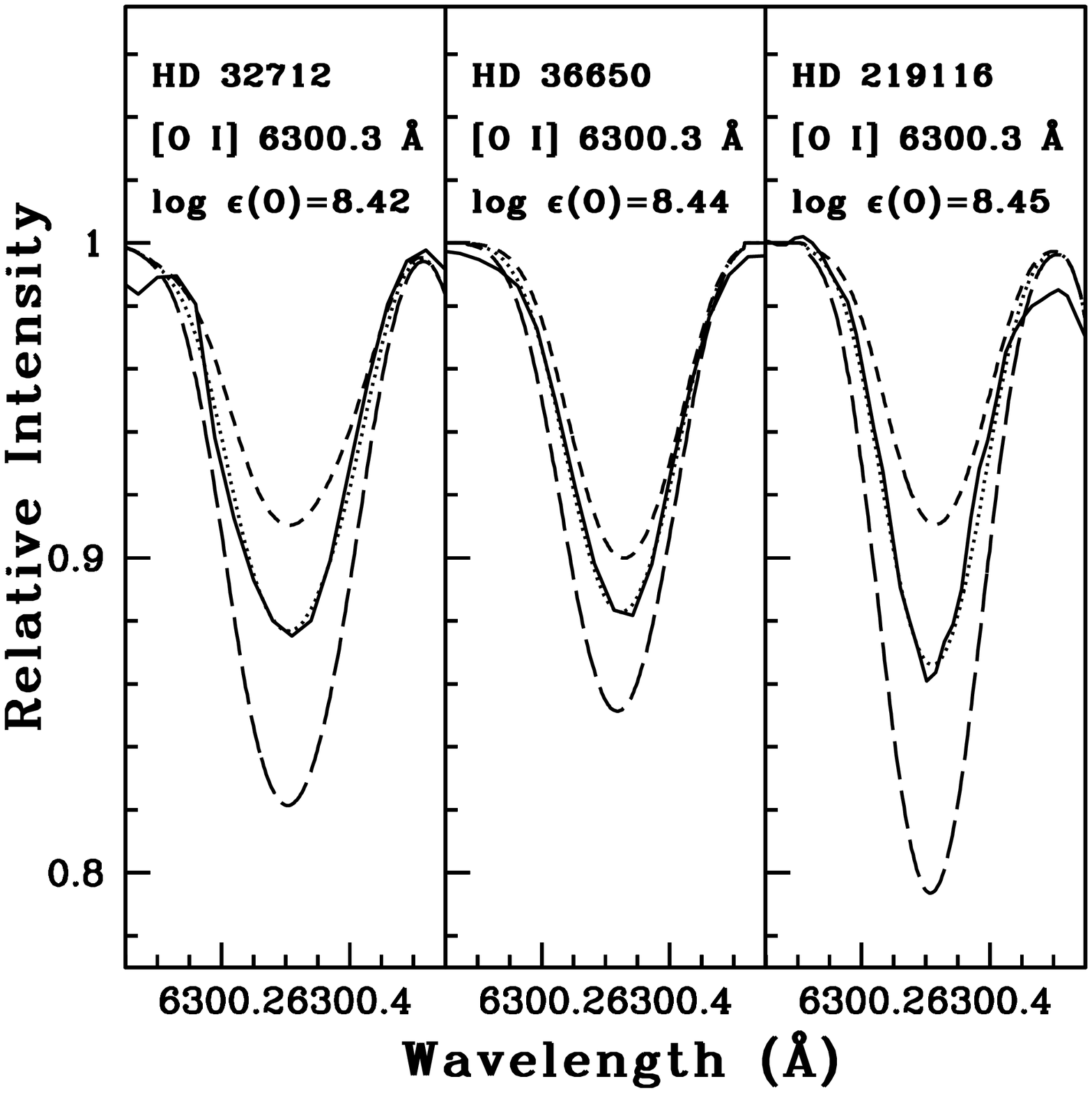}
\includegraphics[width=\columnwidth, height= \columnwidth]{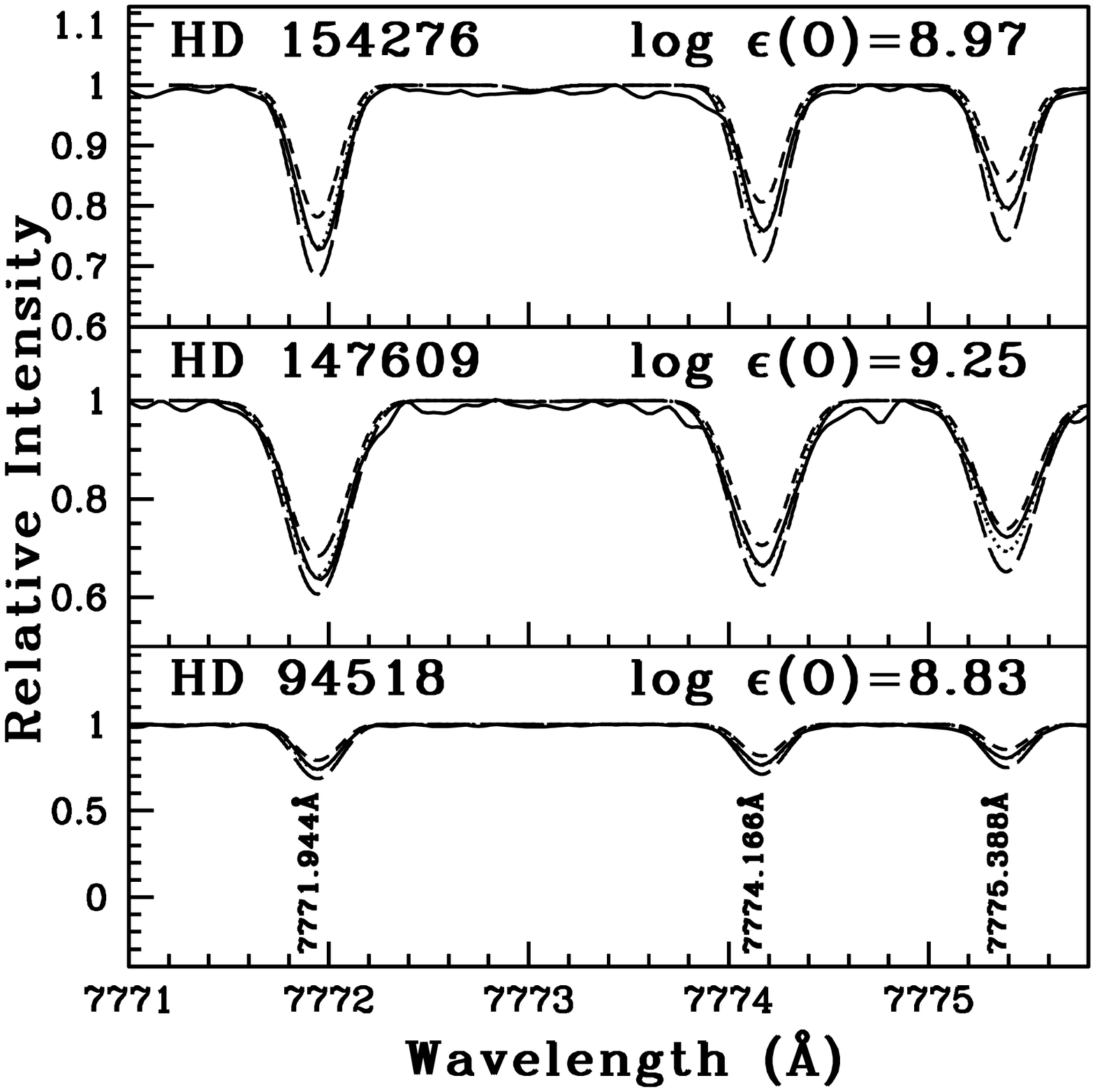}
\caption{ Synthesis of [O I] line around 6300 {\rm \AA} (Top panel) and 
O I triplet around 7770 {\rm \AA} (Bottom panel, LTE abundance estimates). 
Dotted line represents  synthesized spectra and the solid line indicates 
the observed spectra.  Short dashed line represents the synthetic 
spectra corresponding to $\Delta$[O/Fe] = $-$0.3 and long dashed line 
is corresponding  to $\Delta$[O/Fe] = +0.3} \label{OI_synth}
\end{figure}

\begin{figure}
\centering
\includegraphics[width=\columnwidth]{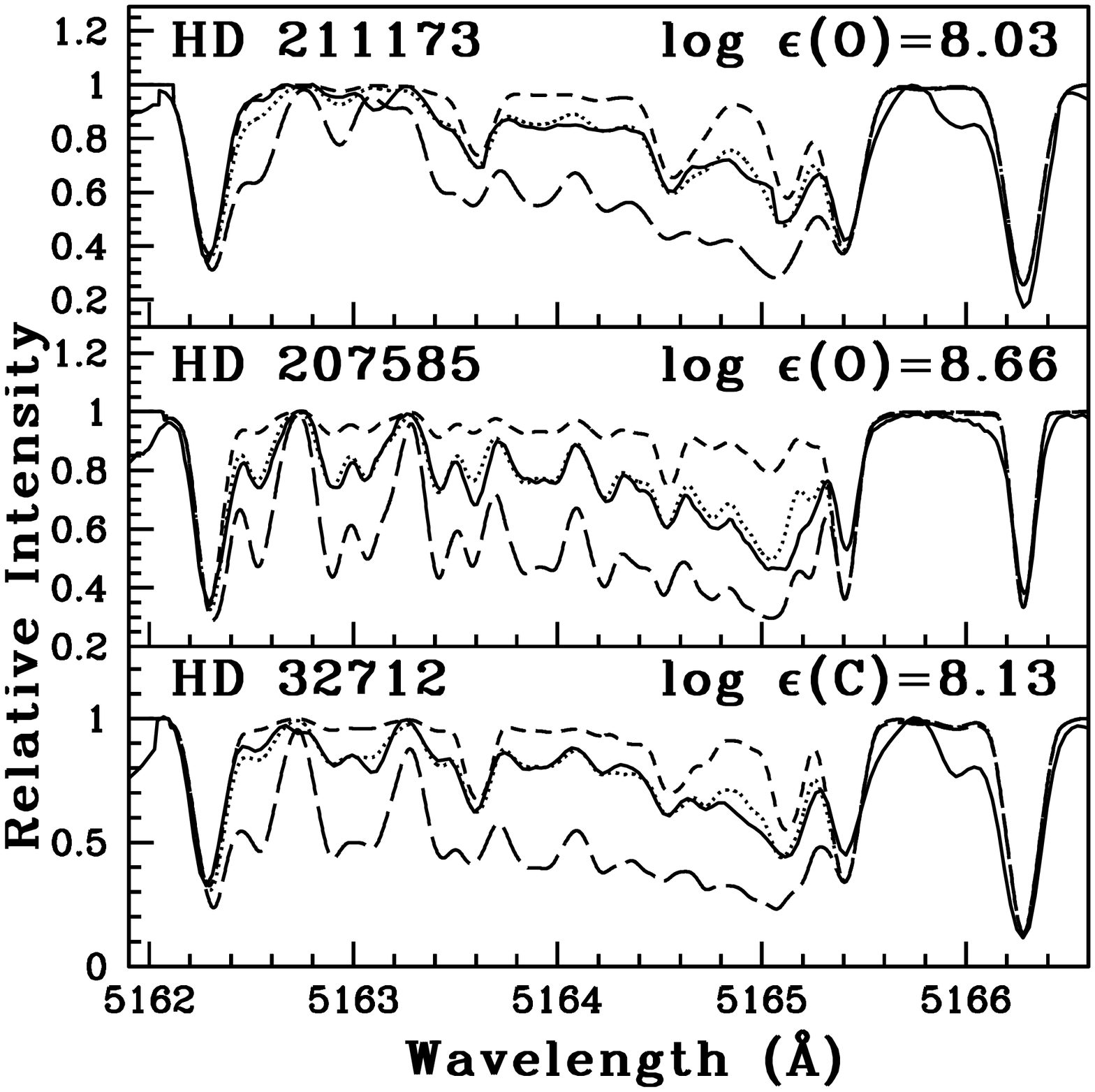}
\caption{ Synthesis of C$_{2}$ band around 5165 {\rm \AA}. Dotted line 
represents synthesized spectra and the solid line indicates the observed 
spectra. Short dashed line represents the synthetic spectra 
corresponding to $\Delta$ [C/Fe] = $-$0.3 and long dashed line is 
corresponding to $\Delta$[C/Fe] = +0.3} \label{carbon_5165}
\end{figure} 

\section{ABUNDANCE UNCERTAINTIES}
The uncertainties in the elemental abundances has two main components: 
random error  and systematic error.  Random error arises from the 
uncertainties in the line parameters such as measured equivalent width, 
line blending and oscillator strength. Since the random error  varies 
inversely as the square-root of the number of lines, we can  reduce 
this error by using  maximum possible number of lines. Systematic 
error is due to the  uncertainties in the adopted stellar atmospheric 
parameters. 
The total uncertainty in the elemental abundance log $\epsilon$ is 
calculated as; \\

$\sigma_{log\epsilon}^{2}$ = $\sigma_{ran}^{2}$ + $(\frac{\partial log \epsilon}{\partial T})^{2}$ $\sigma_{T_{eff}}^{2}$ + $(\frac{\partial log \epsilon}{\partial log g})^{2}$ $\sigma_{log g}^{2}$ + \\
\begin{center}
  $(\frac{\partial log \epsilon}{\partial \zeta})^{2}$ $\sigma_{\zeta}^{2}$ + $(\frac{\partial log \epsilon}{\partial [Fe/H]})^{2}$ $\sigma_{[Fe/H]}^{2}$ \\
\end{center}

\noindent where $\sigma_{ran}$ = $\frac{\sigma_{s}}{\sqrt{N}}$. $\sigma_{s}$ 
is the standard deviation of the abundances derived from the N 
number of lines of the particular species. The $\sigma$'$_{s}$ are 
the typical uncertainties in the stellar atmospheric parameters, which are  
T$_{eff}$$\sim$ $\pm$100 K, log g$\sim$ $\pm$0.2 dex, $\zeta$$\sim$ $\pm$0.2 kms$^{-1}$ and [Fe/H]$\sim$ $\pm$0.1 dex. The abundance uncertainties arising 
from the error of  each stellar atmospheric parameters is estimated by 
varying one parameter at a time by an amount equal to their corresponding 
uncertainty, by keeping others the same and computing the changes in 
the abundances. We have done this procedure for a representative star, 
HD~211173, in our sample with the assumption that the uncertainties 
due to different parameters are independent, following de Castro 
et al. (2016), Karinkuzhi et al. (2018) and Cseh et al. (2018). The 
estimated  differential abundances is given in 
Table \ref{differential_abundance}. The procedure has been applied 
to the abundances estimated from the equivalent width measurement as 
well as the spectral synthesis calculation. 
Finally, the uncertainty in [X/Fe] is calculated as, \\
$\sigma_{[X/Fe]}^{2}$ = $\sigma_{X}^{2}$ + $\sigma_{Fe}^{2}$ .

{\footnotesize
\begin{table*}
\caption{Differential Abundance ($\Delta$log$\epsilon$) of different species due to the variations
in stellar atmospheric parameters for HD~211173}  
\label{differential_abundance}
\resizebox{\textwidth}{!}{\begin{tabular}{lcccccccccccc}
\hline                       
Element & $\Delta$T$_{eff}$ & $\Delta$T$_{eff}$ & $\Delta$log g & $\Delta$log g & $\Delta$$\zeta$ & $\Delta$$\zeta$ & $\Delta$[Fe/H] & $\Delta$[Fe/H] & ($\Sigma \sigma_{i}^{2}$)$^{1/2}$ & ($\Sigma \sigma_{i}^{2}$)$^{1/2}$ & $\sigma_{[X/Fe]}$   & $\sigma_{[X/Fe]}$\\
        & (+100 K)  & ($-$100 K) & (+0.2 dex) & ($-$0.2 dex) & (+0.2 kms$^{-1}$) & ($-$0.2 kms$^{-1}$) & (+0.1 dex) & ($-$0.1 dex)     & (+$\Delta$)    & ($-$$\Delta$) & (+$\Delta$)    & ($-$$\Delta$)    \\
\hline
C           & 0.00    & 0.00               & +0.03   & $-$0.03     & $-$0.03 & +0.03      & +0.01   & $-$0.01    & 0.04 & 0.04   & 0.19 & 0.18 \\
N           & +0.10   & $-$0.10            & 0.00    & 0.00        & +0.02   & $-$0.02    & +0.05   & $-$0.05    & 0.11 & 0.11   & 0.21 & 0.21 \\
O           & $-$0.19 & +0.19              & +0.06   & $-$0.06     & 0.00    & 0.00       & 0.00    & 0.00       & 0.20 & 0.20   & 0.27 & 0.26 \\
Na I        & +0.07   & $-$0.08            & $-$0.02 & +0.02       & $-$0.05 & +0.05      & 0.00    & +0.01      & 0.09 & 0.10   & 0.21 & 0.21 \\
Mg I        & +0.06   & $-$0.05            & 0.00    & +0.01       & $-$0.06 & +0.07      & 0.00    & +0.01      & 0.08 & 0.09   & 0.21 & 0.20 \\
Al I        & +0.06   & $-$0.07            & 0.00    & 0.00        & $-$0.02 & +0.02      & 0.00    & 0.00       & 0.06 & 0.07   & 0.20 & 0.19 \\
Si I        & $-$0.03 & +0.03              & +0.04   & $-$0.04     & $-$0.03 & +0.03      & +0.01   & $-$0.01    & 0.06 & 0.06   & 0.20 & 0.20\\
Ca I        & +0.10   & $-$0.11            & $-$0.04 & +0.03       & $-$0.10 & +0.09      & 0.00    & 0.00       & 0.15 & 0.15   & 0.24 & 0.23 \\
Sc II       & $-$0.02 & +0.02              & +0.09   & $-$0.09     & $-$0.09 & +0.08      & +0.02   & $-$0.03    & 0.13 & 0.13   & 0.22 & 0.21 \\
Ti I        & +0.14   & $-$0.15            & $-$0.01 & +0.01       & $-$0.08 & +0.08      & 0.00    & 0.00       & 0.16 & 0.17   & 0.24 & 0.24 \\
Ti II       & $-$0.02 & 0.00               & +0.07   & $-$0.08     & $-$0.10 & +0.09      & +0.02   & $-$0.03    & 0.13 & 0.12   & 0.23 & 0.22  \\
V I         & +0.16   & $-$0.17            & $-$0.01 & 0.00        & $-$0.07 & +0.07      & $-$0.01 & +0.01      & 0.18 & 0.18   & 0.25 & 0.25 \\
Cr I        & +0.13   & $-$0.13            & $-$0.02 & +0.02       & $-$0.13 & +0.12      & 0.00    & 0.00       & 0.18 & 0.18   & 0.26 & 0.25 \\
Cr II       & $-$0.08 & +0.07              & +0.10   & $-$0.09     & $-$0.08 & +0.09       & +0.01   & $-$0.02    & 0.15 & 0.15   & 0.25 & 0.24 \\
Mn I        & +0.09   & $-$0.10            & $-$0.02 & +0.01       & $-$0.16 & +0.14      & $-$0.01 & 0.00       & 0.18 & 0.17   & 0.26 & 0.24 \\
Fe I        & +0.07   & $-$0.07            & 0.00    & $-$0.01     & $-$0.13 & +0.12      & +0.10   & $-$0.10    & 0.18 & 0.17   & --   & -- \\
Fe II       & $-$0.09 & +0.07              & +0.10   & $-$0.10     & $-$0.10 & +0.09      & +0.10   & $-$0.10    & 0.20 & 0.18   & --   & -- \\
Co I        & +0.07   & $-$0.07            & +0.02   & $-$0.03     & $-$0.06 & +0.06      & +0.01   & $-$0.02    & 0.09 & 0.10   & 0.20 & 0.20 \\
Ni I        & +0.04   & $-$0.03            & +0.02   & $-$0.02     & $-$0.10 & +0.10      & +0.01   & $-$0.01    & 0.11 & 0.11   & 0.21 & 0.20 \\
Cu I        & +0.09   & $-$0.09            & $-$0.01 & 0.00        & $-$0.15 & +0.12      & +0.03   & $-$0.02    & 0.18 & 0.15   & 0.25 & 0.23 \\
Zn I        & $-$0.05 & +0.06              & +0.07   & $-$0.06     & $-$0.08 & +0.09      & +0.02   & $-$0.01    & 0.12 & 0.12   & 0.22 & 0.21 \\
Rb I        & +0.10   & $-$0.10            & 0.00    & 0.00        & $-$0.03 & +0.03      & 0.00    & 0.00       & 0.10 & 0.10   & 0.21 & 0.20 \\
Sr I        & +0.15   & $-$0.16            & $-$0.03 & +0.02       & $-$0.22 & +0.22      & 0.00    & +0.01      & 0.27 & 0.27   & 0.32 & 0.32 \\
Y I         & +0.16   & $-$0.17            & $-$0.01 & 0.00        & $-$0.02 & +0.03      & 0.00    & +0.01      & 0.16 & 0.17   & 0.24 & 0.24 \\
Y II        & $-$0.01 & 0.00               & +0.08   & $-$0.08     & $-$0.14 & +0.14      & +0.02   & $-$0.03    & 0.16 & 0.16   & 0.24 & 0.24 \\
Zr I        & +0.17   & $-$0.19            & $-$0.01 & 0.00        & $-$0.03 & +0.03      & $-$0.01 & 0.00       & 0.17 & 0.19   & 0.25 & 0.26 \\
Zr II       & $-$0.03 & +0.01              & +0.09   & $-$0.09     & $-$0.09 & +0.11      & +0.02   & $-$0.03    & 0.13 & 0.15   & 0.22 & 0.23 \\
Ba II       & +0.02   & $-$0.03            & +0.05   & $-$0.06     & $-$0.19 & +0.15      & +0.03   & $-$0.04    & 0.20 & 0.17   & 0.27 & 0.24 \\
La II       & +0.01   & 0.00               & +0.09   & $-$0.09     & $-$0.06 & +0.07      & +0.03   & $-$0.03    & 0.11 & 0.12   & 0.21 & 0.21 \\
Ce II       & +0.01   & $-$0.01            & +0.09   & $-$0.08     & $-$0.11 & +0.15      & +0.04   & $-$0.03    & 0.15 & 0.17   & 0.23 & 0.25 \\
Pr II       & +0.01   & $-$0.02            & +0.08   & $-$0.09     & $-$0.03 & +0.03      & +0.03   & $-$0.04    & 0.09 & 0.10   & 0.22 & 0.21 \\
Nd II       & +0.01   & $-$0.02            & +0.08   & $-$0.09     & $-$0.09 & +0.09      & +0.03   & $-$0.04    & 0.12 & 0.11   & 0.22 & 0.21 \\
Sm II       & +0.02   & $-$0.02            & +0.09   & $-$0.08     & $-$0.05 & +0.07      & +0.04   & $-$0.03    & 0.11 & 0.11   & 0.21 & 0.21 \\
Eu II       & $-$0.02 & +0.01              & +0.09   & $-$0.09     & $-$0.03 & +0.04      & +0.03   & $-$0.03    & 0.10 & 0.10   & 0.21 & 0.20 \\
\hline
\end{tabular}}

\end{table*}
}

\section{Abundance analysis and DISCUSSION}
\subsection{Light element abundance analysis: C, N, O, $^{12}$C/$^{13}$C, 
Na, Al, $\alpha$- and $Fe$-peak elements}

The [O I] line at 6300.304 {\rm \AA} is used to derive 
the oxygen abundances,
whenever possible, otherwise, the
resonance  O I triplet lines at around 7770 {\rm \AA} are used.
The O I triplet lines are known to be affected by the non-LTE
effects (Eriksson \& Toft 1979, Johnson et al. 1974, Baschek
et al. 1977, Kiselman 1993, Amarsi et al. 2016).
The corrections are made to the LTE abundance obtained with
these lines following  Bensby et al. (2004) and Afsar et al. (2012).
The [O I] line at 6363.776  \AA, is found to be  blended and not
usable for abundance determination in any of the stars. The spectrum
synthesis fits of O I triplet lines for a few program stars are shown
in Figure \ref{OI_synth}. All the three lines of the O I IR triplet gave 
the same abundance values, except for HD 147609. In the case of HD 147609, 
the lines at 7771 and 7774 \AA, gave the same abundance  and the line 
at 7775 \AA,  gave an abundance which is 0.15 dex lower. In  this case, 
we have taken the average of the three values as the final  oxygen abundance.
\par We have estimated the oxygen abundance in all the program stars except 
HD~24035. The derived abundance of oxygen is in the range 
$-$0.26$\leq$[O/Fe]$\leq$0.97. Oxygen is under abundant in HD~36650 
and HD~211173 with [O/Fe]  values $-$0.23 and $-$0.26 respectively.
HD~32712 and HD~179832 show near-solar values. Purandardas et al. (2019) 
found [O/Fe]$\sim$$-$0.33 for a barium star in their sample, somewhat 
closer to our lower limit. A mild overabundance is found in the stars
HD~154276 and HD~219116 with [O/Fe] values 0.38 and 0.21 respectively.
In the other three stars, we found an [O/Fe]$>$0.6, with  
HD~207585 showing the largest enhancement of 0.97. 
The first dredge-up (FDU) is not expected to alter the  
oxygen abundance. 

The carbon abundances are derived using the spectral synthesis 
calculation of C$_{2}$ band at 5165  \AA, (Figure \ref{carbon_5165}) for 
six  objects.  G-band of CH  at 4300  \AA, is used for two stars as the 
C$_{2}$ band at 5165  \AA, are not usable for the abundance determination.
The objects for which we could estimate the carbon abundance using both 
C$_{2}$ and CH bands, we find that the CH band returns a lower value for
carbon by about 0.2 to 0.3 dex.  We could determine the carbon abundance 
in all the objects except for HD 154276 and HD 179832. Carbon is found 
to be under abundant in most of the stars analyzed here. The [C/Fe] value 
ranges from $-$0.28 to 0.61. The stars HD~24035, HD~147609 
and HD~207585  show a mild over abundance of carbon with values 
0.41, 0.38 and 0.61  respectively, whereas  it is near-solar in HD~32712 
and HD~219116. HD~36650, HD~94518 and HD~211173 show mild under abundance 
with  [C/Fe] values, $-$0.22, $-$0.28, $-$0.23  respectively. 
These values are consistent with those generally noticed in barium stars 
(Barbuy et al. 1992,  North et al. 1994a). 

With the estimated carbon abundances, we have derived the 
abundances of nitrogen  using the spectrum synthesis calculation of 
$^{12}$CN lines at 8000 {\rm \AA} region in HD~32712, HD~36650, 
HD~94518 and HD~211173.   In other objects,  where this region is 
not usable or unavailable, CN band at 4215 {\rm \AA} is used.  The 
molecular lines for CN and C$_{2}$ are taken from Brooke et al. (2013),  
Sneden et al. (2014) and Ram et al. (2014). 

\par The nitrogen abundance is estimated in seven of the program stars.
Estimated  [N/Fe] values range from 0.24 to 1.41 dex with HD~24035 and 
HD~94518 showing [N/Fe] $>$ 1.0 dex. Such higher values of nitrogen  
have already been noted in some barium stars by several authors  
(Smith 1984, Luck \& Lambert 1985, Barbuy et al. 1992, Allen 
\& Barbuy 2006a, Smiljanic et al. 2006,  Merle at al. 2016, 
Karinkuzhi et al. 2018).  Nitrogen enhancement with [N/Fe]$>$1 is 
possible if the star is  previously enriched  by the pollution from 
a massive AGB companion  experiencing Hot-Bottom Burning (HBB). 
In super-massive AGB stars nitrogen can be substantially produced 
at the base of the convective envelope  when  the temperature of the 
envelope exceeds 10$^{8}$ K (Doherty et al. 2014a).

\par We could derive the carbon isotopic ratio, $^{12}$C/$^{13}$C, 
using the spectral synthesis calculation of the $^{12}$CN lines at 
8003.292, 8003.553, 8003.910  \AA,  and $^{13}$CN features at 8004.554, 
8004.728, 8004.781 \AA, for four  stars, HD 32712, HD 36650, HD 211173 
and HD 219116.  The values for this ratio  are  20.0, 7.34, 20.0 and 7.34 
respectively. Values in the range 7 - 20 (Barbuy  et al. 1992, 
Smith et al. 1993,  Smith 1984, Harris et al. 1985, Karinkuzhi et al. 2018), 
and 13 -33 (Tomkin \& Lambert 1979, Sneden et al. 1981) are found in 
literature for  barium stars.  

\par The lower level of carbon enrichment and low $^{12}$C/$^{13}$C ratio 
along with  larger over abundance of nitrogen indicate that the matter 
has undergone CN processing and the products have been brought to the 
surface by the FDU. From their locations in  HR diagram,
the three stars for which  we could estimate $^{12}$C/$^{13}$C ratio 
are on the ascent of first giant branch (FGB).
These stars have undergone the FDU at the beginning of FGB. 
It has been noted that less evolved barium stars show higher
 carbon abundance as they have not reached the 
FDU (Barbuy et al. 1992, Allen \& Barbuy 2006a). Among our program stars, 
HD~207585 shows the maximum enhancement of carbon, which is on the 
subgiant branch. However, the star HD~94518 shows
the least enrichment among the program stars despite being the less 
evolved one, dwarf barium star.
According to Vanture (1992), if the accreted material is mixed to 
the hydrogen burning region of the star
either during the main-sequence or the first ascent of the giant branch, 
certain nucleosynthesis  can happen, thereby reducing the carbon 
abundance. Smiljanic et al. (2006) ascribes 
rotational mixing for  reduction in the  surface carbon abundance.
Also, even though the star has not reached the stage of dredge-up, 
the difference in the mean molecular weight of the accreted material 
and the inherent stellar materials in the interiors can induce 
thermohaline mixing, and this  could reduce the surface carbon abundance 
by an order of magnitude compared to the unaltered case 
(Stancliffe et al. 2007).

\par We have estimated the C/O ratios of the program stars except 
for HD~24035, HD~154276 and HD~179832. The estimated  C/O$<$1 as 
normally seen in barium stars (Table \ref{hs_ls}). 

\par The estimated Na abundances  in 
the range 0.05$\leq$[Na/Fe]$\leq$0.42 are similar   to what 
is normally seen for disk stars, normal field giants and some barium 
stars (Antipova et al. 2004, de Castro et al. 2016,  Karinkuzhi et al. 2018). 
 Na, Mg and Al are produced in the carbon burning stages of
 massive stars (Woosely \& Weaver 1995), hence the SN II are the probable 
sources of  these elements in the disk. The thick and thin disk dwarf 
stars  in the Galaxy do not  show any trend in [Na/Fe] ratio with 
metallicity (Edvarsson et al. 1993, Reddy et al. 2003, Reddy et al. 2006). 
 Similar pattern as the dwarf stars is observed in the case of field 
giants (Mishenina et al. 2006, Luck \& Heiter 2007).
 Owing to the common origin, all the stars in the disk is expected 
to show similar abundance. 
An enhanced abundance of Na can be expected in AGB stars during the 
 inter-pulse stage from $^{22}$Ne produced in the previous hot pulses via
 $^{22}$Ne(p, $\gamma$)$^{23}$Na (NeNa chain) (Mowlavi 1999, 
Goriely \& Mowlavi 2000). This Na can be brought to the 
 surface during TDU. Hence an overabundance of Na may be expected 
in the barium stars.
 However Na enrichment can be expected in stars prior to the AGB phase.
 El Eid \& Champagne (1995) and Antipova et al. (2004) related this 
over abundance of Na  to the nucleosynthesis associated with the 
evolutionary stage of the star. According to them, Na is synthesized in the 
 convective H-burning core of the main-sequence stars through NeNa chain. 
Later, this products  are mixed to the surface during the FDU. As a result, 
it is possible to  observe sodium enrichment in giants rather than in dwarfs. 
 Boyarchuk et al. (2001), de Castro et al. (2016) found an anti-correlation of
 [Na/Fe] with log g. We could observe a similar trend in our sample. 
 According to Denissenkov \& Ivanov (1987), a star with a minimum mass 
of 1.5$M_{\odot}$  will be able to raise the Na abundance through the 
NeNa chain even in the main-sequence itself.  Even though the Na enriched 
material can be synthesized in AGB and subsequently transferred to the 
barium stars, there may be a non-negligible contribution to the Na 
enrichment from the  barium star itself.  
  
\par The derived abundance of aluminium in HD~154276 and HD~211173 
return near-solar values for [Al/Fe], $-$0.12 and $-$0.11 respectively. 
Yang et al. (2016) found a range $-$0.22$\leq$[Al/Fe]$\leq$0.56, 
Allen \& Barbuy (2006a) $-$0.1$\leq$[Al/Fe]$\leq$0.1, and de castro 
et al. (2016) $-$0.07$\leq$[Al/Fe]$\leq$0.43 for their sample of barium stars.  

\par The estimated abundances of Mg are in the range 
$-$0.10$\leq$[Mg/Fe]$\leq$0.44. A Mg enrichment is expected to observe 
in the barium stars if the s-process over abundance is resulting from the 
neutrons produced during the convective thermal pulses through the 
reaction $^{22}$Ne($\alpha$,n)$^{25}${Mg}. 
We could not find any enhancement for Mg in our sample when compared with
values from the disk stars and normal giants.
This discards the fact that the origin of neutron is 
$^{22}$Ne($\alpha$,n)$^{25}${Mg} source. 
 
\par The estimated abundances for other elements, from Si, 
to  Zn are found to be well-within the range as normally seen for disk 
stars.

\subsection{Heavy element abundance analysis}
\subsubsection{\textbf{The light s-process elements: Rb, Sr, Y, Zr}}
\par
\begin{figure}
\centering
\includegraphics[width=\columnwidth]{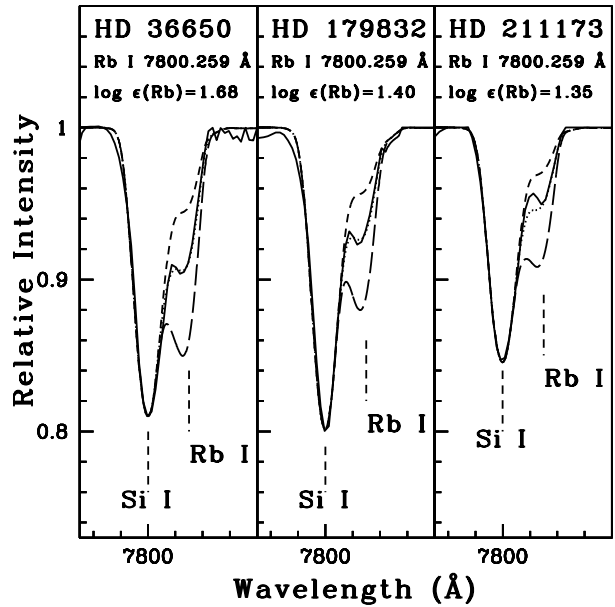}
\caption{ Synthesis of Rb I line around 7800 {\rm \AA}. Dotted line 
represents synthesized spectra and the solid line indicates the 
observed spectra. Short dashed line represents the synthetic spectra 
corresponding to $\Delta$[Rb/Fe] = $-$0.3 and long dashed line is 
corresponding to $\Delta$[Rb/Fe] = +0.3} \label{Rb_7800}
\end{figure}

 The abundance of Rb is derived  using the spectral synthesis 
calculation  of Rb I resonance line at 7800.259 \r{A} in the stars 
HD~32712, HD~36650, HD~179832 and HD~211173. We could not detect the 
Rb I lines in the warmer program stars. The Rb I resonance line 
at 7947.597 \r{A} is not usable for the abundance estimation.
The hyperfine components of Rb is taken from Lambert \& Luck (1976). 
The spectrum synthesis of Rb for the three program stars are shown 
in Figure \ref{Rb_7800}. 
Rubidium is found to be  under abundant in all the four program stars 
with [Rb/Fe] ranging from $-$1.35 to $-$0.82.

\par Strontium abundances are derived from the spectral synthesis 
calculation of  Sr I line at 4607.327 \r{A} whenever possible.
HD~154276 shows a mild under abundance with value [Sr/Fe]$\sim$$-$0.22, 
while  HD~32712 and HD~179832 show near-solar values. Other stars 
show enrichment with value [Sr/Fe]$>$0.66. 

\par The abundance of Y is derived from the spectral synthesis calculation 
of  Y I line at 6435.004 \r{A}  in all the program stars  except for 
HD~94518, HD~147609, HD~154276 and HD~179832 where  no useful Y I line 
were detected. The spectral synthesis of Y II line at 5289.815 \r{A} is 
used in HD~94518 while the equivalent width measurement of several 
lines of Y II  is used in other stars.  The abundances  estimated from 
Y I lines  range from 0.38 to 1.61, and that  from Y II lines,  0.07 to 1.37. 

\par
\begin{figure}
\centering
\includegraphics[width=\columnwidth]{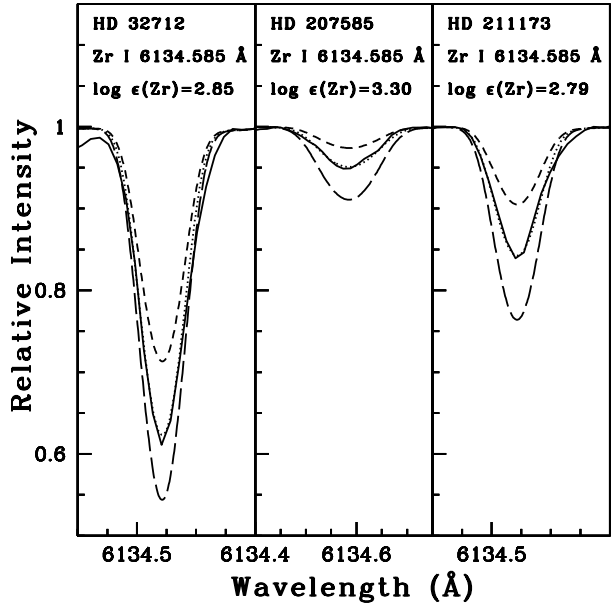}
\caption{ Synthesis of Zr I line at 6134.585 {\rm \AA}. Dotted line 
represents synthesized spectra and the solid line indicates the observed 
spectra. Short dashed line represents the synthetic spectra 
corresponding to $\Delta$[Zr/Fe] = $-$0.3 and long dashed line is 
corresponding to $\Delta$[Zr/Fe] = +0.3} \label{Zr_synth}
\end{figure}

 
The spectral synthesis of Zr I line at 6134.585 \r{A} is used in all 
the stars except HD~94518, HD~147609 and HD~154276 where  this line was 
not detected.  We could detect useful  Zr II lines in all the program 
stars except HD~36650 and HD~219116. In HD~24035, the equivalent 
width measurement of Zr II lines at 4317.321 and 5112.297 \r{A} are used. 
Spectral synthesis  calculation of Zr II line at 4208.977 \r{A} is used 
in HD~94518,  HD~147609 and HD~154276, line at 5112.297 \r{A} is used 
in HD~32712, HD~179832, HD~207585 and HD~211173. The measurement using 
Zr I lines gives the value 0.38$\leq$[Zr I/Fe]$\leq$1.29 and Zr II lines 
return $-$0.08$\leq$[Zr II /Fe]$\leq$1.89.   The spectrum synthesis of 
Zr for a few program stars are shown in Figure \ref{Zr_synth}.

\subsubsection{\textbf{The heavy s-process elements: Ba, La, Ce, Pr, Nd}}
\begin{figure}
\centering
\includegraphics[width=\columnwidth]{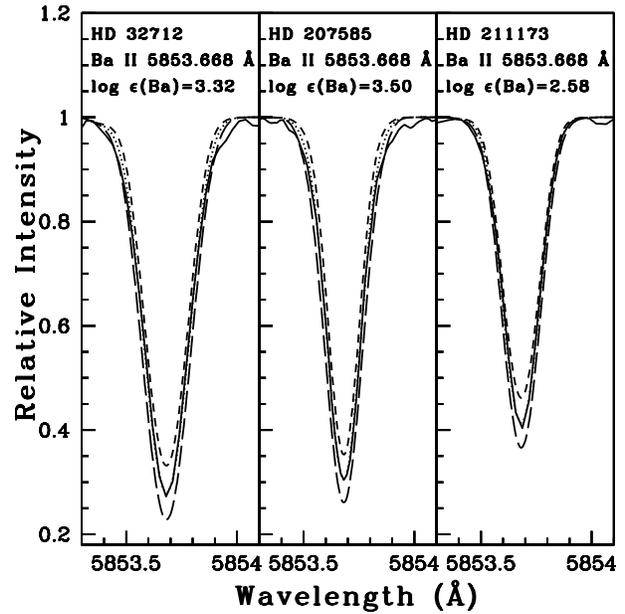}
\caption{ Synthesis of Ba II line at 5853.668 {\rm \AA}. Dotted line 
represents synthesized spectra and the solid line indicates the 
observed spectra. Short dashed line represents the synthetic spectra 
corresponding to $\Delta$[Ba/Fe] = $-$0.3 and long dashed line is 
corresponding to $\Delta$[Ba/Fe] = +0.3} \label{Ba_synth}
\end{figure}

The abundance of Ba is derived from the spectral synthesis.  
Ba II lines at 5853.668, 6141.713 and 6496.897 \r{A} are used 
in HD~154276, whereas,  in all the other stars, we have used the 
line at 5853.668 \r{A}.  The spectrum synthesis fits for Ba for a few 
program stars are  shown in Figure \ref{Ba_synth}. 
Ba shows slight overabundance in HD~154276 with [Ba/Fe]$\sim$0.22, while
HD~179832 and HD~211173 show moderate enhancement with values 0.41 and 
0.57 respectively. All other program stars show the overabundance 
of Ba in the range 0.79 to 1.71. 

\par Lanthanum abundance is obtained from the spectral synthesis 
analysis of La II line at 4322.503 \r{A} in HD~147609. For all other 
stars spectral synthesis  analysis of La II line at 4921.776 \r{A} is used. 
The estimated La abundances are in the range 0.20$\leq$[La/Fe]$\leq$1.70.
HD~154276 shows a mild enhancement of La with [La/Fe]$\sim$0.20.
All other program stars are overabundant in La with [La/Fe] ranging 
from 0.52 to 1.70. 

\par The abundances of Ce is obtained from the equivalent width 
measurement of several  The star HD~154276 show near-solar abundance 
for Ce with [Ce/Fe]$\sim$0.18, whereas all other stars are overabundant 
in Ce with [Ce/Fe]$>$0.7.

\par The abundance of Pr is derived from the the equivalent width 
measurement of  Pr II lines whenever  possible. We could not estimate Pr
abundance in HD~94518 and HD~154276 as there were no useful lines detected. 
HD~179832 is mildly enhanced in Pr with [Pr/Fe]$\sim$0.23 while other stars
show the enrichment in the range 0.85 to 1.98. 

\par Abundance of Nd is estimated from the spectral synthesis 
calculation of Nd II lines at 4177.320 and 4706.543 \r{A} in 
HD~154276. In all other stars, we  have used the equivalent width 
measurement of several Nd II lines.  A near-solar value is obtained 
for the Nd abundance in the star HD~179832  with [Nd/Fe]$\sim$0.04, 
whereas a moderate enhancement is found  in HD~154276 with 
[Nd/Fe]$\sim$0.40. All other objects show an enrichment  in Nd 
with [Nd/Fe]$>$0.81. 

\subsection{The r-process elements: Sm, Eu}
\par Samarium abundance is derived by the spectral synthesis of 
Sm II line  at 4467.341 \r{A} in HD~154276. The equivalent width 
measurement of several  Sm II lines is used to obtain the Sm abundance 
in the rest of the program stars.  All the good Sm lines are found 
in the bluer wavelength region of the spectra. The maximum number of 
Sm II lines used is eight, in HD~207585. The estimated Sm abundances 
give a near-solar value for HD~154276 with  [Sm/Fe]$\sim$0.07 while 
all other stars are enriched in Sm with values ranging from 0.78 to 2.04.

\par The Eu abundance is derived from the spectral synthesis of Eu II line 
at 4129.725 \r{A}
in HD~94518, HD~147609 and HD~207585. In all other stars except HD~154276, 
spectral synthesis calculation of Eu II line at 6645.064 \r{A} is used. In 
HD~154276, no useful lines for abundance analysis is detected. The
estimated Eu abundance covers the range 0.00$\leq$[Eu/Fe]$\leq$0.49. 
The r-process element Eu is not expected to show enhancement in
Ba stars according to their formation scenario.  

\par The observed abundance ratios when compared with their 
counterparts  in other barium stars from literature, the light 
as well as the heavy element Eu are found to follow the Galactic trend.  

\begin{figure}
\centering
\includegraphics[width=\columnwidth]{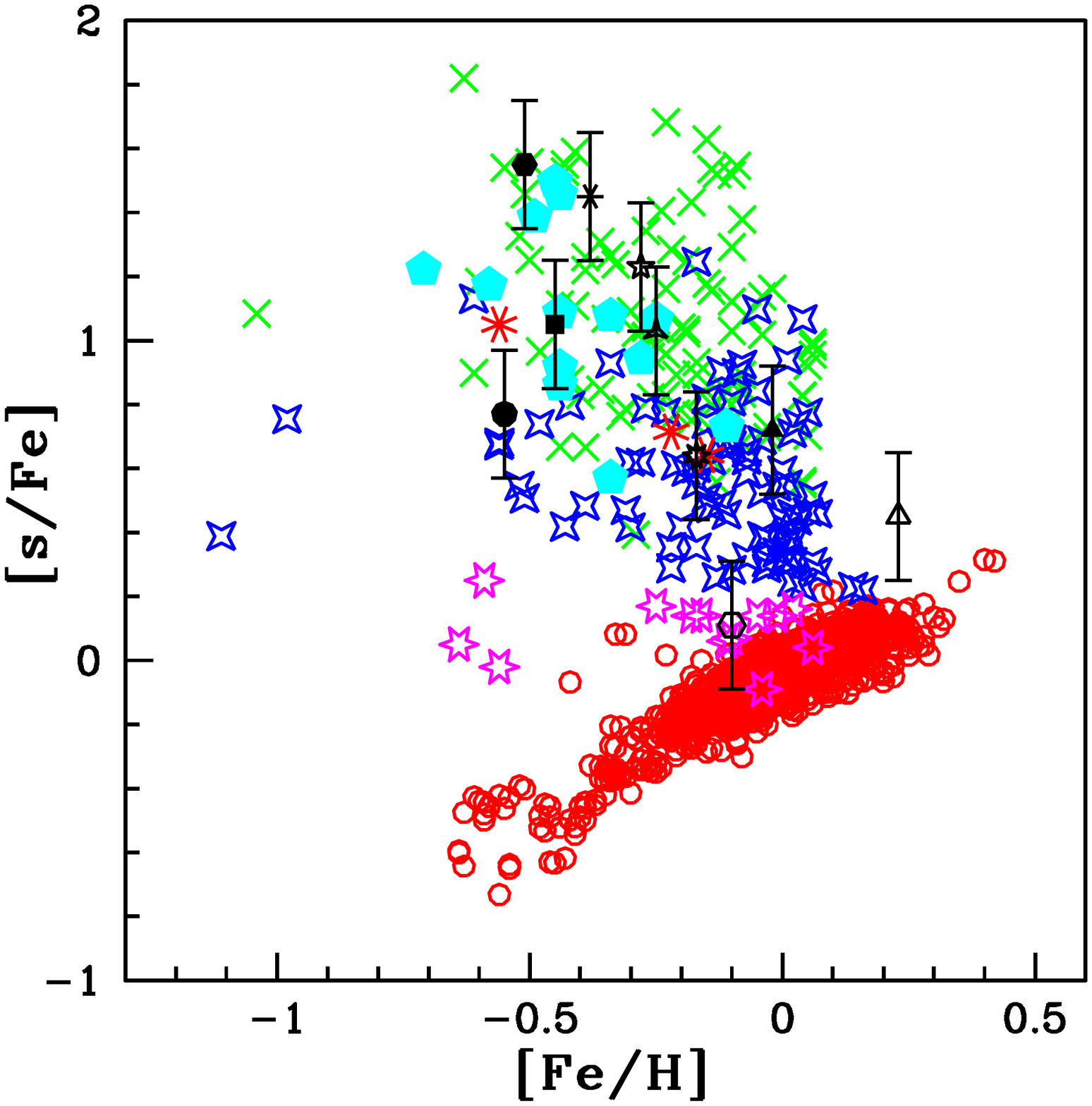}
\caption{\small{Observed [s/Fe] ratios of the 
program stars with respect to metallicity [Fe/H].   
Red open circles represent normal giants from literature (Luck \& Heiter 2007). 
Green crosses, blue four-sided stars, cyan filled pentagons, red 
eight-sided crosses represent strong Ba giants, weak Ba giants, 
Ba dwarfs, Ba sub-giants respectively from literature (de Castro 
et al. 2016, Yang et al. 2016, Allen \& Barbuy 2006a). 
Magenta six-sided stars represent the stars rejected as Ba stars 
by de Castro et al. (2016). HD~24035 (filled hexagon), HD~32712 
(starred triangle), HD~36650 (filled triangle), HD~94518 (filled circle), 
HD~147609 (five-sided star), HD~154276 (open hexagon), HD~179832 
(open triangle), HD~207585 (six-sided cross), HD~211173 (nine-sided star) 
and HD~219116 (filled square).}}\label{rejected_ba_star}
\end{figure}

In order to find the s-process contents in the stars, we have estimated the 
mean abundance ratio of the s-process elements (Sr, Y, Zr, Ba, La, Ce, Nd ), 
[s/Fe], for our stars. The estimated values of [s/Fe] is provided 
in Table \ref{hs_ls}.
The star HD~154276 shows the least value for [s/Fe] ratio.
A comparison of [s/Fe] ratio observed in our program stars with that 
in Ba stars and  normal giants  from literature is shown in 
 Figure \ref{rejected_ba_star}. 
The stars which are rejected as Ba stars from the analysis of 
de Castro et al. (2016) are also shown for a comparison. The [s/Fe] 
value of HD~154276 falls among these rejected stars. 
Most of these rejected Ba stars are listed as marginal Ba stars in 
MacConnell et al. (1972).  There is no clear mention in literature on 
how high should be  the [s/Fe] value for a star to be considered as a 
Ba star. According to de Castro et al. (2018), this value is +0.25, 
while Sneden et al. (1981) found a value +0.21, Pilachowski (1977) 
found +0.50 and Rojas at al. (2013) found a value $>$0.34. If we stick 
on to the values of these authors, the star HD~154276 with [s/Fe] = 0.11, 
can not be consider as a Ba star. However,  if we follow the criteria 
of Yang et al. (2016) that [Ba/Fe] should be atleast 0.17 for the star 
even to be a mild star, HD~154276 can be considered as a mild Ba star 
with [Ba/Fe]$\sim$0.22. 

\par A comparison of the heavy element abundances with the literature 
values whenever available are presented in 
Table \ref{abundance_comparison_literature}. In most of the cases our 
estimates agree within error bars with the  literature values.

{\footnotesize
\begin{table*}
\caption{Elemental abundances in HD 24035, HD 32712 and HD 36650} \label{abundance_table1}
\resizebox{\textwidth}{!}
{\begin{tabular}{lccccccccccccccc}
\hline
       &    &                              &                    & HD 24035 &            &                    & HD 32712 &        &                    & HD 36650 &      \\ 
\hline
       & Z  & solar log$\epsilon^{\ast}$   & log$\epsilon$      & [X/H]    & [X/Fe]     & log$\epsilon$      & [X/H]    & [X/Fe] & log$\epsilon$      & [X/H]    & [X/Fe]\\ 
\hline 
C      & 6  & 8.43                         & 8.33(syn)          & $-$0.10  & 0.41       & 8.13(syn)          & $-$0.3   & $-$0.05 & 8.19(syn)          & $-$0.24  & $-$0.22 \\
N      & 7  & 7.83                         & 8.73(syn)          & 0.90     & 1.41       & 8.10(syn)          & 0.27  & 0.52    & 8.38(syn)          & 0.55     & 0.57 \\
O      & 8  & 8.69                         & --                 & --       & --         & 8.42(syn)          & $-$0.27  & $-$0.02 & 8.20(syn)          & $-$0.49  & $-$0.47 \\
Na I   & 11 & 6.24                         & 6.15$\pm$0.17(4)   & $-$0.09  & 0.42       & 6.20$\pm$0.13(4)   & $-$0.04  & 0.21    & 6.33$\pm$0.08(4)   & 0.09     & 0.11  \\
Mg I   & 12 & 7.60                         & 7.23(1)            & $-$0.37  & 0.14       & 7.25$\pm$0.20(2)   & $-$0.35  & $-$0.10 & 7.69$\pm$0.04(2)   & 0.09     & 0.11 \\
Si I   & 14 & 7.51                         & 7.26$\pm$0.04(2)   & $-$0.25  & 0.14       & 7.60$\pm$0.19(3)   & 0.09     & 0.34    & 7.28$\pm$0.10(3)   & $-$0.23  & $-$0.21 \\
Ca I   & 20 & 6.34                         & 5.91$\pm$0.13(10)  & 0.43     & 0.08       & 5.92$\pm$0.09(11)  & $-$0.42  & $-$0.17 & 6.23$\pm$0.12(16)  & $-$0.11  & $-$0.09 \\
Sc II  & 21 & 3.15                         & 2.65(syn)          & $-$0.50  & 0.01       & 2.95(syn)          & $-$0.20  & 0.05    & 3.08(syn)          & $-$0.12  & $-$0.10 \\
Ti I   & 22 & 4.95                         & 4.76$\pm$0.11(19)  & $-$0.19  & 0.32       & 4.68$\pm$0.11(27)  & $-$0.27  & $-$0.02 & 4.92$\pm$0.09(24)  & $-$0.03  & $-$0.01  \\
Ti II  & 22 & 4.95                         & 4.68$\pm$0.03(2)   & $-$0.27  & 0.24       & 4.75$\pm$0.11(5)   &$-$0.20   & 0.05    & 4.97$\pm$0.14(7)   & 0.02     & 0.04  \\
V I    & 23 & 3.93                         & 3.30(syn)          & $-$0.63  & $-$0.12    & 3.46(syn)          & $-$0.47  & $-$0.22 & 3.92(syn)          & $-$0.53  & $-$0.51  \\
Cr I   & 24 & 5.64                         & 5.11$\pm$0.06(6)   & $-$0.53  & $-$0.02    & 5.33$\pm$0.17(7)   & $-$0.31  & $-$0.06 & 5.60$\pm$0.15(9)   & $-$0.04  & $-$0.02 \\
Cr II  & 24 & 5.64                         & --                 & --       & --         & 5.72$\pm$0.17(4)   & 0.08     & 0.33    & 5.54$\pm$0.08(3)   & $-$0.10  & $-$0.08 \\
Mn I   & 25 & 5.43                         & 4.85(syn)          & $-$0.58  & $-$0.07    & 4.86(syn)          & $-$0.57  & $-$0.32 & 5.08(syn)          & $-$0.40  & $-$0.38 \\
Fe I   & 26 & 7.50                         & 6.99$\pm$0.19(87)  & $-$0.51  & -          & 7.25$\pm$0.12(84)  & $-$0.25  & -       & 7.48$\pm$0.12(92)  & $-$0.02  & -   \\
Fe II  & 26 & 7.50                         & 7.00$\pm$0.16(7)   & $-$0.50  & -          & 7.25$\pm$0.15(9)   & $-$0.25  & -       & 7.48$\pm$0.14(6)   & $-$0.02  & - \\
Co I   & 27 & 4.99                         & 4.67(syn)          & $-$0.32  & 0.19       & 4.43(syn)          & $-$0.56  & $-$0.31 & 4.86(syn)          & $-$0.43  & $-$0.41  \\
Ni I   & 28 & 6.22                         & 6.07$\pm$0.14(13)  & $-$0.15  & 0.36       & 6.04$\pm$0.18(11)  & $-$0.18  & 0.07    & 6.33$\pm$0.10(11)  & 0.11     & 0.13 \\
Cu I   & 29 & 4.19                         & --                 & --       & --         & --                 & --       & --      & 4.57(syn)          & $-$0.09  & $-$0.07  \\
Zn I   & 30 & 4.56                         & 3.92(1)            & $-$0.64  & $-$0.13    & 4.42(1)            & $-$0.14  & 0.11    & --                 & --       & --   \\
Rb I   & 37 & 2.52                         & --                 & --       & --         & 1.14(syn)          & $-$1.38  & $-$1.13 & 1.68(syn)          & $-$0.84  & $-$0.82   \\
Sr I   & 38 & 2.87                         & --                 & --       & --         & 2.65(syn)          & $-$0.22  & 0.03    & 3.78(syn)          & 0.64     & 0.66    \\
Y I    & 39 & 2.21                         & 3.31(syn)          & 1.1      & 1.61       & 2.52(syn)          & 0.31     & 0.56    & 2.70(syn)          & 0.49     & 0.51    \\
Y II   & 39 & 2.21                         & --                 & --       & --         & 3.00$\pm$0.15(6)   & 0.79     & 1.04    & 2.89$\pm$0.13(10)  & 0.68     & 0.70  \\
Zr I   & 40 & 2.58                         & 3.28(syn)          & 0.7      & 1.21       & 2.85(syn)          & 0.27     & 0.52    & 3.07(syn)          & 0.49     & 0.51 \\
Zr II  & 40 & 2.58                         & 3.96$\pm$0.07(2)   & 1.38     & 1.89       & 3.15(syn)          & 0.57     & 0.82    & --                 & --       & --    \\
Ba II  & 56 & 2.18                         & 3.38(syn)          & 1.20     & 1.71       & 3.32(syn)          & 1.14     & 1.39    & 2.95(syn)          & 0.77     & 0.79   \\
La II  & 57 & 1.10                         & 2.22(syn)          & 1.12     & 1.63       & 2.10(syn)          & 1.00     & 1.25    & 1.81(syn)          & 0.60     & 0.62  \\
Ce II  & 58 & 1.58                         & 2.77$\pm$0.12(9)   & 1.19     & 1.70       & 3.01$\pm$0.11(11)  & 1.43     & 1.68    & 2.55$\pm$0.13(11)  & 0.97     & 0.99  \\
Pr II  & 59 & 0.72                         & 2.19$\pm$0.18(6)   & 1.47     & 1.98       & 2.14$\pm$0.18(6)   & 1.42     & 1.67    & 1.55$\pm$0.13(3)   & 0.83     & 0.85 \\
Nd II  & 60 & 1.42                         & 2.32$\pm$0.17(15)  & 0.90     & 1.41       & 2.93$\pm$0.18(11)  & 1.51     & 1.76    & 2.34$\pm$0.16(15)  & 0.92     & 0.94 \\
Sm II  & 62 & 0.96                         & 2.48$\pm$0.13(4)   & 1.52     & 2.03       & 2.42$\pm$0.17(6)   & 1.46     & 1.71    & 1.93$\pm$0.12(6)   & 0.97     & 0.99  \\
Eu II  & 63 & 0.52                         & 0.50(syn)          & $-$0.02  & 0.49       & 0.61(syn)          & 0.09  & 0.34    & 0.59(syn)          & 0.07 & 0.09 \\
\hline
\end{tabular}}

$\ast$  Asplund (2009), The number inside the paranthesis shows
the number of lines used for the abundance determination.
\end{table*}
}

{\footnotesize
\begin{table*}
\caption{Elemental abundances in HD 94518, HD 147609 and HD 154276} \label{abundance_table2}
\resizebox{\textwidth}{!}
{\begin{tabular}{lccccccccccccccc}
\hline
       &    &                              &                    & HD 94518 &            &                    & HD 147609 &            &                    & HD 154276 &  \\ 
\hline
       & Z  & solar log$\epsilon^{\ast}$   & log$\epsilon$      & [X/H]    & [X/Fe]     & log$\epsilon$      & [X/H]    & [X/Fe]      & log$\epsilon$      & [X/H]    & [X/Fe]\\ 
\hline 
C      & 6  & 8.43                         & 7.60(syn)          & $-$0.83  & $-$0.28    & 8.53(syn)          & 0.10   & 0.38       & --                 & --       & -- \\
N      & 7  & 7.83                         & 8.63(syn)          & 0.80     & 1.35       & -          & -  & -    & -          & -     & - \\
O      & 8  & 8.69                         & 8.79(syn)          & 0.10     & 0.65       & 9.05(syn)          & 0.36     & 0.64       & 8.91(syn)          & 0.22     & 0.32 \\
Na I   & 11 & 6.24                         & 5.88$\pm$0.06(4)   & $-$0.44  & 0.11       & 6.26$\pm$0.19(2)   & 0.02     & 0.30       & 6.20$\pm$0.05(3)   & $-$0.04  & 0.06 \\
Mg I   & 12 & 7.60                         & 7.37$\pm$0.12(4)   & $-$0.23  & 0.32       & 7.49$\pm$0.08(4)   & $-$0.11  & 0.17       & 7.81$\pm$0.06(3)   & 0.21     & 0.31 \\
Al I   & 13 & 6.45                         & --                 & --       & --         & --                 & --       & --         & 6.23$\pm$0.08(2)   & $-$0.22  & $-$0.12 \\
Si I   & 14 & 7.51                         & 6.90$\pm$0.20(4)   & $-$0.61  & $-$0.06    & 7.37$\pm$0.05(4)   & $-$0.14  & 0.14       & 7.54$\pm$0.05(5)   & 0.03     & 0.13  \\
Ca I   & 20 & 6.34                         & 6.01$\pm$0.13(18)  & $-$0.33  & 0.22       & 6.04$\pm$0.21(21)  & $-$0.30  & 0.21       & 6.22$\pm$0.20(27)  & $-$0.12  & $-$0.02\\
Sc II  & 21 & 3.15                         & 2.72(syn)          & $-$0.63  & $-$0.08    & 2.90(syn)          & $-$0.25  & 0.03       & 3.25(syn)          & 0.10     & 0.20 \\
Ti I   & 22 & 4.95                         & 4.64$\pm$0.13(12)  & $-$0.31  & 0.24       & 4.66$\pm$0.06(10)  & $-$0.29  & $-$0.01    & 5.02$\pm$0.17(27)  & 0.07     & 0.17 \\
Ti II  & 22 & 4.95                         & 4.81$\pm$0.14(13)  & $-$0.14  & 0.41       & 4.67$\pm$0.16(8)   & $-$0.28  & 0.00       & 5.06$\pm$0.21(18)  & 0.11     & 0.21 \\
V I    & 23 & 3.93                         & 3.15(syn)          & $-$0.78  & $-$0.23    & 3.53(syn)          & $-$0.40  & $-$0.12    & 3.90(syn)          & $-$0.03  &  0.07 \\
Cr I   & 24 & 5.64                         & 5.07$\pm$0.14(15)  & $-$0.57  & $-$0.02    & 5.29$\pm$0.15(9)   & $-$0.35  & $-$0.07    & 5.51$\pm$0.20(11)  & $-$0.13  & $-$0.03 \\
Cr II  & 24 & 5.64                         & 5.06$\pm$0.19(4)   & $-$0.58  & $-$0.03    & 5.29$\pm$0.11(3)   & $-$0.35  & $-$0.07    & 5.53$\pm$0.07(5)   & $-$0.11  & $-$0.01 \\
Mn I   & 25 & 5.43                         & 4.61(syn)          & $-$1.1   & $-$0.55    & 5.03(syn)          & $-$0.40  & $-$0.12    & 5.17$\pm$0.15(6)   & $-$0.26  & $-$0.16 \\
Fe I   & 26 & 7.50                         & 6.95$\pm$0.10(110) & $-$0.55  & -          & 7.22$\pm$0.16(151)  & $-$0.28  & -          & 7.41$\pm$0.13(150) & $-$0.09  & - \\
Fe II  & 26 & 7.50                         & 6.95$\pm$0.12(15)  & $-$0.55  & -          & 7.22$\pm$0.12(20)   & $-$0.28  & -          & 7.40$\pm$0.14(15)  & $-$0.10  & - \\
Co I   & 27 & 4.99                         & 4.52(syn)          & $-$0.64  & $-$0.09    & --                 & --       & --         & 4.85$\pm$0.10(2)   & $-$0.14  & $-$0.04 \\
Ni I   & 28 & 6.22                         & 5.70$\pm$0.15(26)  & $-$0.52  & 0.03       & 5.88$\pm$0.11(14)  & $-$0.34  & $-$0.06    & 6.13$\pm$0.15(19)  & $-$0.09  & 0.01 \\
Cu I   & 29 & 4.19                         & 3.67(syn)          & $-$0.82  & $-$0.27    & --                 & --       & --         & --                 & --       & -- \\
Zn I   & 30 & 4.56                         & 4.15$\pm$0.10(2)   & $-$0.41  & 0.14       & 4.30 $\pm$0.07(2)  & $-$0.26  & 0.02       & 4.64$\pm$0.00(2)   & 0.08     & 0.18   \\
Sr I   & 38 & 2.87                         & 3.59(syn)          & 0.63     & 1.18       & 4.10(syn)          & 1.23     & 1.51       & 2.55(syn)          & $-$0.32  & $-$0.22 \\
Y II   & 39 & 2.21                         & 2.16(syn)          & $-$0.05  & 0.50       & 3.00$\pm$0.14(9)   & 0.79     & 1.07       & 2.18$\pm$0.19(4)   & $-$0.03  & 0.07  \\
Zr II  & 40 & 2.58                         & 2.32(syn)          & $-$0.26  & 0.29       & 3.30(syn)          & 0.72     & 1.00       & 2.40(syn)          & $-$0.18  & $-$0.08 \\
Ba II  & 56 & 2.18                         & 2.58(syn)          & 0.35     & 0.90       & 3.30(syn)          & 1.12     & 1.40       & 2.30(syn)          & 0.12     & 0.22 \\
La II  & 57 & 1.10                         & 2.12(syn)          & 0.08     & 0.58       & 2.09(syn)          & 0.99     & 1.27       & 1.20(syn)          & 0.10     &  0.20 \\
Ce II  & 58 & 1.58                         & 1.94$\pm$0.09(9)   & 0.36     & 0.91       & 2.56$\pm$0.11(8)   & 0.98     & 1.26       & 1.66$\pm$0.03(2)   & 0.08     & 0.18 \\
Pr II  & 59 & 0.72                         & --                 & --       & --         & 1.79(1)            & 1.07     & 1.35       & --                 & --       & -- \\
Nd II  & 60 & 1.42                         & 1.92$\pm$0.20(9)   & 0.50     & 1.05       & 2.21$\pm$0.17(8)   & 0.79     & 1.07       & 1.72$\pm$0.12(syn)(2) & 0.30  & 0.40  \\
Sm II  & 62 & 0.96                         & 1.80$\pm$0.08(4)   & 0.84     & 1.39       & 1.97$\pm$ 0.19(4)  & 1.01     & 1.29       & 0.93(syn)          & $-$0.03  & 0.07 \\
Eu II  & 63 & 0.52                         & 0.12(syn)       & $-$0.40  & 0.15    & 0.37(syn)          & $-$0.15  & 0.13       & --                 & --       & --  \\
\hline
\end{tabular}}

 $\ast$  Asplund (2009), The number inside the parenthesis shows
the number of lines used for the abundance determination.
\end{table*}
}

{\footnotesize
\begin{table*}
\caption{Elemental abundances in HD 179832, HD 207585, HD 211173 and HD 219116} \label{abundance_table3}
\resizebox{\textwidth}{!}
{\begin{tabular}{lccccccccccccccc}
\hline
       &    &                              &                    & HD 179832 &           &                    & HD 207585 &        &                    & HD 211173 &    &                    & HD 219116 & \\ 
\hline
       & Z  & solar log$\epsilon^{\ast}$   & log$\epsilon$      & [X/H]    & [X/Fe]     & log$\epsilon$      & [X/H]    & [X/Fe]  & log$\epsilon$      & [X/H]    & [X/Fe]       & log$\epsilon$      & [X/H]    & [X/Fe]\\ 
\hline  
C      & 6  & 8.43                         & --                 & --       & --         & 8.66(syn)          & 0.23     & 0.61    & 8.03(syn)          & $-$0.40  & $-$0.23    & 8.03(syn)          &  $-$0.43 & 0.02  \\
N      & 7  & 7.83                         & --                 & --       & --         & 8.20(syn)          & 0.37     & 0.75    & 8.20(syn)          & 0.37     & 0.54       & 7.85(syn)          & 0.02     & 0.47 \\
O I    & 8  & 8.69                         &  8.93(syn)         & 0.24     & 0.01       & 9.28(syn)          & 0.59     & 0.97    & 8.26(syn)          & $-$0.43  & $-$0.26    & 8.45(syn)          & $-$0.24  & 0.21   \\
Na I   & 11 & 6.24                         &  6.52$\pm$0.13(2)  & 0.28     & 0.05       & 6.11$\pm$0.11(4)   & $-$0.13  & 0.25    & 6.30$\pm$0.14(4)   & 0.06     & 0.23       & 6.05$\pm$0.16(4)   & $-$0.19  & 0.26  \\
Mg I   & 12 & 7.60                         &  7.83$\pm$0.02(2)  & 0.23     & 0.00       & 7.29$\pm$0.12(3)   & $-$0.31  & 0.07    & 7.66$\pm$0.08(2)   & 0.06     & 0.23       & 7.47$\pm$0.02(3)   & $-$0.11  & 0.34 \\
Al I   & 13 & 6.45                         & --                 & --       & --         & --                 & --       & --      & 6.17$\pm$0.07(2)   & $-$0.28  & $-$0.11    & --                 & --       & -- \\
Si I   & 14 & 7.51                         &  7.71$\pm$0.08(4)  & 0.20     & $-$0.03    & 7.25$\pm$0.02(2)   & $-$0.26  & 0.12    & 6.97$\pm$0.11(2)   & $-$0.54  & $-$0.37    & 7.08$\pm$0.20(2)   & $-$0.43  & 0.02  \\
Ca I   & 20 & 6.34                         &  6.27$\pm$0.06(9)  & $-$0.07  & $-$0.30    & 6.22$\pm$0.17(11)  & $-$0.12  & 0.26    & 6.23$\pm$0.17(15)  & $-$0.11  & 0.06       & 6.02$\pm$0.14(16)  & $-$0.34  & 0.13  \\
Sc II  & 21 & 3.15                         &  3.35(syn)         & 0.20     & $-$0.03    & 2.63(syn)          & $-$0.52  & $-$0.14   & 2.79(syn)          & $-$0.36  & $-$0.19    & 2.66(syn)          & $-$0.49  & $-$0.04 \\
Ti I   & 22 & 4.95                         &  5.06$\pm$0.07(4)  & 0.11     & $-$0.12    & 4.58$\pm$0.15(7)   & $-$0.37  & 0.01    & 4.77$\pm$0.11(27)  & $-$0.18  & $-$0.01    & 4.73$\pm$0.09(21)  & $-$0.22  & 0.23 \\
Ti II  & 22 & 4.95                         &  5.39$\pm$0.08(4)  & 0.44     & 0.21       & 4.82$\pm$0.12(9)   & $-$0.13  & 0.25    & 4.76$\pm$0.16(9)   & $-$0.19  & $-$0.02    & 4.65$\pm$0.15(6)   & $-$0.3   & 0.15 \\
V I    & 23 & 3.93                         &  4.03(syn)         & 0.10     & $-$0.13    & 3.11(syn)          & $-$0.82  & $-$0.44 & 3.47(syn)          & $-$0.46  & $-$0.29    & 3.67(syn)          & $-$0.26  & 0.19 \\
Cr I   & 24 & 5.64                         &  5.78$\pm$0.12(2)  & 0.14     & $-$0.09    & 5.39$\pm$0.15(11)  & $-$0.25  & 0.13    & 5.45$\pm$0.16(11)  & $-$0.19  & $-$0.02    & 5.34$\pm$0.16(9)   & $-$0.3   & 0.15 \\
Cr II  & 24 & 5.64                         &  5.71$\pm$0.02(2)  & 0.07     & $-$0.16    & 5.38$\pm$0.15(4)   & $-$0.26  & 0.12    & 5.24$\pm$0.16(4)   & $-$0.4   & $-$0.23    & 5.20$\pm$0.09(2)   & $-$0.44  & 0.01 \\
Mn I   & 25 & 5.43                         &  5.25$\pm$0.09(3)  & $-$0.18  &  $-$0.41   & 4.60(syn)          & $-$0.83  & $-$0.45 & 5.13(syn)          & $-$0.30  & $-$0.13    & 4.78(syn)          & $-$0.65  & 0.2  \\
Fe I   & 26 & 7.50                         &  7.73$\pm$0.01(68) & 0.23     & -          & 7.12$\pm$0.12(107) & $-$0.38  & -       & 7.33$\pm$0.10(109) & $-$0.17  & -          & 7.05$\pm$0.11(92)  & $-$0.45  & - \\
Fe II  & 26 & 7.50                         &  7.72$\pm$0.04(9)  & 0.22     & -          & 7.12$\pm$0.11(12)  & $-$0.38  & -       & 7.33$\pm$0.09(11)  & $-$0.17  & -          & 7.06$\pm$0.12(9)   & $-$0.44  & - \\
Co I   & 27 & 4.99                         &  5.24$\pm$0.08(6)  & 0.25     & 0.02       & 4.55(syn)          & $-$0.44  & $-$0.06 & 4.58(syn)          & $-$0.41  & $-$0.24    & 4.69$\pm$0.10(7)   & $-$0.3   & 0.15   \\
Ni I   & 28 & 6.22                         &  6.43$\pm$0.07(10) & 0.21     & $-$0.02    & 5.83$\pm$0.14(16)  & $-$0.39  & $-$0.01 & 6.14$\pm$0.17(27)  & $-$0.08  & 0.09       & 5.91$\pm$0.12(11)  & $-$0.31  & 0.14  \\
Cu I   & 29 & 4.19                         & --                 & --       & --         & 4.36(syn)          & $-$0.63  & $-$0.25 & 3.92(syn)          & $-$0.27  & $-$0.10    & 4.09(syn)          & $-$0.12  & 0.33 \\
Zn I   & 30 & 4.56                         &  4.94$\pm$0.11(2)  & 0.38     & 0.15       & --                 & --       & --      & 4.43$\pm$0.02(2)   & $-$0.13  & 0.04       & 4.04(1)            & $-$0.56  & 0.11 \\
Rb I   & 37 & 2.52                         &  1.40(syn)         & $-$1.12  & $-$1.35    & --                 & --       & --      & 1.35(syn)          & $-$1.17  & $-$1.00    & --                 & --       & --   \\
Sr I   & 38 & 2.87                         &  3.12(syn)         & 0.25     & 0.02       & --                 & --       & --      & 3.40(syn)          & 0.53     & 0.70       & 3.13(syn)          & 0.26     & 0.71  \\
Y I    & 39 & 2.21                         & --                 & --       & --         & 2.77(syn)          & 0.56     & 0.94    & 2.42(syn)          & 0.21     & 0.38       & 2.49(syn)          & 0.28     & 0.73 \\
Y II   & 39 & 2.21                         &  2.55$\pm$0.05(5)  & 0.34     & 0.11       & 3.20$\pm$0.08(9)   & 0.99     & 1.37    & 2.69$\pm$0.07(8)   & 0.48     & 0.65       & 2.51$\pm$0.09(3)   & 0.30     & 0.75 \\
Zr I   & 40 & 2.58                         &  4.10(syn)         & 1.52     & 1.29       & 3.30(syn)          & 0.72     & 1.10    & 2.79(syn)          & 0.21     & 0.38       & 2.79(syn)          & 0.21     & 0.66   \\
Zr II  & 40 & 2.58                         &  4.25(syn)         & 1.67     & 1.44       & 3.74(syn)          & 0.82     & 1.20    & 2.80(syn)          & 0.22     & 0.39       & --                 & --       & --    \\
Ba II  & 56 & 2.18                         &  2.82(syn)         & 0.64     & 0.41       & 3.50(syn)          & 1.22     & 1.60    & 2.58(syn)          & 0.40     & 0.57       & 2.90(syn)          & 0.77     & 1.22  \\
La II  & 57 & 1.10                         &  1.85(syn)         & 0.75     & 0.52       & 2.47(syn)          & 1.32     & 1.70    & 1.88(syn)          & 0.78     & 0.95       & 2.00(syn)          & 0.9      & 1.35   \\
Ce II  & 58 & 1.58                         &  2.55$\pm$0.11(2)  & 0.97     & 0.74       & 2.92$\pm$0.16(14)  & 1.34     & 1.72    & 2.15$\pm$0.10(11)  & 0.57     & 0.74       & 2.70$\pm$0.20(10)  & 1.12     & 1.57   \\
Pr II  & 59 & 0.72                         &  1.18$\pm$0.05(2)  & 0.46     & 0.23       & 1.93$\pm$0.11(3)   & 1.21     & 1.59    & 1.93$\pm$0.11(2)   & 1.21     & 1.59       & 1.54$\pm$0.18(3)   & 0.82     & 1.27 \\
Nd II  & 60 & 1.42                         &  1.69$\pm$0.04(2)  & 0.27     & 0.04       & 2.66$\pm$0.10(19)  & 1.24     & 1.62    & 1.98$\pm$0.17(14)  & 0.56     & 0.73       & 2.10$\pm$0.12(10)  & 0.68     & 1.13   \\
Sm II  & 62 & 0.96                         &  1.97$\pm$0.04(5)  & 1.01     & 0.78       & 2.62$\pm$0.17(8)   & 1.66     & 2.04    & 1.66$\pm$0.07(4)   & 0.70     & 0.87       & 2.09$\pm$0.18(7)   & 1.13     & 1.58  \\
Eu II  & 63 & 0.52                         &  0.75(syn)         & 0.23     & 0.00       & 0.42(syn)          & $-$0.10  & 0.28 & 0.48(syn)          & $-$0.04  & 0.13    & 0.50(syn)          & $-$0.02  & 0.43     \\
\hline
\end{tabular}}

 $\ast$  Asplund (2009), The number inside the parenthesis shows the 
number of lines used for the abundance determination. 
\end{table*}
}

\subsection{\textbf{The [hs/ls] ratio} as an indicator of neutron source}
In Table \ref{hs_ls}, we have presented the  estimated [ls/Fe], [hs/Fe] and 
[hs/ls] ratios  for the program stars, where ls refers to the light 
s-process elements (Sr, Y and Zr) and hs to the heavy s-process elements 
(Ba, La, Ce and Nd). 

The [hs/ls] ratio is a useful indicator of neutron source in the 
former AGB star. As the metallicity decreases, the neutron exposure 
increases. As a result, lighter  s-process elements are bypassed in 
favour of heavy elements. Hence, [hs/ls] ratio  increases with 
decreasing metallicity. The models of Busso et al. (2001) have shown 
the behaviour of this ratio with metallicity for AGB stars of mass 1.5 
and 3.0 M$_{\odot}$  for different $^{13}$C pocket efficiencies. 
According to these models, the maximum value of [hs/ls] is $\sim$ 1.2 
which is at  metallicities $\sim$ $-$1.0  and $\sim$ $-$0.8  for 
the 3 and 1.5 M$_{\odot}$ models respectively for the standard 
$^{13}$C pocket efficeincy. In their models, Goriely \& Mowlavi (2000), 
have shown the run of [hs/ls]  ratio with meatallicity for different 
thermal pulses for AGB stars in the range 1.5-3M$_{\odot}$. The maximum 
value  of [hs/ls]$\sim$0.6 occurs at metallicity $\sim$ $-$0.5. It was 
noted that,  in all these models, the [hs/ls] ratio does not follow a linear 
anti-correlation with metallicity, rather  exhibits a loop like behaviour. 
The ratio increases with decreasing  metallicity upto a particular value  
of [Fe/H] and then  starts to drop.  Our [hs/ls] ratio has a maximum 
value of $\sim$ 1.15 which occurs at a metallicity of $\sim$ $-$0.25. 
The anti-correlation of [hs/ls] suggest the operation of 
$^{13}$C($\alpha$, n)$^{16}$O neutron source,  since 
$^{13}$C($\alpha$, n)$^{16}$O is found to be anti-correlated with 
metallicity (Clayton 1988, Wallerstein 1997). 

As seen from the Table \ref{hs_ls}, all the stars show positive values 
for [hs/ls] ratio. At metallicities higher than solar, a negative value 
is expected  for this ratio and at lower metallicities, a positive 
value is expected for low-mass AGB stars where 
$^{13}$C($\alpha$, n)$^{16}$O is the neutron source (Busso et al. 2001, 
Goriely \& Mowlavi 2000). However, it is possible that AGB stars with 
masses in the  range 5-8M$_{\odot}$ can also exhibit low [hs/ls] 
ratios considering the $^{22}$Ne($\alpha$, n)$^{25}$Mg neutron source 
(Karakas \& Lattanzio 2014). The models of Karakas \& Lattanzio (2014) 
predicted that the ls elements are predominantly produced over the 
hs elements for AGB stars of mass 5 and 6 M$_{\odot}$. The [hs/ls] 
ratio is correlated to the neutron exposure.
The $^{22}$Ne($\alpha$, n)$^{25}$Mg source has smaller neutron 
exposure compared to the $^{13}$C($\alpha$, n)$^{16}$O source.
Hence, in the stars where $^{22}$Ne($\alpha$, n)$^{25}$Mg operates, 
we expect a lower [hs/ls] ratio. The lower neutron exposure of the 
neutrons produced from the $^{22}$Ne source together with the 
predictions of low [hs/ls] ratio in massive AGB star models 
have been taken as the evidence of operation of  
$^{22}$Ne($\alpha$, n)$^{25}$Mg in massive AGB stars. A Mg 
enrichment is expected in the stars where this reaction takes place.
As  none of our stars shows such an enrichment, we  discard 
the possibility of $^{22}$Ne($\alpha$, n)$^{25}$Mg reaction 
as a possible neutron source for any of our program stars,
with respect to [hs/ls] ratio. This is also supported by our 
estimates of Rb and Zr as discussed in the following section.

\subsection{Rb as a probe to the neutron density at the s-process site}
In addition to the [hs/ls] ratio, the abundance of rubidium can 
also provide clues to the mass of the companion  AGB stars. 
The AGB star models predict higher Rb abundances for massive AGB stars 
where the neutron source is $^{22}$Ne($\alpha$,n)$^{25}${Mg} 
reaction (Abia et al. 2001, van Raai et al. 2012). In the s-process 
nucleosynthesis path, the  branching points at the  unstable 
nuclei $^{85}$Kr and $^{86}$Rb controls the  Rb production. The 
amount of Rb produced along this s-process path  is determined by 
the probability of these unstable nuclei to  capture the neutron 
before $\beta$-decaying,  which in turn depends on the neutron density 
at the s-process site (Beer \& Macklin 1989, Tomkin \& Lambert 1983, 
Lambert et al. 1995).

The production of $^{87}$Rb from $^{85}$Kr and 
$^{86}$Rb is possible only at higher neutron densities, 
N$_{n}$ $>$ 5$\times$10$^{8}$ n/cm$^{3}$, Sr, Y, Zr etc. are produced 
otherwise (Beer 1991, Lugaro \& Chieffi 2011). The $^{87}$Rb 
isotope has magic number of neutrons and hence  it is fairly stable 
against neutron capture. Also, the neutron capture cross-section of 
$^{87}$Rb is very small ($\sigma$ $\sim$ 15.7 mbarn at 30 KeV) 
compared to that of $^{85}$Rb ($\sigma$ $\sim$ 234 mbarn) (Heil 
et al. 2008a). Hence, once the nucleus $^{87}$Rb is produced, it 
will be accumulated. Therefore, the isotopic ratio $^{87}$Rb/$^{85}$Rb
could be  a direct indicator of the neutron density at 
the s-process site, as a consequence help  to infer the mass 
of the AGB star. But, it is impossible to distinguish the lines 
due to these two isotopes of Rb in the stellar spectra 
(Lambert \& Luck 1976, Garc\'ia-Hern\'andez et al. 2006).
However, the abundance of Rb relative to other elements in 
this region of the s-process path, such as Sr, Y, and Zr, 
can be used to estimate the average neutron density of the s-process.
Detailed nucleosynthesis models for the stars with masses 
between 5 - 9 M$_{\odot}$ at solar metallicity 
predict [Rb/(Sr,Zr)]$>$0 (Karakas et al. 2012). A positive value 
of [Rb/Sr] or [Rb/Zr] ratio  indicates a higher neutron 
density, whereas a negative value indicates a low neutron density.
This fact has been used an evidence to conclude that 
$^{13}$C($\alpha$, n)$^{16}$O reaction act as the 
neutron source in M, MS and S stars (Lambert et al. 1995) and 
C stars must be low-mass AGB stars with M$<$3M$_{\odot}$ 
(Abia et al. 2001). The observed [Rb/Zr] ratios in the AGB 
stars both in our Galaxy and the Magellanic Clouds show a 
value $<$ 0 for low-mass AGB stars and a value $>$ 0 for intermedaite-mass 
(4-6 M$_{\odot}$) AGB stars (Plez et al. 1993, Lambert et al. 1995, 
Abia et al. 2001,  Garc\'ia-Hern\'andez et al. 2006, 2007, 2009, 
van Raai et al. 2012).

\begin{figure}
\centering
\includegraphics[width=\columnwidth]{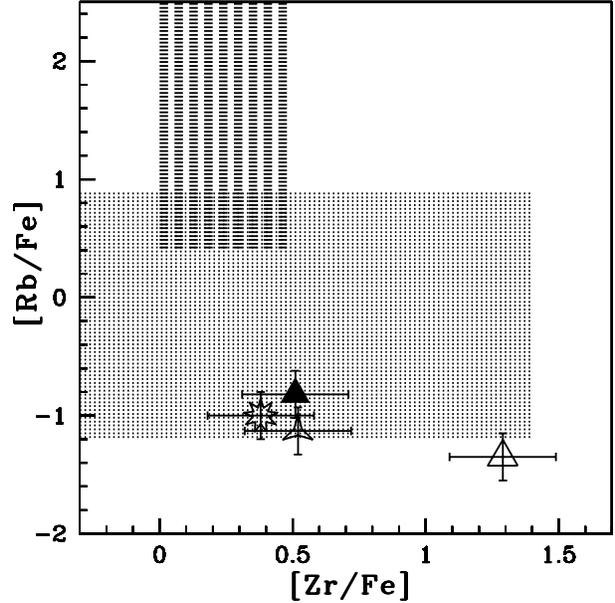}
\caption{The observed abundances [Rb/Fe] vs [Zr/Fe].
HD~32712 (starred triangle), HD~36650 (filled triangle), 
HD~179832 (open triangle), and HD~211173 (nine-sided star). 
The region shaded with short-dashed line  and dots 
corresponds to the observed range of Zr and Rb in 
intermediate-mass  and low-mass AGB stars respectively 
in the Galaxy and the Magellanic Clouds 
(van Raai et al. 2012). The four stars occupy the region 
of low-mass AGB stars except for HD~179832 (open triangle) 
which lie marginally below this region.} \label{Rb_Zr}
\end{figure}

The estimated [Rb/Zr] and [Rb/Sr] ratios (Table 13) give 
negative values for our stars for which we could estimate 
these ratios. The observed [Rb/Fe] and [Zr/Fe] ratios are  shown 
in Figure \ref{Rb_Zr}. The observed ranges of Rb and Zr in 
low- and intermediate-mass AGB stars (shaded regions) 
in the Galaxy and Magellanic Clouds are also shown for a 
comparison. It is clear that the abundances of Rb and Zr 
are consistent with the range normally observed in the 
low-mass AGB stars.  

\subsection{Comparison with FRUITY models and a parametric model based 
analysis}
A publicly available  (http://fruity.oa-teramo.inaf.it/, Web sites of 
the Teramo Observatory (INAF))  data set for the s-process in AGB stars 
is the FRANEC Repository of Updated Isotopic Tables \& Yields (FRUITY) 
models (Cristallo et al. 2009, 2011, 2015b). These models cover the 
whole range  of metallicity observed for Ba stars from z = 0.001 to 
z = 0.020 for the mass range 1.3 - 6.0 M$_{\odot}$.  The  computations 
comprise of the evolutionary models  starting from the pre-main 
sequence to the tip of AGB  phase through the core He-flash. During 
the core H-burning, no core overshoot has been considered, a 
semi-convection is  assumed during the core He-burning. The only mixing 
considered in this model is arising from the convection, additional 
mixing phenomena such as rotation is not considered here.  The 
calculations are based on a full nuclear network considering  all the 
stable and relevant unstable isotopes from hydrogen to bismuth. 
This includes 700 isotopes and about 1000 nuclear processes such as 
charged particle reactions,  neutron captures, and $\beta$-decays 
(Straniero et al. 2006, G\"orres et al. 2000, Jaeger et al. 2001, 
Abbondanno et al. 2004, Patronis et al. 2004, Heil et al. 2008b). 
The details of the input  physics and the computational algorithms 
are provided in Straniero et al. (2006).  In this model, $^{13}$C 
pocket is formed through time-dependent  overshoot mechanisms, which 
is controlled by a free overshoot  parameter ($\beta$) in the  
exponentially declining convective  velocity function. This parameter 
is set in  such a way that  the neutrons released are enough to 
maximise the production  of s-process  elements. For the low-mass 
AGB star models (initial mass $<$4 M$_{\odot}$),  neutrons are 
released by  the $^{13}$C($\alpha$, n)$^{16}$O reaction during the 
interpulse phase in radiative conditions, when temperatures 
within the pockets reaches T $\sim$ 1.0 $\times$ 10$^{8}$ K, 
with typical densities of 10$^{6}$ - 10$^{7}$ neutrons cm$^{-3}$. 
However, in the case of the metal-rich models (z = 0.0138, z = 0.006 
and z = 0.003), $^{13}$C is only partially burned during the interpulse; 
surviving part  is ingested in  the convective zone generated by the 
subsequent  thermal pulse (TP) and then burned at 
T $\sim$ 1.5 $\times$ 10$^{8}$ K,  producing a neutron density 
of 10$^{11}$ neutrons cm$^{-3}$. For larger z, 
$^{22}$Ne($\alpha$, n)$^{25}$Mg neutron source 
is marginally activated during the TPs;  but for low z, it 
becomes an important source when most of the $^{22}$Ne is 
primary (Cristallo et al. 2009, 2011).   For the intermediate-mass 
AGB star models, the s-process distributions are dominated by 
the $^{22}$Ne($\alpha$, n)$^{25}$Mg neutron source, which is 
efficiently activated during TPs. The contribution from the 
$^{13}$C($\alpha$, n)$^{16}$O reaction is  strongly reduced in 
the massive stars. 
This is due to the lower extent of the $^{13}$C pocket in them. 
It is shown that the extent of $^{13}$C pocket decreases 
with increasing core mass  of the AGB, due to the shrinking 
and compression of He-intershell (Cristallo et al. 2009). 
These massive  models experience Hot Bottom Burning and Hot-TDUs 
at lower  metallicities (Cristallo et al. 2015b). 

\begin{figure}
\centering
\includegraphics[width=0.8\columnwidth, height= 0.8\columnwidth]{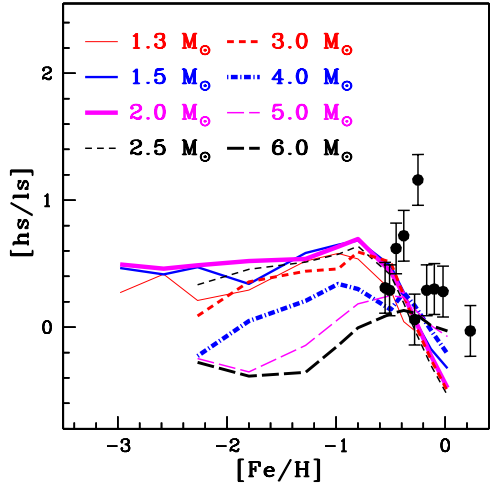}
\caption{Comparison of predicted and observed values of [hs/ls] ratios.} 
\label{fruity_ls_hs_hsls}
\end{figure}

We have compared our observational data with the FRUITY model.
The model predictions are unable to reproduce the [hs/ls] ratios 
characterizing  the surface composition of the stars. 
A comparison of the observed [hs/ls] ratios with metallicity 
shows a large spread (Figure \ref{fruity_ls_hs_hsls}), somewhat similar 
to the  comparison between the model and observational data as shown  
in Cristallo et al. (2011, Figure 12). 

The observed discrepancy may be explained considering  different  
$^{13}$C pocket efficiencies in the AGB models. In the FRUITY models  
a  standard $^{13}$C pocket is being considered, however,  it needs to 
be checked   if a variation in  the  amount of $^{13}$C pocket would 
give a  better match with the observed spread.  Absence of stellar 
rotation in the current FRUITY  models   may also be a cause for the 
observed discrepancy.  The rotation induced mixing alters the extend 
of $^{13}$C pocket (Langer et al. 1999), which in turn affects 
the s-process abundance pattern. However, a study made by 
Cseh et al. (2018) using the rotating star models available for the 
metallicity range of  Ba stars (Piersanti et al. 2013) could not 
reproduce the observed abundance  ratios of stars studied in 
de Castro et al. (2016). 

\begin{figure}
\begin{center}
\includegraphics[width=\columnwidth]{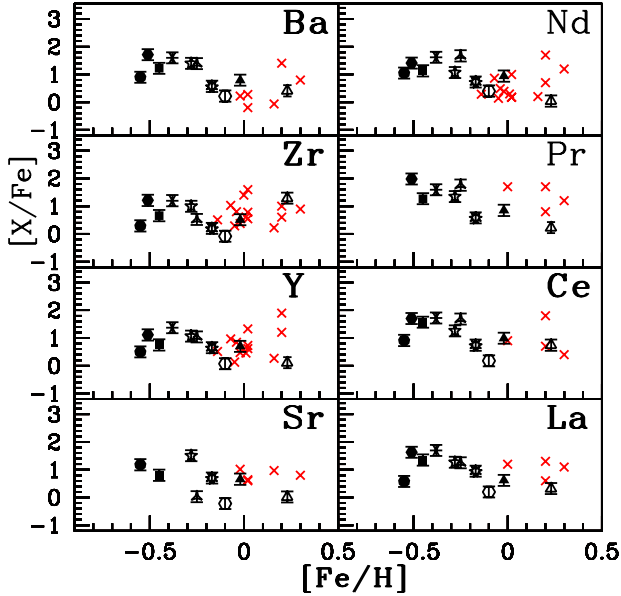}
\caption{\small{Comparison of abundance ratios of neutron-capture elements 
observed in the program stars and the AGB stars with respect to 
metallicity [Fe/H].  Red crosses represent the AGB stars from literature 
(Smith \& Lambert 1985, 1986b, 1990, Abia \& wallerstein 1998).}} 
\label{agb_heavy_comparison}
\end{center}
\end{figure}

The observed  abundance ratios for eight neutron-capture elements are 
compared with their counterparts in the low-mass AGB stars from literaure, 
that are found to be associated with $^{13}$C($\alpha$, n)$^{16}$O 
neutron source (Figure \ref{agb_heavy_comparison}).  As discussed in 
de Castro et al. (2016) the  scatter observed in the  ratios may be a 
consequence of different  dilution factors during the mass transfer, 
as well as the orbital  parameters, metallicity and initial mass.

\begin{figure}
\centering
\includegraphics[width=0.8\columnwidth, height= 0.8\columnwidth]{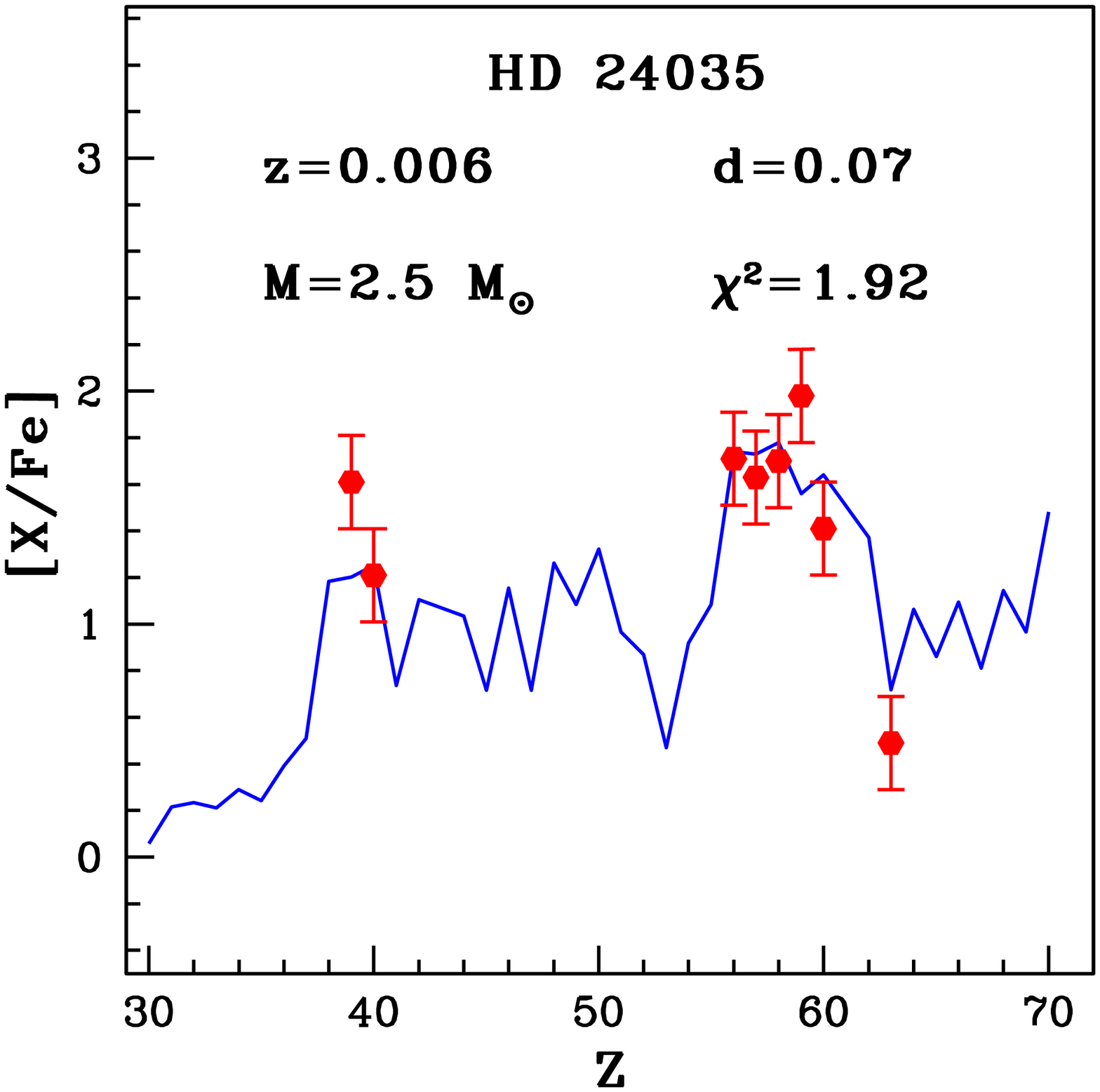}
\includegraphics[width=0.8\columnwidth, height= 0.8\columnwidth]{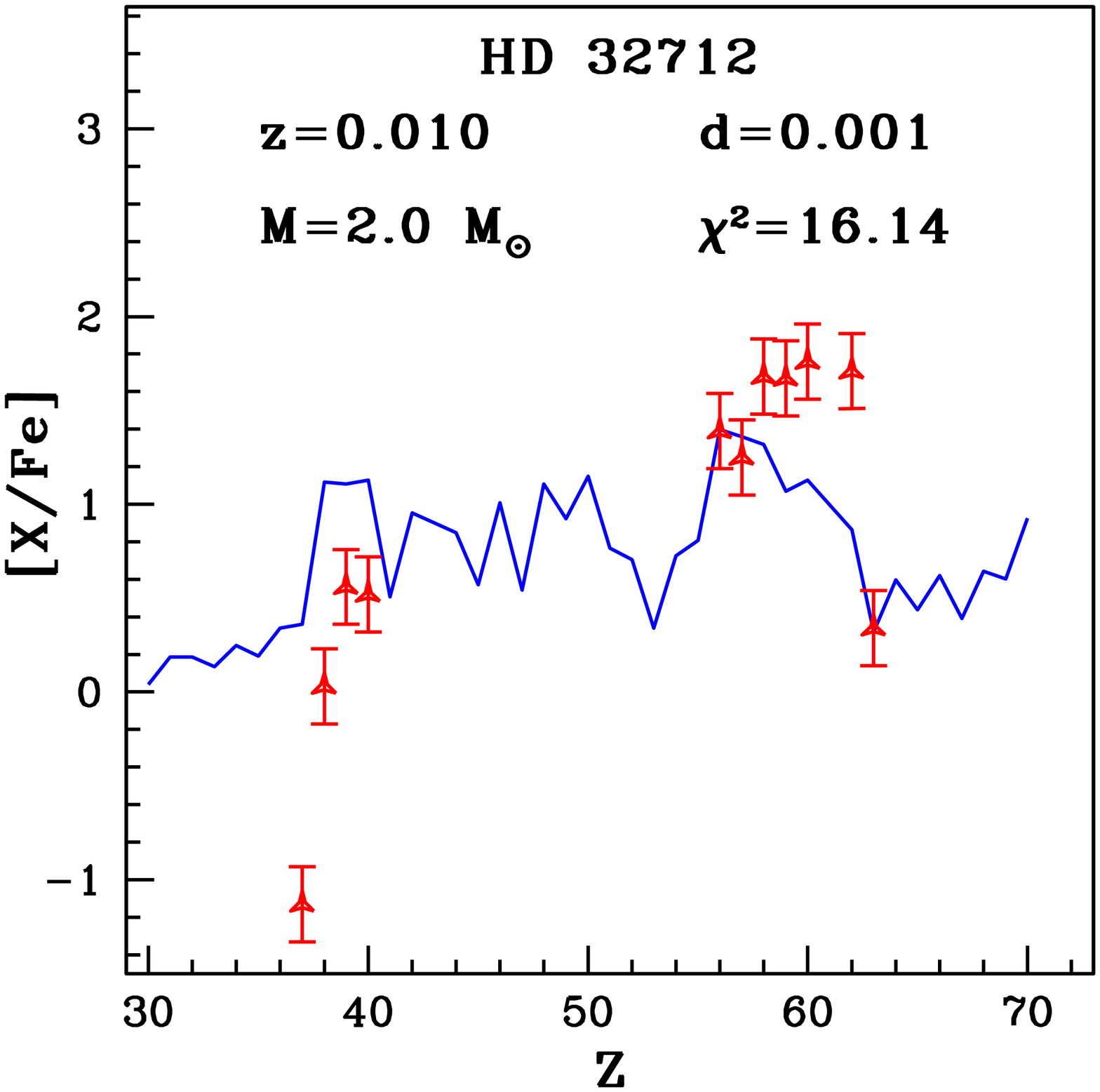}
\includegraphics[width=0.8\columnwidth, height= 0.8\columnwidth]{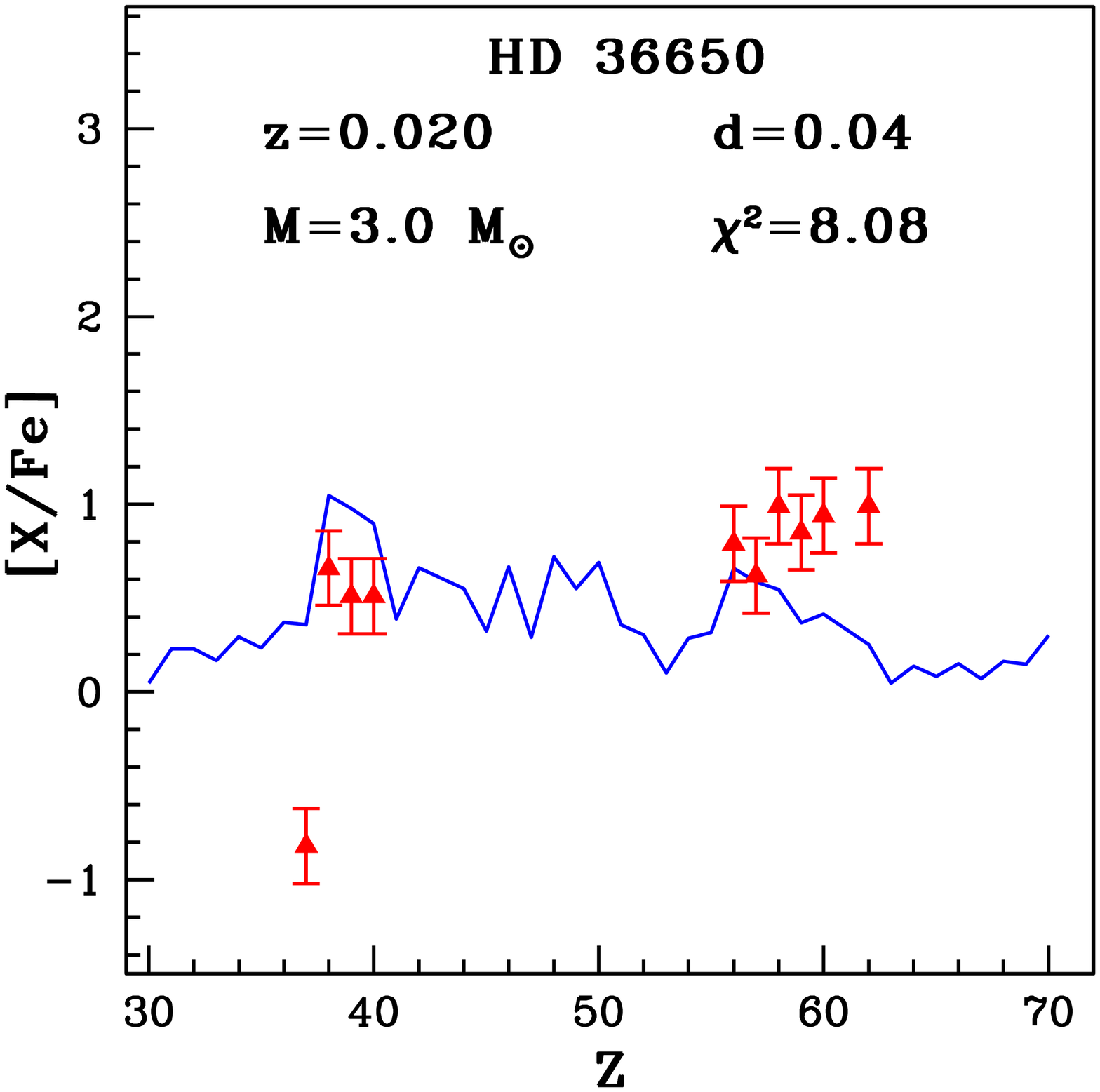}
\caption{Solid curve represent the best fit for the parametric model function.
The points with error bars indicate the observed abundances in 
(i) \textit{Top panel:} HD~24035 (ii) \textit{Middle panel:} HD~32712 
(iii) \textit{Bottom panel:} HD~36650.} 
\label{parametric1}
\end{figure}

\begin{figure}
\centering
\includegraphics[width=0.8\columnwidth, height= 0.8\columnwidth]{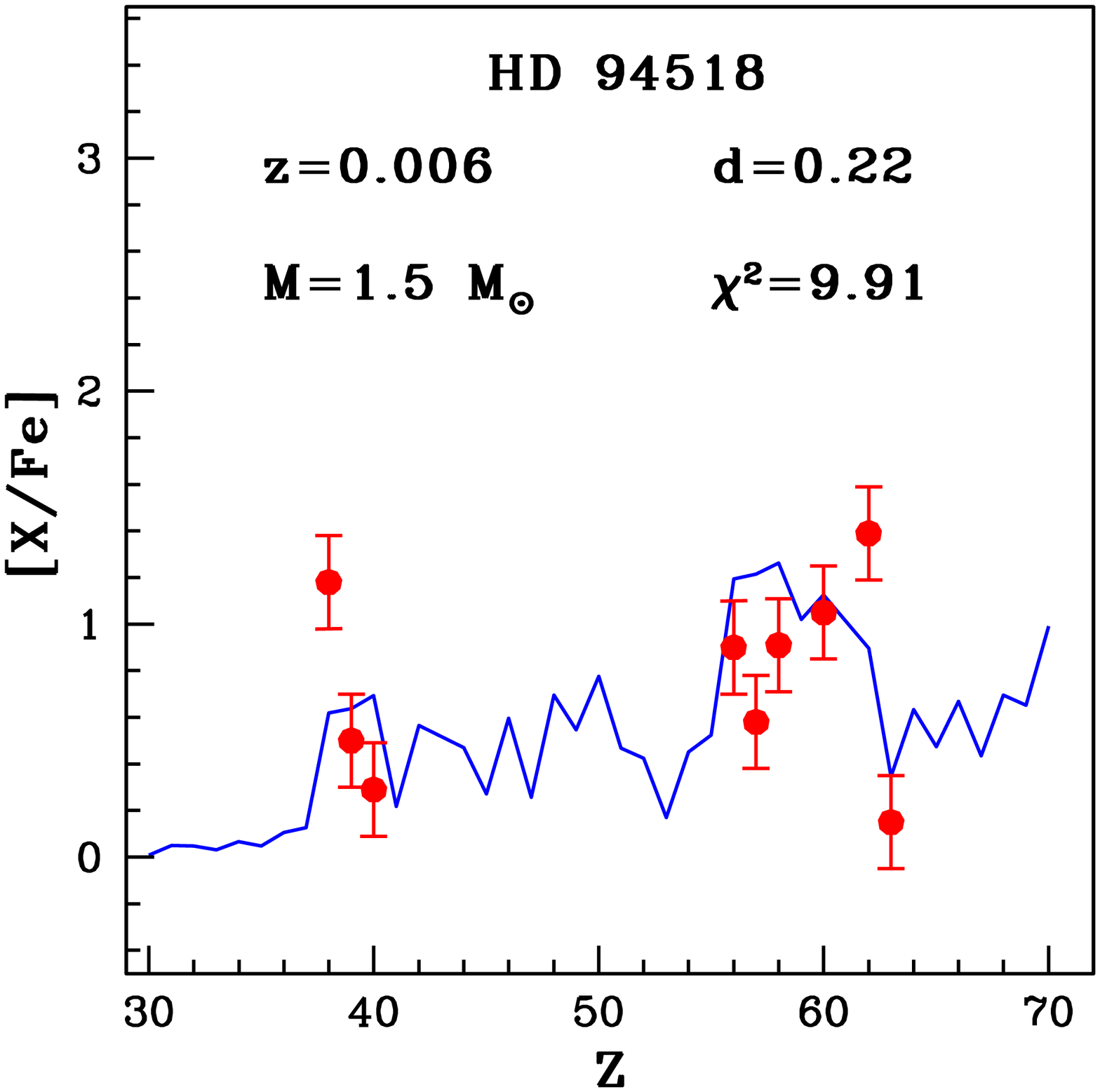}
\includegraphics[width=0.8\columnwidth, height= 0.8\columnwidth]{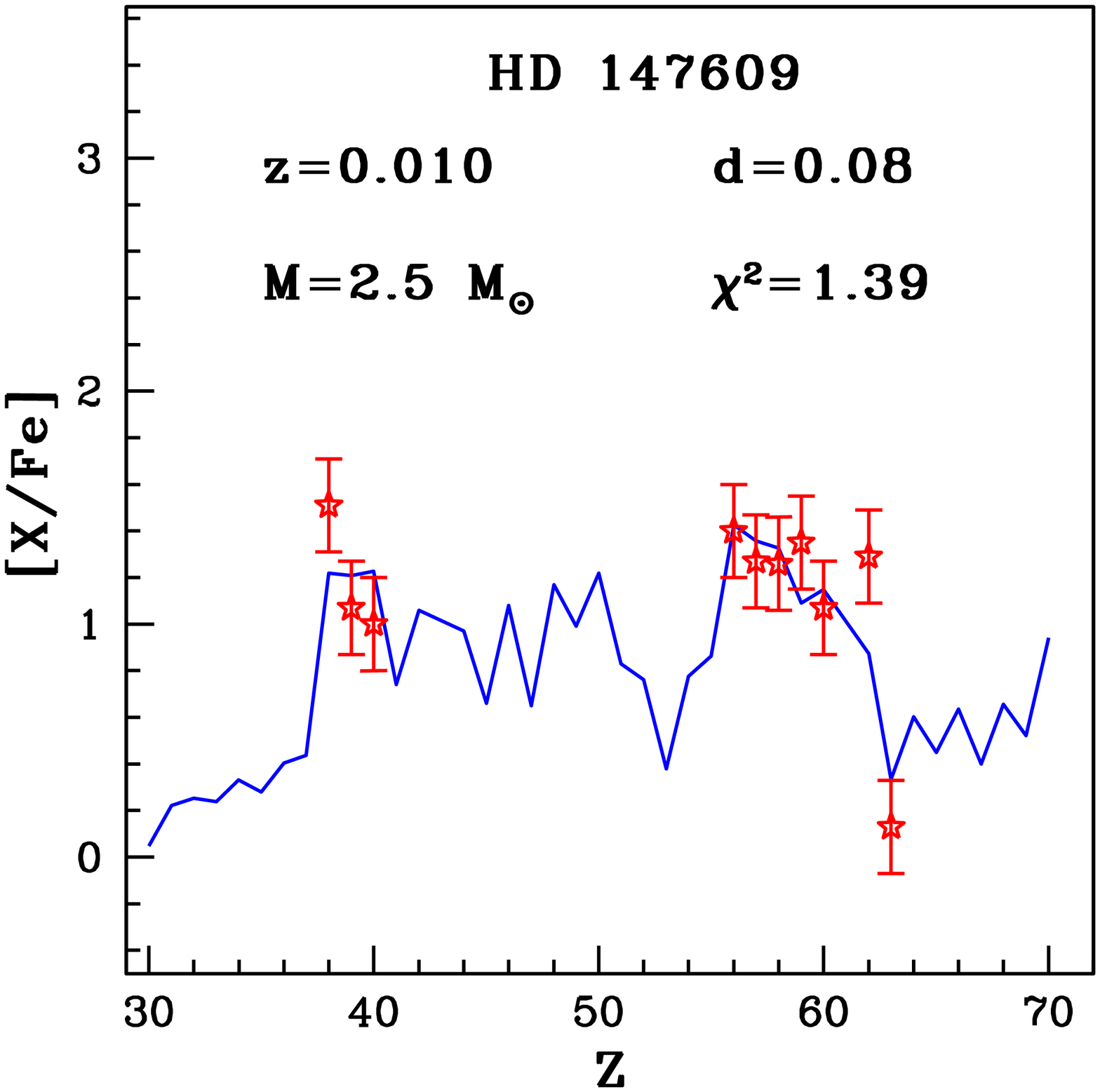}
\includegraphics[width=0.8\columnwidth, height= 0.8\columnwidth]{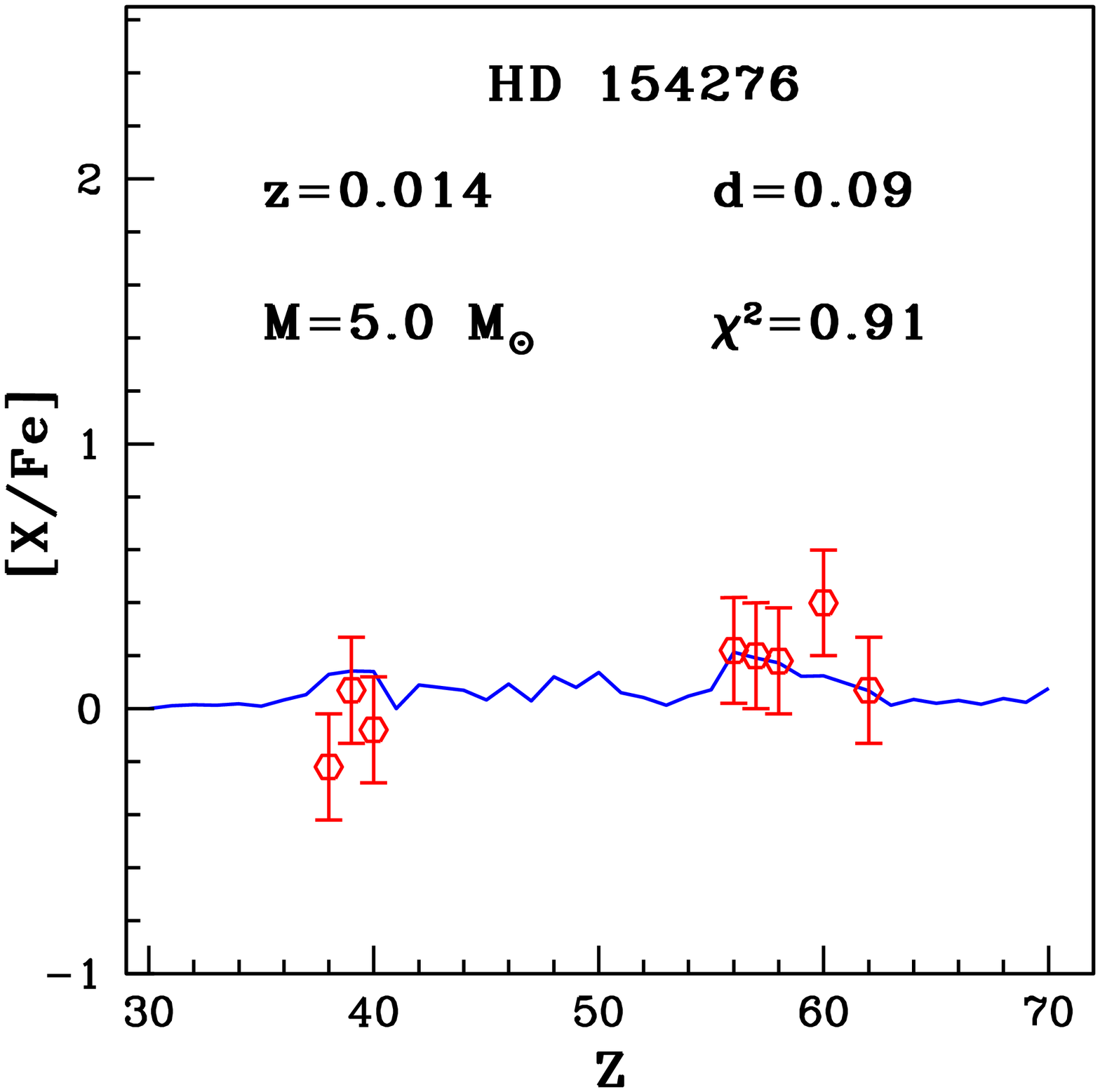}
\caption{Solid curve represent the best fit for the parametric model function.
The points with error bars indicate the observed abundances in 
(i) \textit{Top panel:} HD~94518 (ii) \textit{Middle panel:} HD~147609 
(iii) \textit{Bottom panel:} HD~154276.} 
\label{parametric2}
\end{figure}

\begin{figure}
\centering
\includegraphics[width=0.8\columnwidth, height= 0.8\columnwidth]{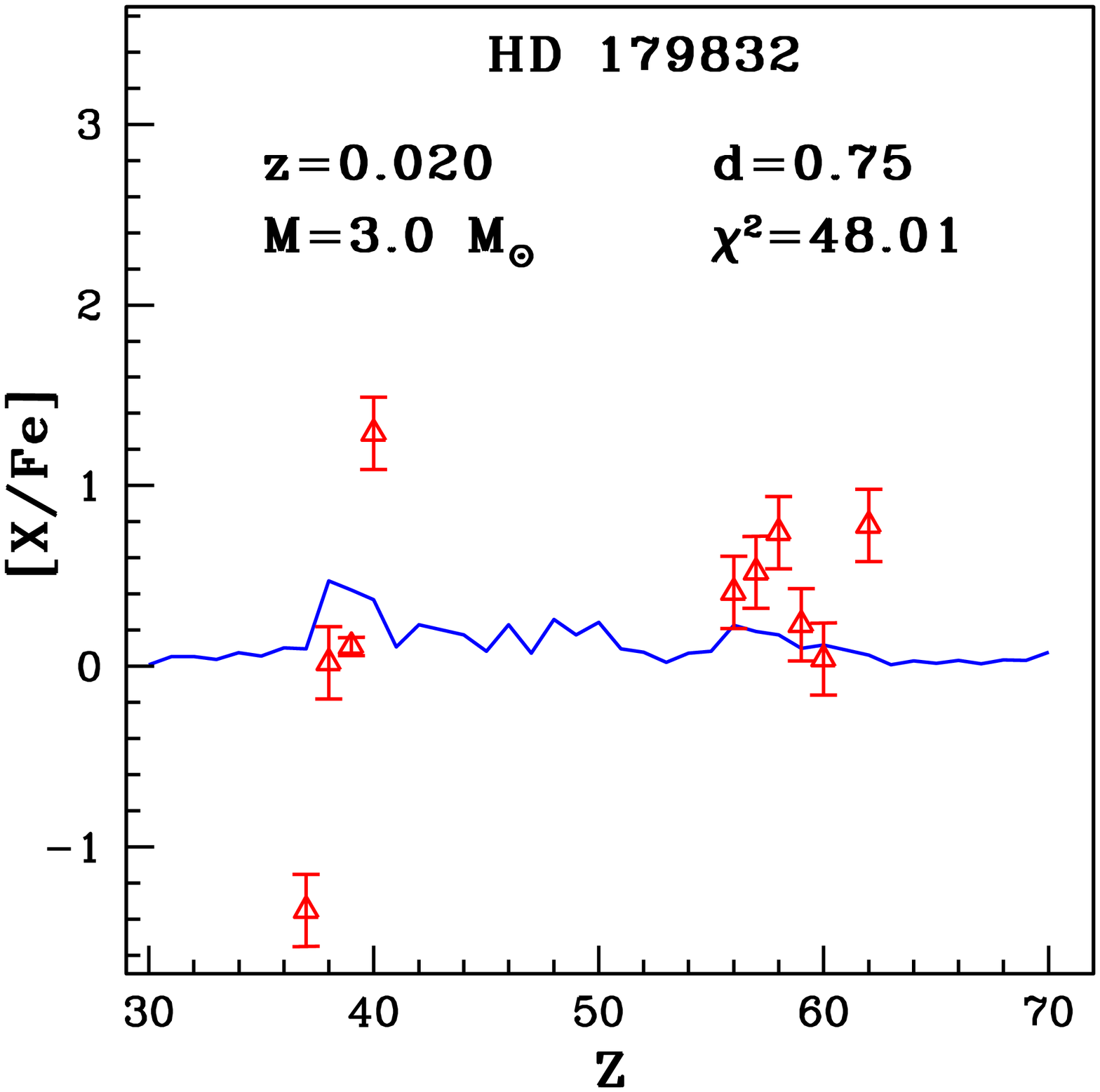}
\includegraphics[width=0.8\columnwidth, height= 0.8\columnwidth]{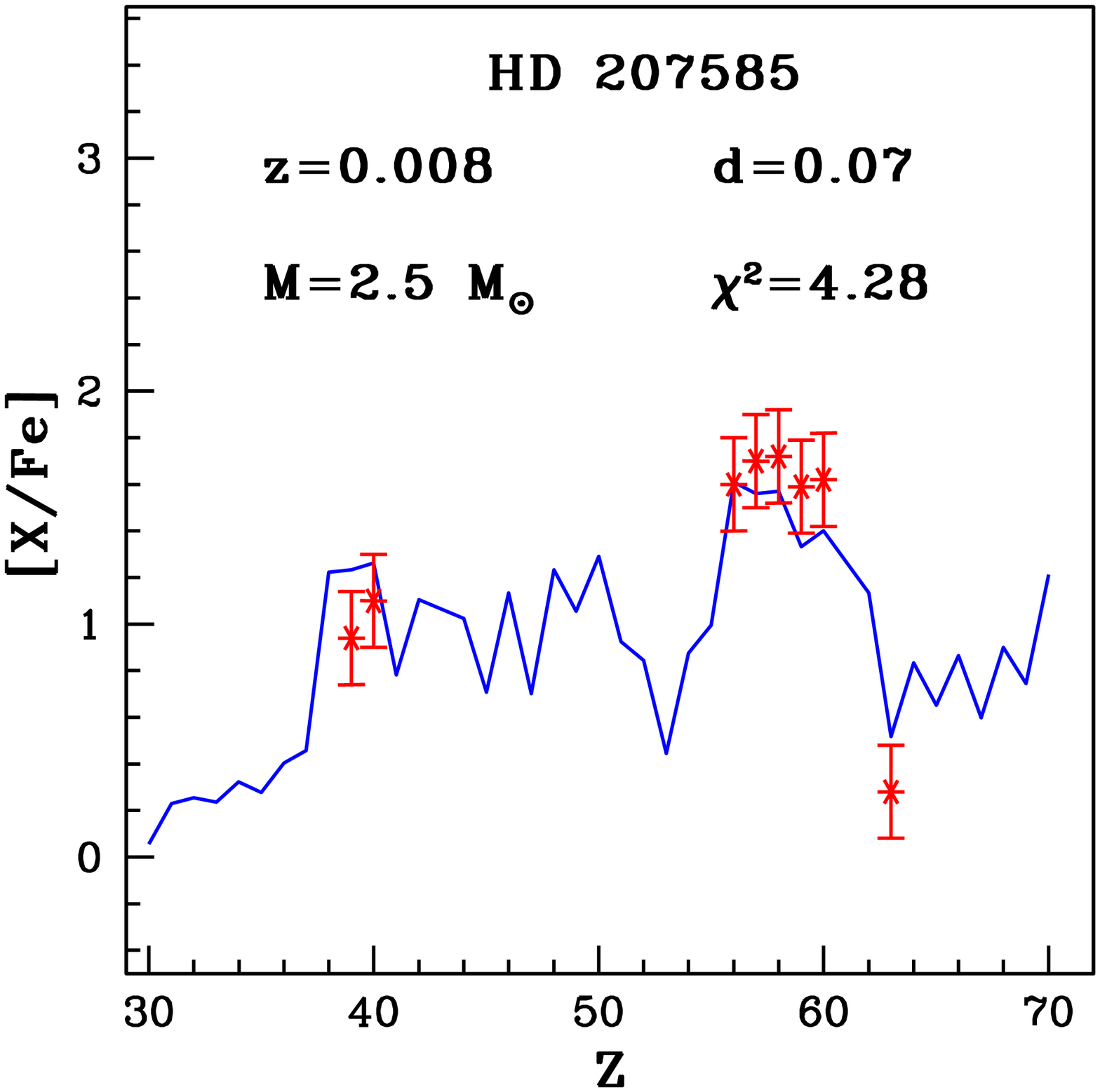}
\caption{Solid curve represent the best fit for the parametric model function.
The points with error bars indicate the observed abundances in 
(i) \textit{Top panel:} HD~179832  
(ii) \textit{Bottom panel:} HD~207585.} 
\label{parametric3}
\end{figure}

\begin{figure}
\centering
\includegraphics[width=0.8\columnwidth, height= 0.8\columnwidth]{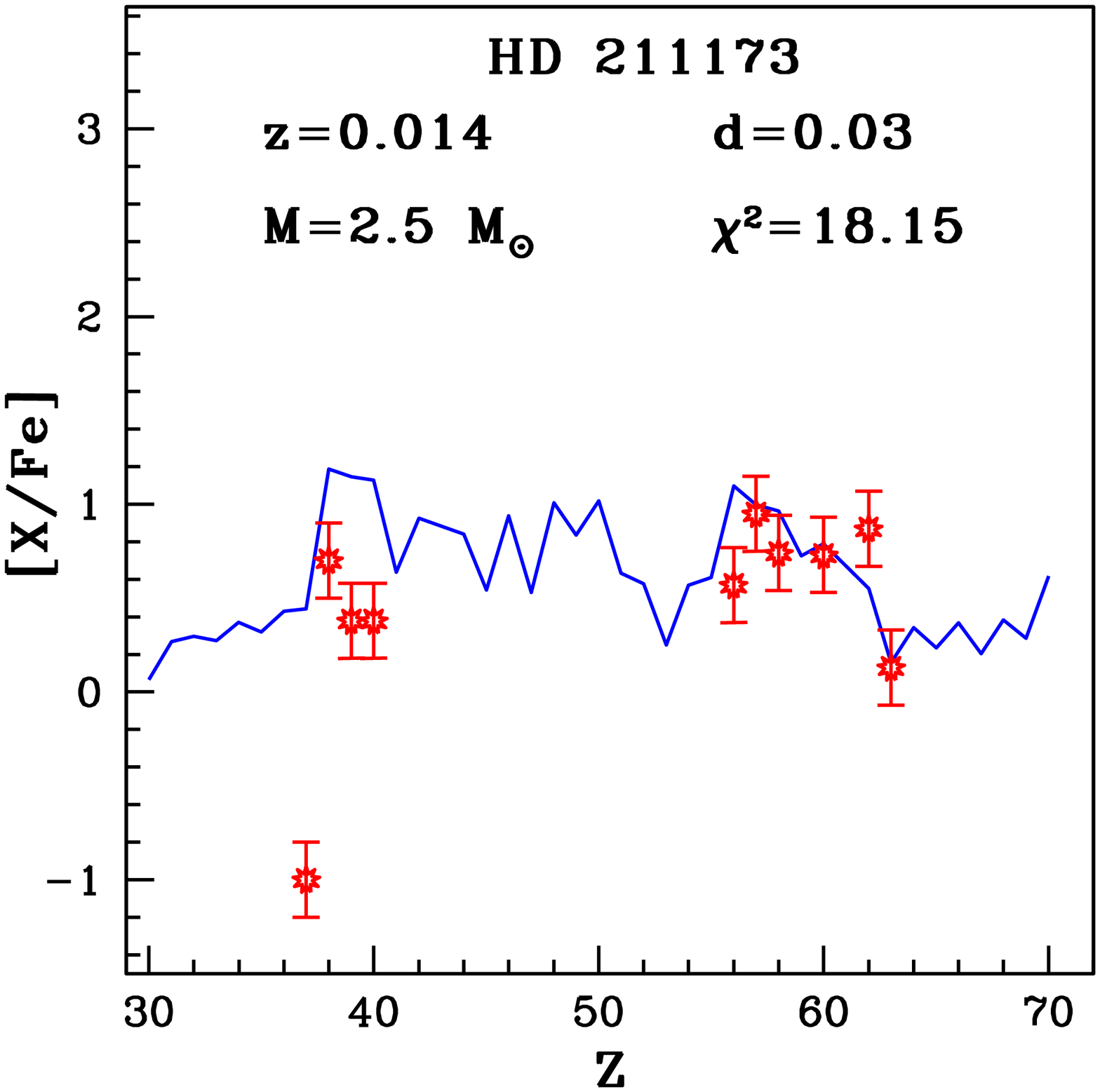}
\includegraphics[width=0.8\columnwidth, height= 0.8\columnwidth]{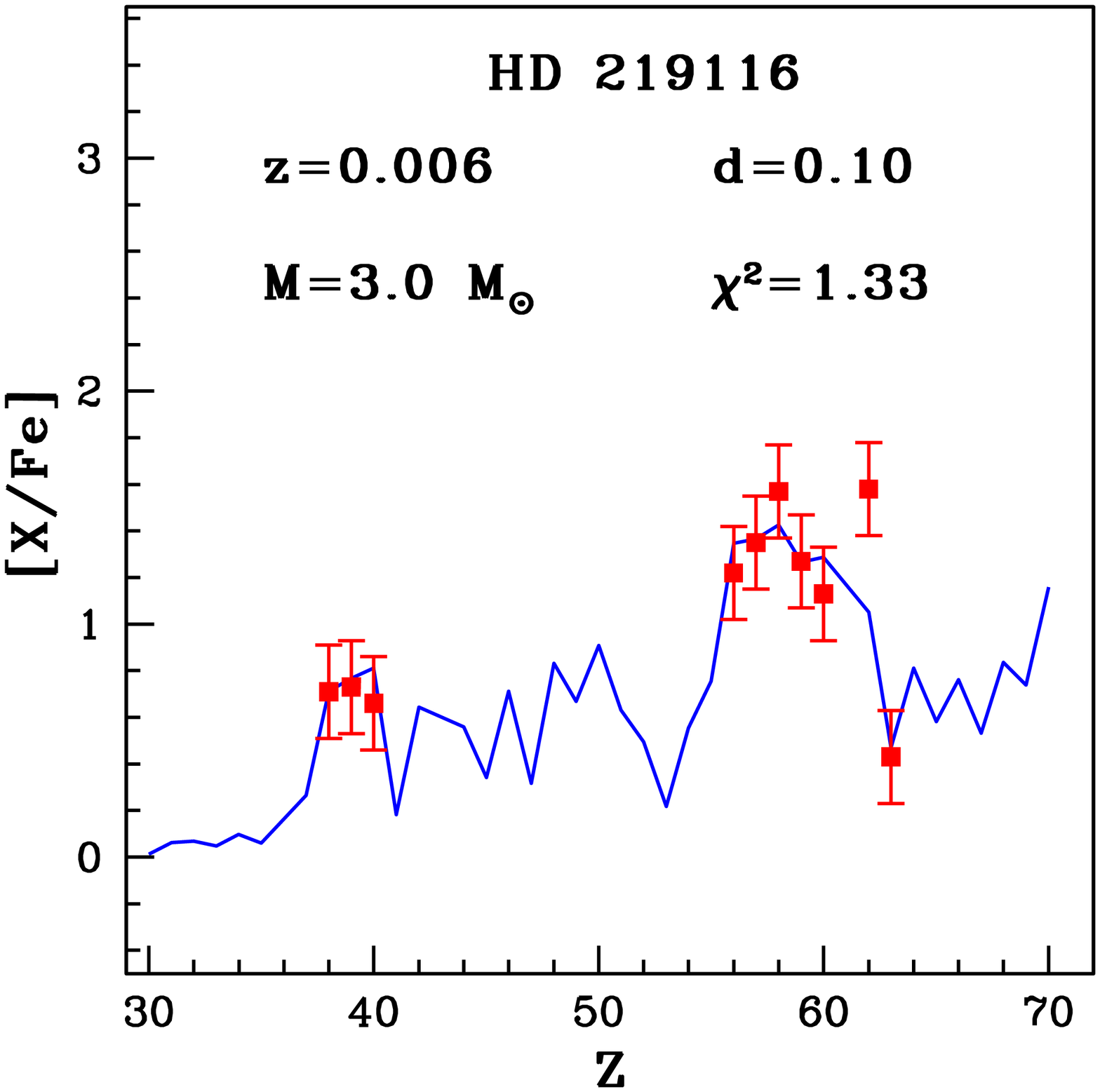}
\caption{Solid curve represent the best fit for the parametric model function.
The points with error bars indicate the observed abundances in 
(i) \textit{Top panel:} HD~211173  
(ii) \textit{Bottom panel:} HD~219116.} 
\label{parametric4}
\end{figure}

We have performed a parametric analysis in order to find the 
dilution experienced by the s-rich material after the mass transfer.
The  dilution factor, d, is defined as M$_{\star}^{env}$/M$_{AGB}^{transf}$ = 10$^{d}$,
where M$_{\star}^{env}$ is the mass of the envelope of the observed star 
after the mass transfer, M$_{AGB}^{transf}$ is the mass transferred from the AGB.
The dilution factor is derived by comparing the observed abundance 
with the predicted abundance from FRUITY model for the heavy elements 
(Rb, Sr, Y, Zr, Ba, La, Ce, Pr, Nd, Sm and Eu). The solar
values has been taken as the initial composition. The observed
elemental abundances are fitted with the parametric model function
(Husti et al. 2009). The best fits masses and corresponding dilution 
factors along with the $\chi^{2}$ values are given in 
Table \ref{parametric model}. The goodness  of fit of the parametric 
model function is determined by the uncertainty in the observed 
abundance. The best  fits obtained are shown in Figures 
\ref{parametric1} - \ref{parametric4}. All the Ba stars are found 
to have low-mass AGB companions with M $\leq$ 3 M$_{\odot}$.
Among our stars, HD~147609 is found to have a companion of 3 M$_{\odot}$ by 
Husti et al. (2019), whereas our estimate is 2.5 M$_{\odot}$.

{\footnotesize
\begin{table*}
\caption{The best fitting mass, dilution factor and reduced chi-square values.}  \label{parametric model}
\resizebox{\textwidth}{!}{\begin{tabular}{lcccccccccccc}
\hline                       
Star name/  &     & HD~24035  & HD~32712  & HD~36650  & HD~94518  & HD~147609 & HD~154276 & HD~179832 & HD~207585 & HD~211173 & HD~219116  \\
mass (M/M$_{\odot}$) & & & & & & & & & &  \\
\hline
\hline   
1.5   & d          & - & - & - & 0.22 & - & 0.71 & 0.21  & - & - & 0.04 \\
      & $\chi^{2}$ & - & - & - & 9.91 & - & 1.46 & 51.39 & - & - & 1.40 \\
\hline   
2.0   & d          & - & 0.001 & - & 0.52  & - & 1.18 & 0.65  & - &  - & 0.36 \\
      & $\chi^{2}$ & - & 16.14 & - & 10.14 & - & 1.43 & 48.04 & - &  - & 1.55 \\
\hline   
2.5   & d          & 0.07 & 0.08  & 0.10 & 0.62  & 0.08 & 1.31 & 0.82  & 0.07 & 0.03  & 0.46 \\
      & $\chi^{2}$ & 1.92 & 17.64 & 8.15 & 10.31 & 1.39 & 1.60 & 48.06 & 4.28 & 18.15 & 1.66 \\
\hline   
3.0   & d          & - & - & 0.04 & 0.27 & - & 1.20 & 0.75  & - & - & 0.10 \\
      & $\chi^{2}$ & - & - & 8.08 & 9.92 & - & 1.52 & 48.01 & - & - & 1.33 \\
\hline   
4.0   & d          & - & - & & - & - & 0.58 & -  & - &  - & - \\
      & $\chi^{2}$ & - & - & & - & - & 1.01 & -  & - &  - & -  \\
\hline   
5.0   & d          & - & - & & - & - & 0.09 & - & - &  - & - \\
      & $\chi^{2}$ & - & - & & - & - & 0.91 & - & - &  - & - \\
\hline   
\end{tabular}}
 \end{table*}
}

\subsection{Discussion on individual stars}
{\small\textbf{HD~24035,  HD~219116, HD~32712, HD~36650, HD~207585 
and HD~211173:}}\\
These  objects are listed in  the CH star catalogue of
Bartkevicius (1996) as well as in  the barium star catalogue 
of L\"u (1991). While Smith et al. (1993) classified  HD~219116 as 
a CH subgiant, MacConnell et al. (1972) and Mennessier et al. (1997)  
suggested  these objects  to be giant barium stars.
Based on our temperature and luminosity estimates, their locations 
in the H-R diagram correspond to the giant phase of evolution, except
for HD~207585 which is found to be a strong Ba sub-giant. 
Earlier studies on these objects include  Masseron et al. (2010) and
de Castro et al. (2016) on abundance analysis. We have estimated  
the abundances of all the important  s-process  elements and  Eu in 
these objects except Sr in HD~24035. Based  on our abundance annalysis
we find these objects to satisfy the criteria for 
s-process enriched stars (Beers \& Christlieb 2005) with 
[Ba/Fe]$>$1 and  [Ba/Eu]$>$0.50 respectively.  Following  
Yang et al. (2016), they can also be  included in strong 
Ba giant category while HD~211173 is a mild Ba giant. 
They show the characteristics of Ba stars
with  estimated   C/O  $\sim$0.95, 0.51. 0.56, 0.24 and 0.59 for
HD~219116, HD~32712, HD~36650, HD~207585 and HD~211173 respectively. 
HD~24035 shows the largest enhancement of Ba among our 
program stars with [Ba/Fe]$\sim$1.71
and largest mean s-process abundance with [s/Fe]$\sim$1.55. 
A comparison with FRUITY models shows that the former AGB 
companions  of HD~24035, HD~32712, HD~36650, HD~219116, HD~207585 
and HD~211173  are low mass objects  with masses 2.5 M$_{\odot}$, 
2.0 M$_{\odot}$, 3.0 M$_{\odot}$, 3.0 M$_{\odot}$, 2.5 M$_{\odot}$, 
and  2.5 M$_{\odot}$ respectively.
From kinematic analysis we find  these objects to  belong 
to the thin disk population with probability $\geq$0.97.
The estimated  spatial velocities $<$ 85 km/s, also satisfies 
the  criterion of  Chen et al. (2004) for  stars to be thin 
disk objects. From radial velocity monitoring, HD~24035 
and HD~207585 are confirmed to be binaries with orbital periods 
of 377.82$\pm$0.35 days (Udry et al. 1998a) and 672$\pm$2 days 
(Escorza et al. 2019) respectively.    \\

\noindent {\small\textbf{HD~94518:}}  
This object belongs to the CH star catalogue of Bartkevicius (1996).
Our abundance analysis places this object in the strong Ba star 
category with C/O$\sim$ 0.06.
The abundance pattern observed in this star resembles with 
that of a 1.5 M$_{\odot}$ AGB star. 
The position of this stars in H-R diagram shows this object to be a 
subgiant star. Kinematic analysis shows this object to belong to  
thick disk population  with a  probability $\sim$0.95.  \\

\noindent {\small\textbf{HD~147609:}} 
This star is listed in CH star catalogue of Bartkevicius (1996).
This star is a strong Ba dwarf with C/O$\sim$ 0.27. 
Comparison of the observed abundance in HD~147609 with the 
FRUITY model shows  close resemblance with that seen in 
2.5 M$_{\odot}$ AGB star.
Kinematic analysis has shown that  this object belongs to thin disk 
population with  characteristic spatial velocity of thin disk objects.
The radial velocity monitoring study by Escorza et al. (2019) 
has confirmed this object to be a binary with an orbital period of
1146$\pm$1.5 days. \\

\noindent {\small\textbf{HD~154276:}} 
This star is listed in CH star catalogue of Bartkevicius (1996). 
Our analysis have shown that this star is a dwarf.  
Our analysis based on mean s-process abundance, [s/Fe], revealed 
that this object can not be considered as a Ba star. \\
 
\noindent {\small\textbf{HD~179832:}} 
This object belongs to the CH star catalogue of Bartkevicius (1996).
We have presented a first time detailed abundance analysis for 
this object.  Our analyses have shown that this object is a mild Ba giant. 
The abundance trend observed in this star suggest that the former
companion AGB star's mass to be 3 M$_{\odot}$ .  
From kinematic analysis, HD~179832 is found to be a thin disk object 
with probability of 0.99. The spatial velocity is estimated to be 
11.97 km s$^{-1}$, as expected for thin disk stars (Chen at al. 2004). \\

{\footnotesize
\begin{table*}
\caption{Comparison of the heavy elemental abundances of our program stars with the literature values.}  \label{abundance_comparison_literature}
\resizebox{0.6\textheight}{!}{\begin{tabular}{lcccccccccc}
\hline                       
Star name     & [Fe I/H]  & [Fe II/H] & [Fe/H]   & [Rb I/Fe] & [Sr/Fe] & [Y I/Fe] & [Y II/Fe] & [Zr I/Fe]    & Ref \\
\hline
HD~24035      & $-$0.51   & $-$0.50   & $-$0.51  & -         & -       & 1.61     & -         & 1.21         & 1 \\
              & $-$0.23   & $-$0.28   & $-$0.26  & -         & -       & 1.35     & -         & 1.20         & 2  \\
              & -         & -         & $-$0.14  & -         & -       & -        & -         & -            & 3 \\
HD~32712      & $-$0.25   & $-$0.25   & $-$0.25  & $-$1.13   & 0.03    & 0.56     & 1.04      & 0.52         & 1 \\ 
              & $-$0.24   & $-$0.25   & $-$0.25  & -         & -       & 0.74     & -         & 0.56         & 2  \\ 
HD~36650      & $-$0.02   & $-$0.02   & $-$0.02  & $-$0.82   & 0.66    & 0.51     & 0.70      & 0.51         & 1  \\
              & $-$0.28   & $-$0.28   & $-$0.28  & -         & -       & 0.55     & -         & 0.46         & 2  \\ 
HD~94518      & $-$0.55   & $-$0.55   & $-$0.55  & -         & 1.18    & -        & 0.50      & -            & 1  \\
              & $-$0.49   & $-$0.50   & $-$0.50  & -         & 0.55    & -        & -         & -            & 4   \\
HD~147609     & $-$0.28   & $-$0.28   & $-$0.28  & -         & 1.51    & -        & 1.07      & -            & 1  \\ 
              & $-0.45$   & $+0.08$   & $-$0.45  & -         & 1.32    & -        & 1.57      & 0.89         & 5  \\  
              & -         & -         & -        & -         & -       & 0.96     & -         & 0.80         & 6  \\
HD~154276     & $-$0.09   & $-$0.10   & $-$0.10  & -         & $-$0.22 & -        & 0.07      & -            & 1  \\
              & -         & -         & $-$0.29  & -         & -       & $-$0.07  & -         & -            & 4  \\       
HD~179832     & +0.23     & +0.23     & +0.22    & $-$1.35   & 0.02    & -        & 0.11      & 1.29         & 1  \\ 
HD~207585     & $-$0.38   & $-$0.38   & $-$0.38  & -         & -       & 0.94     & 1.37      & 1.10         & 1  \\
              & -         & -         & $-$0.50  & -         & -       & -        & -         & -            & 3  \\
              & -         & -         & $-$0.57  & -         & -       & 1.29     & -         & 1.50         & 7  \\
HD~211173     & $-$0.17   & $-$0.17   & $-$0.17  & $-$1.00   & 0.70    & 0.38     & 0.65      & 0.38         & 1  \\ 
              & -         & -         & $-$0.12  & -         & -       & -        & -         & -            & 3  \\
HD~219116     & $-$0.45   & $-$0.44   & $-$0.45  & -         & 0.71    & 0.73     & 0.75      & 0.66         & 1  \\
              & $-$0.61   & $-$0.62   & $-$0.62  & -         & -       & 0.59     & -         & 0.65         & 2  \\                
              & -         & -         & $-$0.34  & -         & -       & -        & -         & -            & 3  \\ 
              & -         & -         & $-$0.30  & -         & -       & -        & -         & -            & 8  \\           
\hline                       
Star name     & [Zr II/Fe] & [Ba II/Fe] & [La II/Fe] & [Ce II/Fe] & [Pr II/Fe] & [Nd II/Fe] & [Sm II/Fe] & [Eu II/Fe] & Ref \\
\hline
HD~24035      & 1.89       & 1.71       & 1.63       & 1.70       & 1.98       & 1.41       & 2.03       & 0.19       & 1 \\
              & -          & -          & 2.70       & 1.58       & -          & 1.58       & -          & -          & 2  \\
              & -          & 1.07       & 1.01       & 1.63       & -          & -          & -          & 0.32       & 3  \\
HD~32712      & 0.82       & 1.39       & 1.25       & 1.68       & 1.67       & 1.76       & 1.71       & 0.04       & 1   \\
              & -          & -          & 1.53       & 1.16       & -          & 1.19       & -          & -          & 2  \\
HD~36650      & -          & 0.79       & 0.62       & 0.99       & 0.85       & 0.94       & 0.99       & $-$0.21    & 1  \\
              & -          & -          & 0.83       & 0.68       & -          & 0.57       & -          & -          & 2  \\
HD~94518      & 0.29       & 0.90       & 0.58       & 0.91       & -          & 1.05       & 1.39       & $-$0.17    & 1  \\
              & -          & 0.77       & -          & -          & -          & -          & -          & -          & 4  \\ 
HD~147609     & 1.00       & 1.40       & 1.27       & 1.26       & 1.35       & 1.07       & 1.29       & 0.13       & 1  \\
              & 1.56       & 1.57       & 1.63       & 1.64       & 1.22       & 1.32       & 1.09       & 0.74       & 5  \\
              & -          & -          & -          & -          & -          & 0.98       & -          & -          & 6  \\
HD~154276     & $-$0.08    & 0.22       & 0.20       & 0.18       & -          & 0.40       & 0.07       & -          & 1  \\
              & -          & $-$0.03    & -          & -          & -          & -          & -          & -          & 4  \\          
HD~179832     & 1.44       & 0.41       & 0.52       & 0.74       & 0.23       & 0.04       & 0.78       & 0.00       & 1  \\  
HD~207585     & 1.20       & 1.60       & 1.70       & 1.72       & 1.59       & 1.62       & 2.04       & $-$0.02    & 1 \\
              & -          & 1.23       & 1.37       & 1.41       & -          & -          & -          & 0.58       & 3  \\
              & -          & -          & 1.60       & 0.84       & 0.61       & 0.93       & 1.05       & -          & 7  \\
HD~211173     & 0.39       & 0.57       & 0.95       & 0.74       & 1.59       & 0.73       & 0.87       & $-$0.17    & 1  \\
              & -          & 0.35       & 0.29       & 0.73       & -          & -          & -          & 0.15       & 3  \\               
HD~219116     & -          & 1.22       & 1.35       & 1.57       & 1.27       & 1.13       & 1.58       & 0.13       & 1  \\
              & -          & -          & 1.21       & 1.07       & -          & -          & -          & -          & 2   \\   
              & -          & 0.77       & 0.56       & 0.80       & -          & -          & -          & 0.17       & 3  \\
              & -          & 0.90       & -          & -          & -          & 1.43       & -          & -          & 8  \\
\hline   
\end{tabular}}
 
References: 1. Our work, 2. de Castro (2016), 3. Masseron et al. (2010),  
4. Bensby et al. (2014), 5. Allen \& Barbuy (2006a), 6. North et al. (1994a), 
7. Luck \& Bond (1991), 8. Smith et al. (1993) \\

\end{table*}
}                    

\section{CONCLUSIONS}
Results from high resolution spectroscopic analysis of ten objects are 
presented. All the objects are listed in the CH star catalog of 
Bartkevicius (1996). Six of them are also listed in the barium star catalog
of L\"u (1991). Except for one object HD~154276, all other objects are shown 
to be bonafied  barium stars from our analysis.  Although it
satisfies the criteria of Yang et al. (2016) to be a mild
barium star, our detailed abundance analysis shows this object to 
 be  a normal metal-poor star. An analysis based on the mean s-process 
abundance clearly
shows that this particular star lies among the stars rejected as
barium stars  by de Castro et al. (2016).

\par Some of the objects analysed here, although  common to the sample  
analysed by different authors,   abundances of important  heavy elements  
such as Rb, and  C, N, O are not found in  literature. New results for 
these elements  are presented  in this work. 

\par We have presented  first time abundance results for HD~179832 and 
shown it to be a mild barium giant. Kinematic analysis have shown it to be
a thin disk object. A parametric model based analysis have shown the object's 
former  companion AGB star's mass  to be about 3M$_{\odot}$.
  
\par The sample of  stars analysed here covers a metallicity range 
from $-$0.55 to +0.23, and a kinematic analysis has shown that all of them
 belong to the Galactic disk, as expected for barium stars. 

\par  The estimated mass  of the barium stars  are consistent 
with that observed for other barium  stars (Allen \& Barbuy 2006a, 
Liang et al. 2003, Antipova et  al. 2004, de Castro et al. 2016). 
The abundance estimates are consistent with the operation of
$^{13}$C($\alpha$, n)$^{16}$O source in the former low-mass AGB companion. 

\par We did not find any enhancement
of Mg  in our sample, that discards  the source of neutron as the 
$^{22}$Ne($\alpha$, n)$^{25}$Mg reaction. An enhancement of Mg 
abundances when compared with their counterparts in disk stars and 
normal giants would have indicated the operation 
of $^{22}$Ne($\alpha$, n)$^{25}$Mg.
 
\par The detection of Rb I line at 7800.259 {\rm \AA} 
in the spectra of HD~32712, HD~36650, HD~179832 and HD~211173 
allowed us to determine [Rb/Zr] ratio. This ratio gives an indication of 
the neutron source at the s-process site 
and in turn provides  clues to the mass of the star.
We have obtained  negative values for this ratio in all the four stars.
 The negative values obtained for these  stars indicate the 
operation of $^{13}$C($\alpha$, n)$^{16}$O reaction.
As this reaction occurs in the low-mass AGB stars, we confirm 
that the former companions 
of these stars are low-mass AGB stars with M $\leq$ 3 M$_{\odot}$. 

\par Distribution of abundance patterns and [hs/ls] ratios also indicate 
low-mass companions for the  objects for which [Rb/Zr] could not be estimated.
A comparison of observed abundances  with the predictions from  FRUITY 
models, and with those that are observed  in  low-mass AGB 
stars from literature, confirms  low-mass for the former companion AGB stars.

\section{ACKNOWLEDGMENT}
 We thank the staff at IAO and at the remote control station at 
CREST, Hosakotte for assisting during the observations.
Funding from the DST SERB project No. EMR/2016/005283 is gratefully 
acknowledged. 
This work made use of the SIMBAD astronomical database, operated
at CDS, Strasbourg, France, and the NASA ADS, USA.
This work has made use of data from the European Space Agency (ESA) 
mission Gaia (https://www.cosmos.esa.int/gaia), processed by the Gaia 
Data Processing and Analysis Consortium 
(DPAC, https://www.cosmos.esa.int/web/gaia/dpac/consortium).
T.M. acknowledges support provided by the Spanish Ministry
of Economy and Competitiveness (MINECO) under grant AYA-
2017-88254-P. Based on data collected using  HESP, UVES
and FEROS. The authors would like to thank the  referee for useful suggestions
that had improved the readability of the paper.

{\footnotesize
\begin{table*}
\caption{Equivalent widths (in m\r{A}) of Fe lines used for deriving 
atmospheric parameters.} \label{linelist1}
\resizebox{\textwidth}{!}{\begin{tabular}{lccccccccccccccc}
\hline                       
Wavelength(\r{A}) & El & $E_{low}$(eV) & log gf & HD~24035 & HD~32712 & HD~36650 & HD~94518 & HD~147609 & HD~154276 & HD~179832 & HD~207585 & HD~211173 & HD~219116 & Ref  \\ 
\hline 
4114.445 & Fe I & 2.832 & $-$1.220 & - & - & - & - & - & - & - & -& - & - &  1 \\
4132.899 &      & 2.850 & $-$1.010 & -  & - & -  & 88.7(6.94) & - & - & - & 83.4(6.96) & 132.2(7.31) & - & 1  \\
4153.900 &      & 3.400 & $-$0.320 & 129.9(6.79) & - & - & - & - & - & - & 90.8(6.90)  & - & -  &  1            \\
4154.499 &      & 2.830 & $-$0.690 & - & - & - & - & - & - & - & - & - & - &     2       \\
4184.891 &      & 2.832 & $-$0.860 & - & - & - & - & - & - & - & -& - & - &  1 \\
\hline
\end{tabular}}

The numbers in the paranthesis in columns 5-14 give the derived 
abundances from the respective line. \\
References: 1. F\"uhr et al. (1988) 2. Kurucz (1988)\\
\textbf{Note:} This table is available in its entirety in online only.
A portion is shown here for guidance regarding its form and content.
\end{table*}
}

{\footnotesize
\begin{table*}
\caption{Equivalent widths (in m\r{A}) of lines used for deriving 
elemental abundances.} \label{linelist2}
\resizebox{\textwidth}{!}{\begin{tabular}{lccccccccccccccc}
\hline                       
Wavelength(\r{A}) & El & $E_{low}$(eV) & log gf & HD~24035 & HD~32712 & HD~36650 & HD~94518 & HD~147609 & HD~154276 & HD~179832 & HD~207585 & HD~211173 & HD~219116 & Ref  \\ 
\hline 
5682.633 & Na I & 2.102 & $-0.700$ & 126.7(6.29) & 133.5(6.31) & 126.0(6.44) & 63.5(5.86) & 66.2(6.13) & 83.5(6.23) & 132.2(6.55) & 79.0(6.17) & 121.5(6.43) & 104.5(6.13) & 1 \\
5688.205 &      & 2.105 & $-$0.450 & 137.1(6.21) & 151.3(6.31) & 130.1(6.26) & 85.2(5.94) & 98.2(6.39) & - & 145.1(6.50) & 98.0(6.22) & 136.2(6.41) & 126.8(6.24) & 1 \\
6154.226 &      & 2.102 & $-$1.560 & 49.4(5.91) & 66.9(6.04) & 67.7(6.30) & 20.5(5.90) & - & 27.4(6.13) & - & 22.2(6.00) & 60.1(6.21) & 37.9(5.92) & 1 \\
6160.747 &      & 2.104 & $-$1.260 & 88.3(6.20) & 92.7(6.16) & 87.5(6.31) & 29.2(5.81) & - & 47.9(6.22) & - & 38.0(6.03) & 74.8(6.15) & 57.2(5.93) & 1 \\
4571.096 & Mg I & 0.000 & $-$5.691 & - & - & - & - & - & 108.9(7.88) & - & - & - & - & 2 \\
4702.991 &      & 4.346 & $-$0.666 & - & 179.6(7.10) & - & 179.6(7.27) & 156.2(7.56) & - & - & 149.0(7.16) & - & 180.0(7.48) & 3 \\
4730.029 &      & 4.346 & $-$2.523 & -        &-          & 85.1(7.71) & 45.5(7.46) & 32.3(7.48) & 60.2(7.80) & 87.1(7.81) & - & 76.8(7.60) & - & 3 \\

\hline
\end{tabular}}

The numbers in the paranthesis in columns 5-14 give the derived abundances from the respective line. \\
References: 1. Kurucz et al. (1975), 2. Laughlin et al. (1974), 3. Lincke et al. (1971)\\
\textbf{Note:} This table is available in its entirety in online only.
A portion is shown here for guidance regarding its form and content.
\end{table*}
}

{\footnotesize
\begin{table*}
\caption{Estimates of  [ls/Fe], [hs/Fe], [s/Fe], [hs/ls], [Rb/Sr], [Rb/Zr],  C/O} \label{hs_ls}
\begin{tabular}{lcccccccc}
\hline                       
Star name    & [Fe/H]   & [ls/Fe] & [hs/Fe]  & [s/Fe] & [hs/ls] & [Rb/Sr]  & [Rb/Zr]  & C/O\\ 
\hline
HD 24035     & $-$0.51  & 1.41    & 1.61     & 1.55   & 0.20    &  --     & --       & --   \\  
HD 32712     & $-$0.25  & 0.37    & 1.52     & 1.03   & 1.15    & $-$1.06 & $-$1.65  & 0.51 \\
HD 36650     & $-$0.02  & 0.56    & 0.84     & 0.72   & 0.28    & $-$1.48 & $-$1.33  & 0.56  \\
HD 94518     & $-$0.55  & 0.67    & 0.86     & 0.77   & 0.19    & --      & --       & 0.06 \\
HD 147609    & $-$0.28  & 1.19    & 1.25     & 1.23   & 0.06    & --      & --       & 0.27  \\
HD 154276    & $-$0.10  & $-$0.08 & 0.25     & 0.11   & 0.33    & --      & --       & -- \\
HD 179832    & +0.23    & 0.47    & 0.66     & 0.45   & 0.19    & $-$1.37 & $-$2.64  & -- \\
HD 207585    & $-$0.38  & 1.02    & 1.66     & 1.45   & 0.64    & --      & --       & 0.24 \\
HD 211173    & $-$0.17  & 0.49    & 0.75     & 0.64   & 0.26    & $-$1.70 & $-$1.38  & 0.59 \\   
HD 219116    & $-$0.45  & 0.70    & 1.32     & 1.05   & 0.62    &--       & --       & 0.95 \\         
\hline
\end{tabular}
\end{table*}
}

{}

\newpage
\section*{Appendix}

\begin{figure}
\centering
\includegraphics[width=0.8\columnwidth]{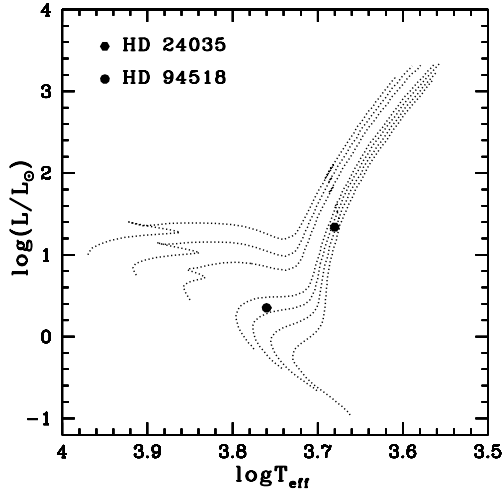}
\caption{The locations of HD~24035 and HD~94518. The evolutionary tracks
for 0.6, 0.7, 0.8, 0.9, 1.2, 1.4, and 1.6 M$_{\odot}$ are shown from bottom to top for z = 0.004.} \label{track_004}
\end{figure}

\begin{figure}
\centering
\includegraphics[width=0.8\columnwidth]{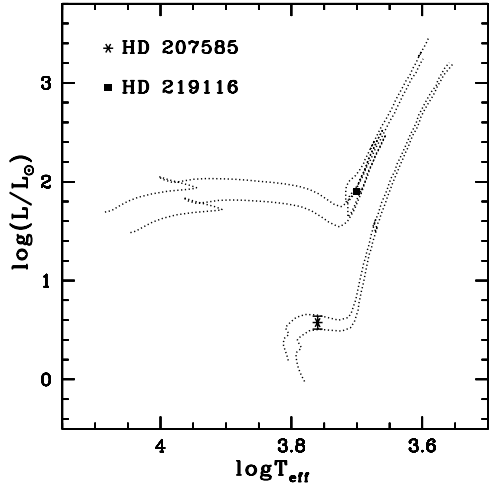}
\caption{The locations of HD~207585 and HD~219116. The evolutionary tracks
for 1.0, 1.1, 2.2, and 2.5 M$_{\odot}$ are shown from bottom to top for z = 0.008.} \label{track_008}
\end{figure}

\begin{figure}
\centering
\includegraphics[width=0.8\columnwidth]{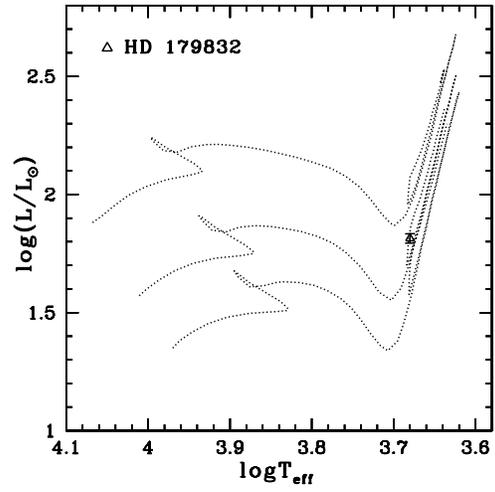}
\caption{The location of HD~179832. The evolutionary tracks
for 2.2, 2.5, and 3.0 M$_{\odot}$ are shown from bottom to top for z = 0.030.} \label{track_030}
\end{figure}

{\footnotesize
\begin{table*}
\caption{Spatial velocity and probability estimates for 
the program stars} \label{kinematic_analysis} 
\begin{tabular}{lcccccccc} 
\hline                       
Star name                   & U$_{LSR}$             & V$_{LSR}$          & W$_{LSR}$ & V$_{spa}$  & p$_{thin}$ & p$_{thick}$ & p$_{halo}$  & Population\\
                            & (kms$^{-1}$)          & (kms$^{-1}$)       & (kms$^{-1}$)  & (kms$^{-1}$) &          &             &         &      \\
\hline
HD 24035      &        $-$11.18$\pm$0.52   &   27.32$\pm$0.38   &    $-$16.35$\pm$0.58  &   33.74$\pm$0.15   &  0.99    &     0.01    &      0.00   &       Thin disk \\
HD 32712      &        76.78$\pm$0.66   &   $-$11.17$\pm$0.16   &    17.13$\pm$0.17  &   79.46$\pm$0.65   &  0.97    &     0.03    &       0.00  &         Thin disk   \\
HD 36650      &        21.44$\pm$0.13   &   $-$18.64$\pm$0.17   &    $-$10.09$\pm$0.13  &   30.15$\pm$0.06   &  0.99    &     0.01    &       0.00  &         Thin disk   \\
HD 94518      &        46.24$\pm$0.11   &   $-$119.85$\pm$0.43   &    $-$62.49$\pm$0.44  &   142.85$\pm$0.52   &  0.00    &     0.95    &      0.05   &        Thick disk  \\
HD 147609     &        19.45$\pm$0.87   &   $-$15.64$\pm$0.90   &    $-$4.11$\pm$1.03  &   25.30$\pm$0.06   &  0.99    &     0.01    &      0.00   &        Thin disk   \\
HD 154276     &        1.75$\pm$0.96   &   $-$116.16$\pm$0.76   &    22.48$\pm$0.76  &   118.33$\pm$0.58   &  0.07    &     0.92    &      0.01   &        Thick disk   \\
HD 179832     &        9.21$\pm$0.15   &   $-$5.80$\pm$0.32   &    $-$4.98$\pm$0.21  &   11.97$\pm$0.41   &  0.99    &     0.01    &      0.00   &        Thin disk   \\
HD 207585     &        $-$27.10$\pm$0.32   &   $-$41.91$\pm$2.68   &    44.05$\pm$1.04  &   66.57$\pm$1.13   &  0.72    &     0.28    &      0.00   &        Thin disk   \\
HD 211173     &        $-$48.53$\pm$0.88   &   27.71$\pm$0.42   &    4.79$\pm$0.54  &   56.09$\pm$0.51   &  0.99    &     0.01    &      0.00   &        Thin disk    \\
HD 219116     &        $-$46.22$\pm$1.31   &   $-$17.18$\pm$0.49   &    25.18$\pm$0.59  &   55.37$\pm$0.98   &  0.97    &     0.03    &      0.00   &        Thin disk    \\
\hline
\end{tabular} 
\end{table*}
}

\begin{figure}
\centering
\includegraphics[width=\columnwidth]{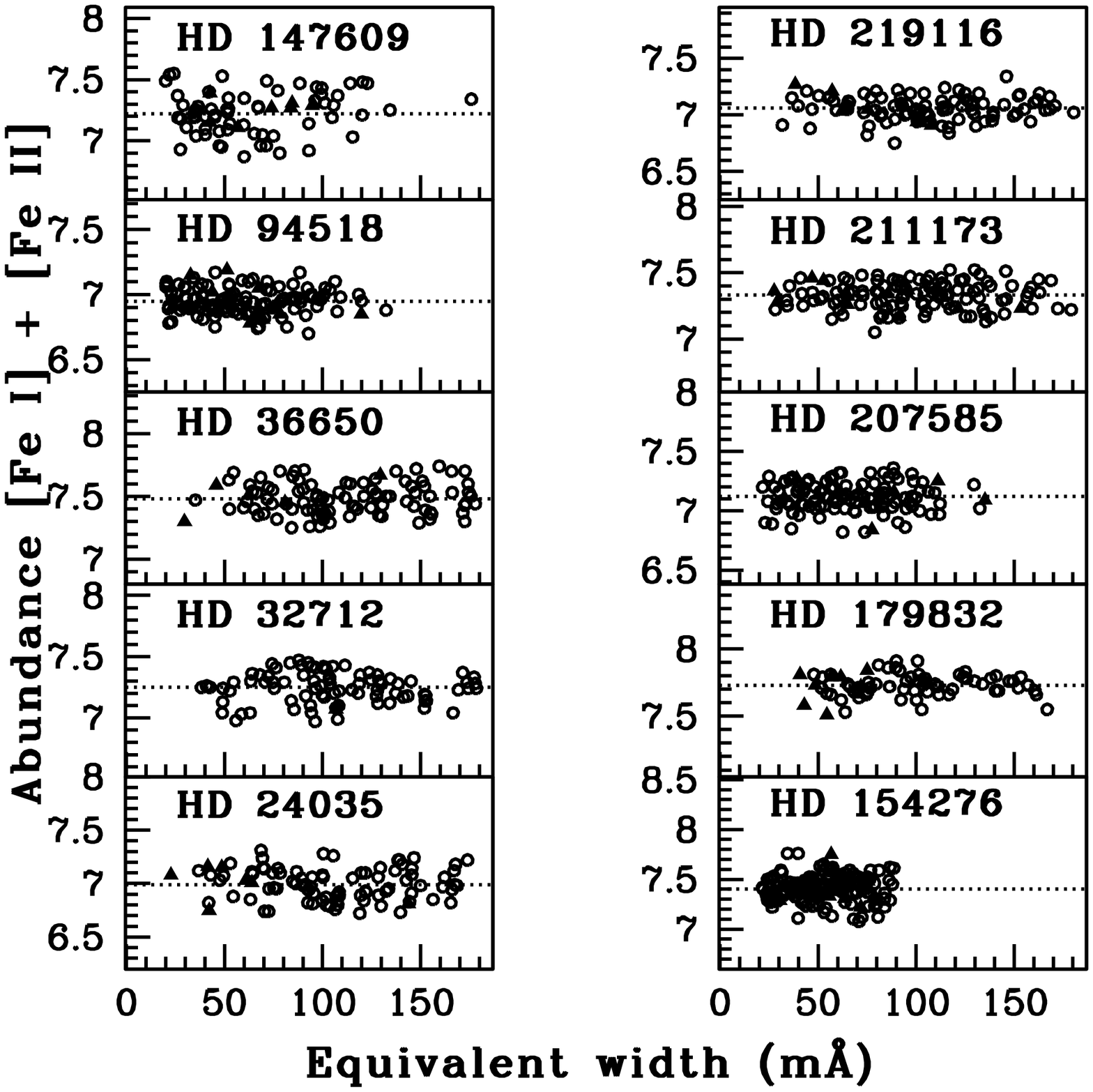}
\includegraphics[width=\columnwidth]{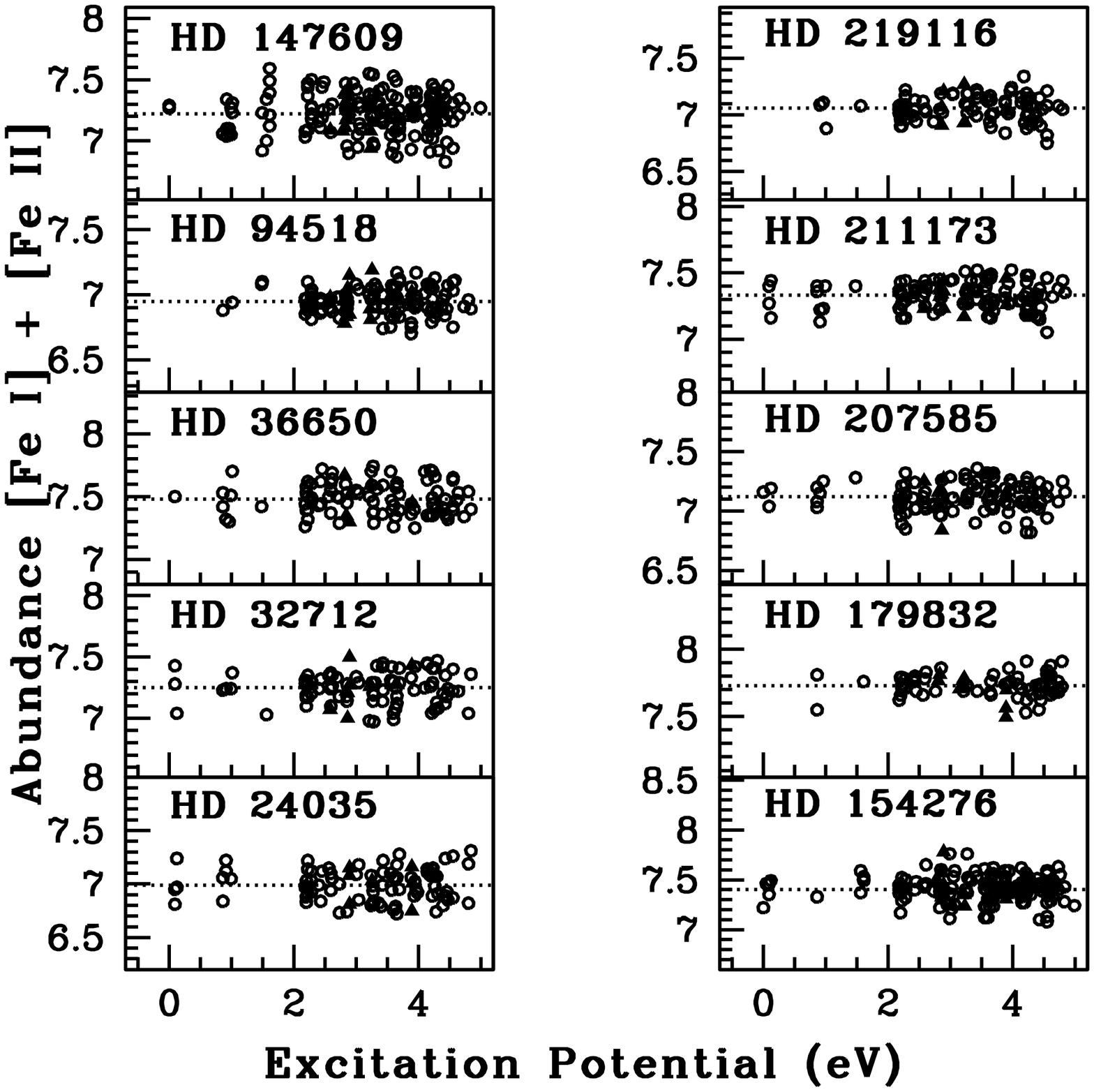}
\caption{Iron abundances of the program stars for individual Fe I and Fe II 
lines as a function of Excitation potential (lower panel) and equivalent width (upper panel). 
Open circles correspond to Fe I and solid 
triangles correspond to Fe II lines. } \label{ep_ew}
\end{figure}

\begin{figure}
\centering
\includegraphics[width=0.8\columnwidth, height= 0.8\columnwidth]{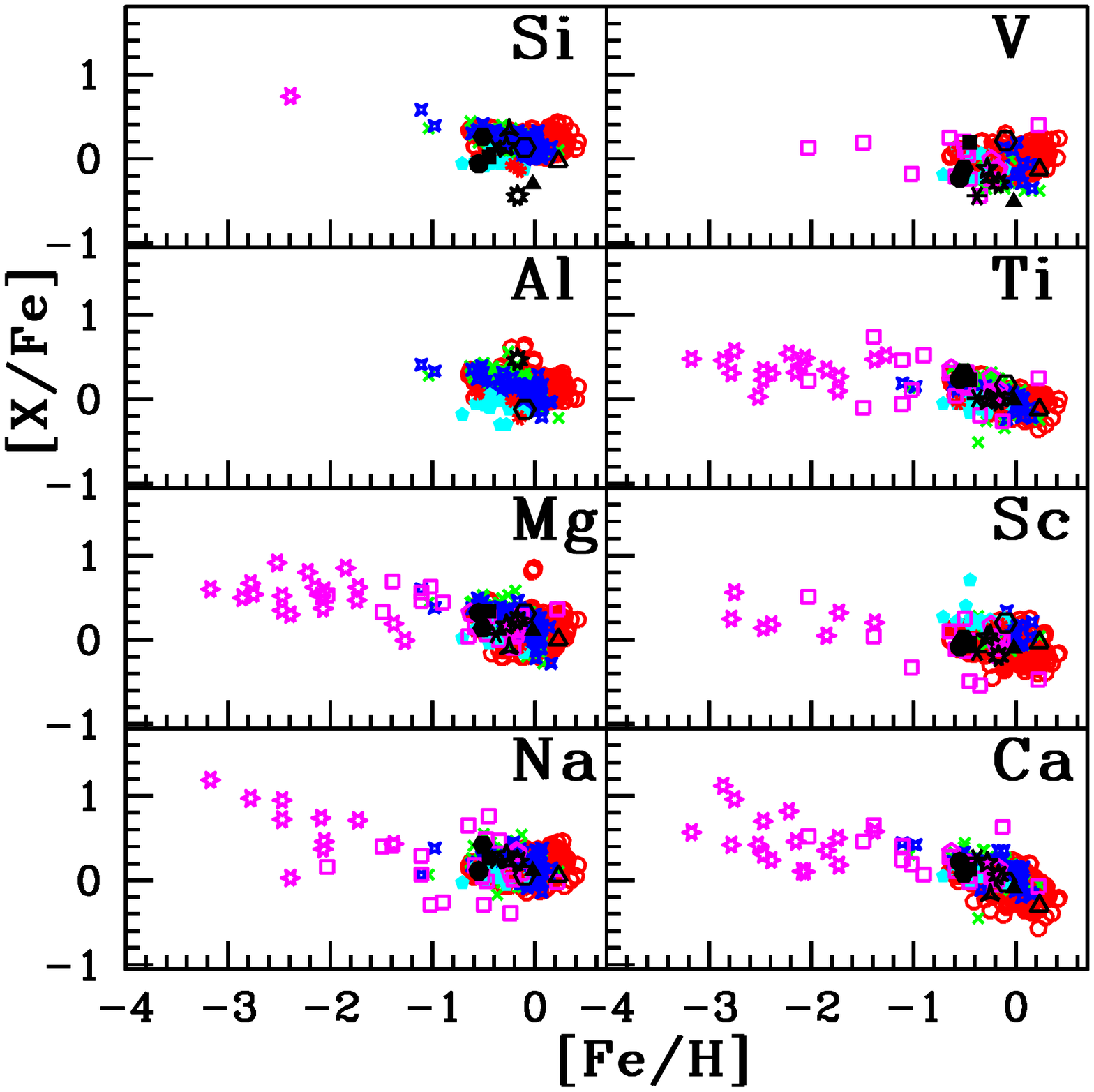}
\includegraphics[width=0.8\columnwidth, height= 0.8\columnwidth]{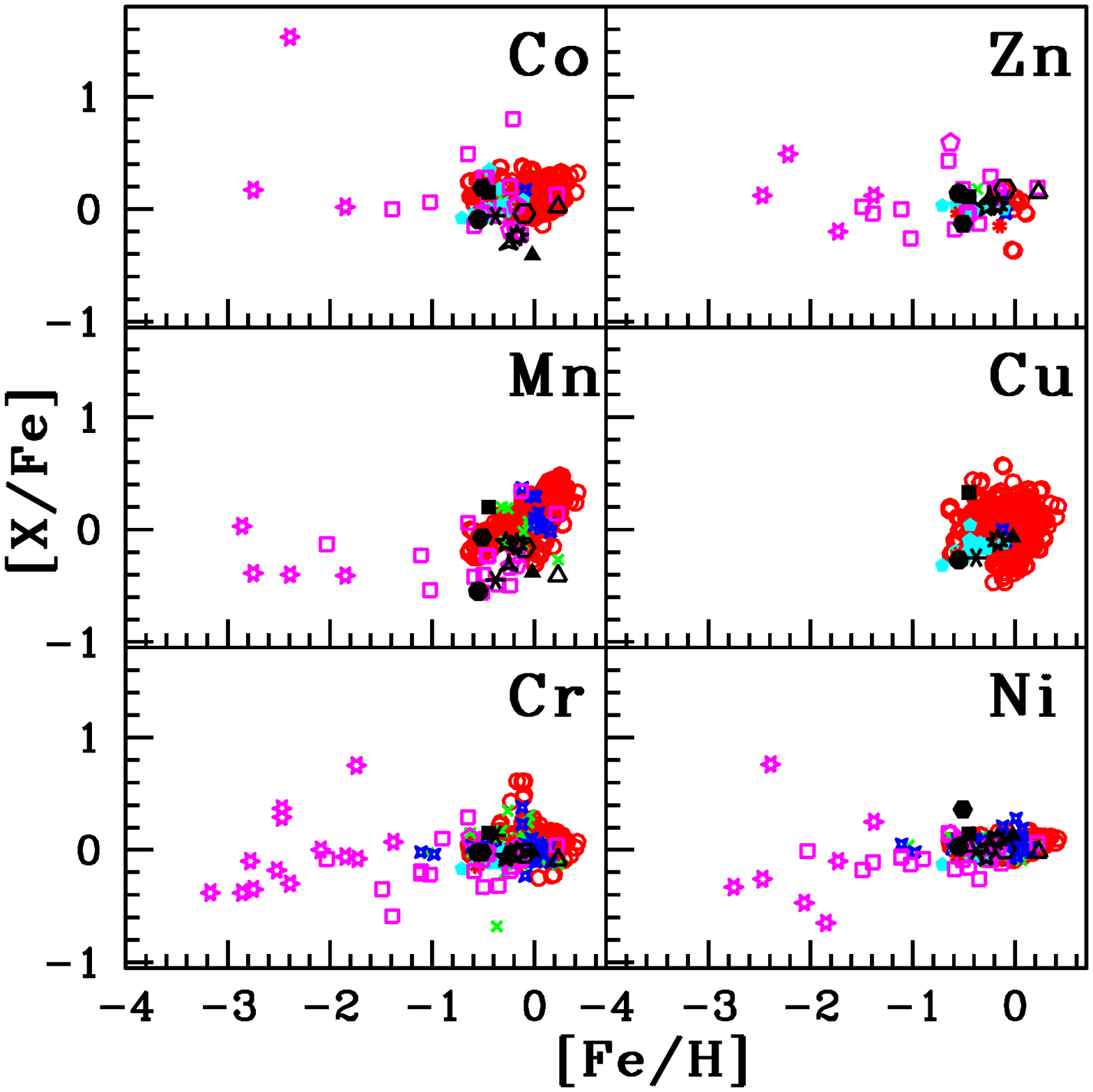}
\caption{Observed [X/Fe] ratios of the light elements in the  
program stars with respect to metallicity [Fe/H].   
Red open circles represent normal giants from literature (Luck \& Heiter 2007).
Green crosses, blue four-sided stars, cyan filled pentagons, red 
eight-sided crosses represent strong Ba giants, weak Ba giants, 
Ba dwarfs, Ba sub-giants respectively from literature (de Castro 
et al. 2016, Yang et al. 2016, Allen \& Barbuy 2006a). 
Magenta starred triangles represent CEMP stars from litertaure (Masseron et al. 2010). 
Magenta open squares and open pentagons represent CH giants and sub-giants respectively from literature 
(Karinkuzhi \& Goswami 2014, 2015, Goswami et al. 2006, 2016, Sneden \& Bond 1976, 
Vanture 1992, Goswami \& Aioki 2010, Jonsell et al. 2006, Masseron et al. 2010).  HD~24035 (filled hexagon), HD~32712 
(starred triangle), HD~36650 (filled triangle), HD~94518 (filled circle), 
HD~147609 (five-sided star), HD~154276 (open hexagon), HD~179832 
(open triangle), HD~207585 (six-sided cross), HD~211173 (nine-sided star) 
and HD~219116 (filled square).} \label{light_elements}
\end{figure}

\begin{figure}
\centering
\includegraphics[width=\columnwidth]{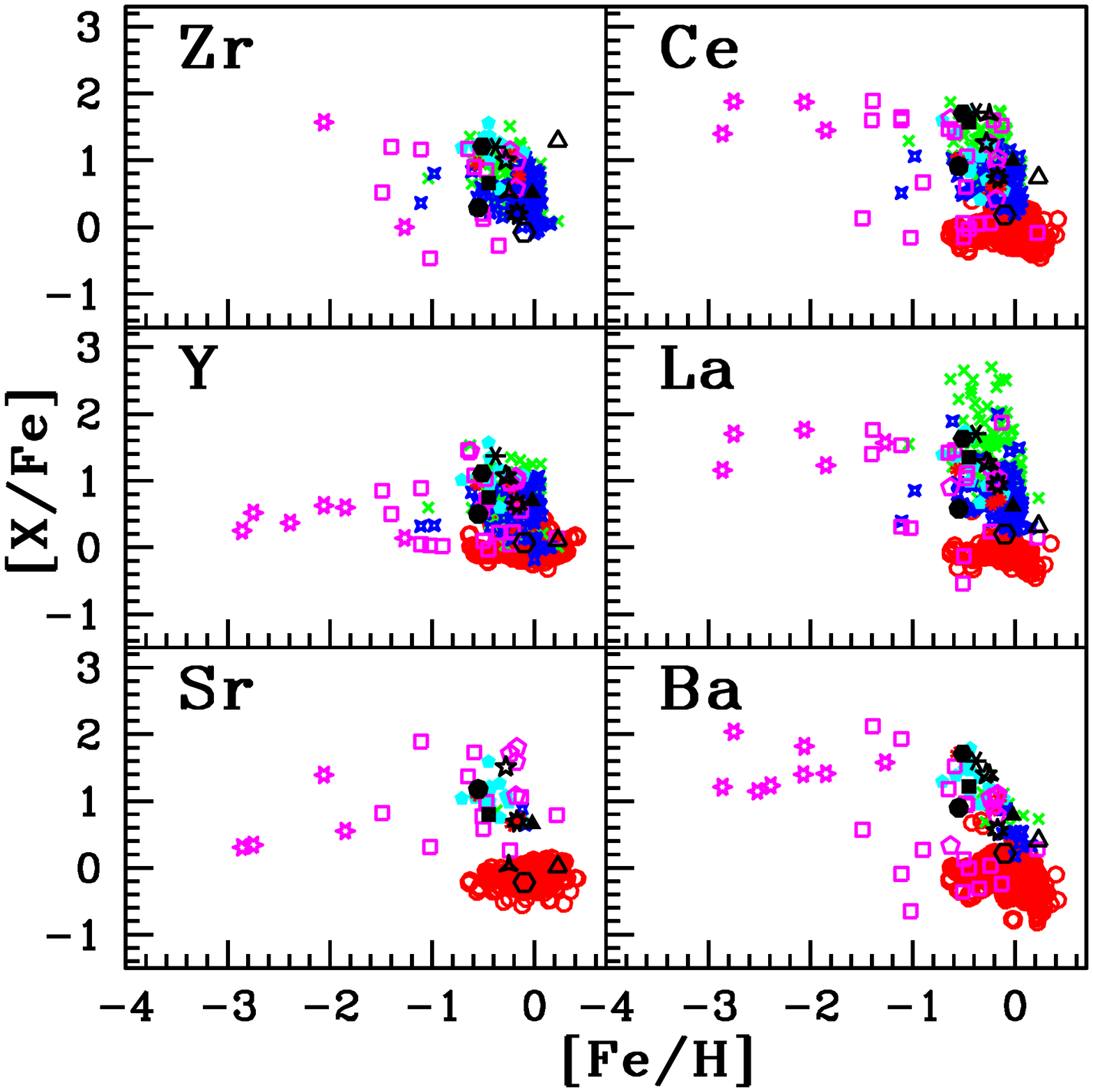}
\includegraphics[width=\columnwidth]{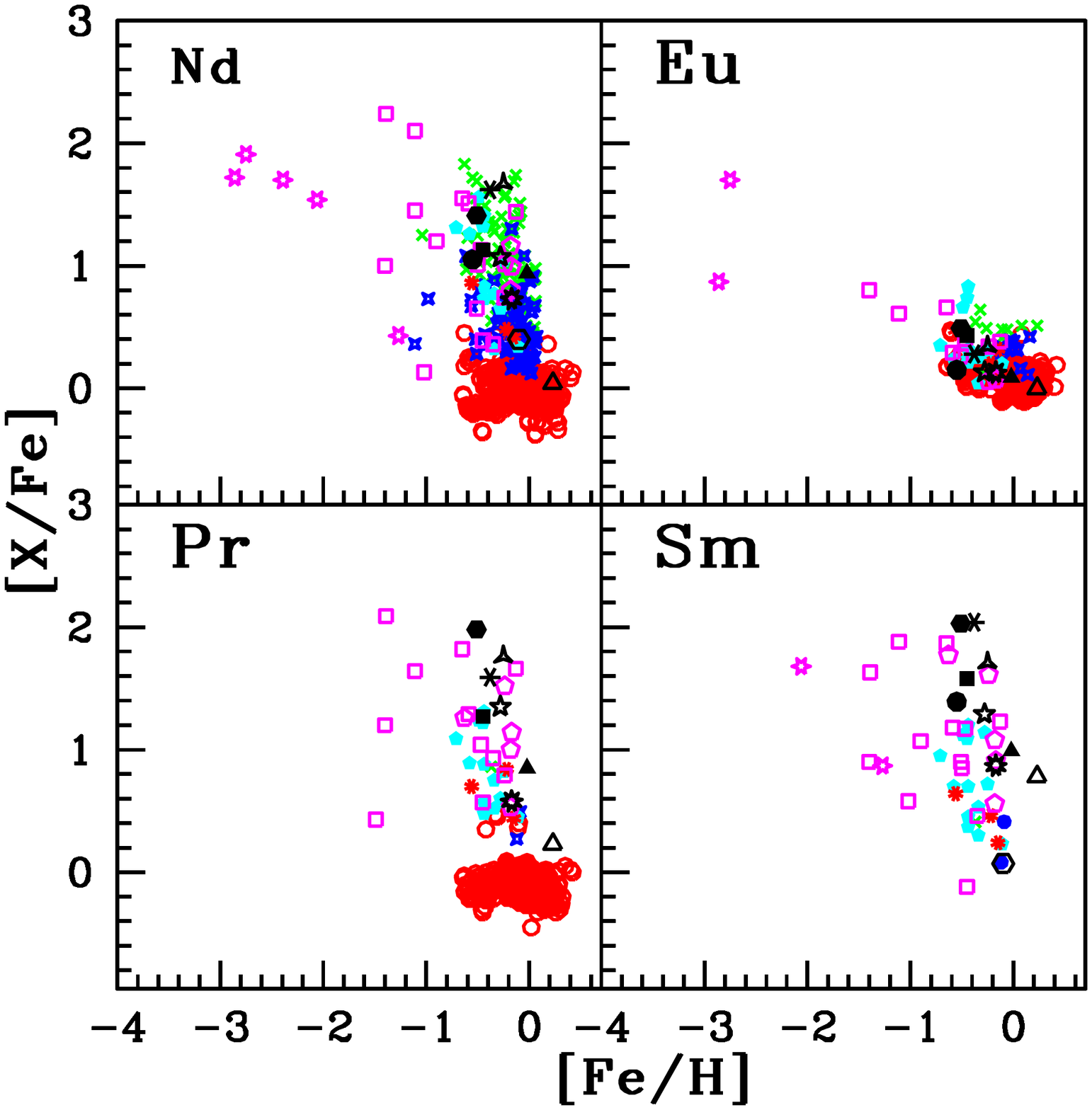}
\caption{\small{Observed [X/Fe] ratios of heavy elements in the 
program stars with respect to metallicity [Fe/H]. 
Symbols have same meaning as in Figure \ref{light_elements}.}}\label{heavy_elements}
\end{figure}

\end{document}